{% ****** Start of file apssamp.tex ******
\documentclass[superscriptaddress,
 amsmath,amssymb,
 aps,
 pre,
]{revtex4}
\usepackage{dcolumn}% Align table columns on decimal point
\usepackage{bm}% bold math
\usepackage{graphicx}
\usepackage{bbm}
\usepackage{hyperref}% add hypertext capabilities
\usepackage{comment}
\usepackage{tikz}
\usetikzlibrary{calc}
\usetikzlibrary{shapes}
\usetikzlibrary{calc,decorations.markings,decorations.pathmorphing}
\tikzset{cross/.style={cross out, draw=black, minimum size=2*(#1-\pgflinewidth), inner sep=0pt, outer sep=0pt},cross/.default={1pt}}

\newcommand{\tm}{t_{\rm m}}
\newcommand{\tmax}{t_{\rm m}}
\newcommand{\Pt}{P(t_{\rm m}|T)}
\newcommand{\PT}{P(t_{\rm m}|T)}
\newcommand{\erf}{\operatorname{erf}}

\begin{document}

%\preprint{}

\title{Time to reach the maximum for a stationary stochastic process}% Force line breaks with \\

\author{Francesco Mori}
\affiliation{LPTMS, CNRS, Universit\'e  Paris-Sud,  Universit\'e Paris-Saclay,  91405 Orsay,  France}

\author{Satya N. Majumdar}
\affiliation{LPTMS, CNRS, Universit\'e  Paris-Sud,  Universit\'e Paris-Saclay,  91405 Orsay,  France}

\author{Gr\'egory Schehr}
\affiliation{Sorbonne Universit\'e, Laboratoire de Physique Th\'eorique et Hautes Energies, CNRS, UMR 7589 4 Place Jussieu, 75252 Paris Cedex 05, France}

\date{\today}

\begin{abstract}
We consider a one-dimensional stationary time series of fixed duration $T$. We investigate the time $\tm$ at which the process reaches the global maximum within the time interval $[0,T]$. By using a path-decomposition technique, we compute the probability density function $\PT$ of $\tm$ for several processes, that are either at equilibrium (such as the Ornstein-Uhlenbeck process) or out of equilibrium (such as Brownian motion with stochastic resetting). We show that for equilibrium processes the distribution of $\PT$ is always symmetric around the midpoint $\tm=T/2$, as a consequence of the time-reversal symmetry. This property can be used to detect nonequilibrium fluctuations in stationary time series. Moreover, for a diffusive particle in a confining potential, we show that the scaled distribution $\PT$ becomes universal, i.e., independent of the details of the potential, at late times. This distribution $\PT$ becomes uniform in the ``bulk'' $1\ll\tm\ll T$ and has a nontrivial universal shape in the ``edge regimes'' $\tm\to0$ and $\tm\to T$. Some of these results have been announced in a recent Letter [Europhys. Lett. {\bf 135}, 30003 (2021)].
\end{abstract}

\maketitle

%\tableofcontents
\newpage

\section{Introduction}
\label{sec:intro}

\begin{figure}[t]
\includegraphics[scale=1]{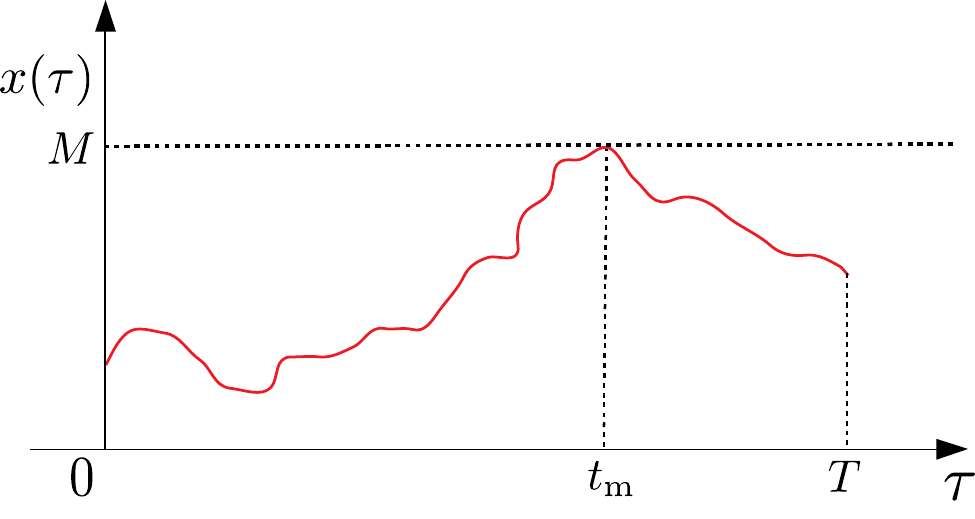}  
\caption{\label{fig:tmax_schem} Schematic representation of a stochastic process $x(\tau)$ as a function of $\tau$, for $0\leq \tau\leq T$. The global maximum $M=x(\tm)$ is reached at time $\tm$.}
\end{figure}

Understanding the statistical properties of the extremes of stochastic processes is a task of paramount importance in a wide range of contexts, including the physics of disordered systems \cite{D81,BBP07}, computer science \cite{KM00,MK02,MK03}, and evolutionary biology \cite{SBA98,KJ05}. During the last century, these properties have been investigated systematically within the field of Extreme Value Statistics (EVS) -- for a recent review, see \cite{MP20}. Given a one-dimensional time series $x(\tau)$, where $0\leq \tau\leq T$ indicates time, one of the central quantities in EVS is the global maximum $M$ of the process up to time $T$, defined as
\begin{equation}
M=\max_{0\leq \tau\leq T}x(\tau)\,.
\end{equation}
A schematic representation of a stochastic process $x(\tau)$ is shown in Fig.~\ref{fig:tmax_schem}, where the global maximum $M$ is highlighted.

Even though computing the distribution of $M$ is generally quite nontrivial, a few exactly solvable cases exist. In particular, one of the fundamental results in EVS deals with the case where the positions of the process at different times are independent and identically distributed (i.i.d.) random variables (meaning that $x(\tau)$ and $x(\tau')$ are i.i.d. if $\tau\neq \tau'$). In this i.i.d. case, one can show that for large $T$ the distribution of $M$ always belongs to one of three universality classes, independently of the specific distribution of the random variables $x(\tau)$ \cite{Gumbel_book}. This universal result can also be extended to the case where the process $x(\tau)$ is weakly correlated, meaning that the autocorrelation function of $x(\tau)$ decays exponentially in time as
\begin{equation}
\langle x(\tau) x(\tau')\rangle -\langle x(\tau)  \rangle\langle x(\tau')\rangle \sim f\left(\frac{|\tau-\tau'|}{\xi}\right)\,,
\label{autocorrelation_function}
\end{equation}
where $\xi$ is the correlation time of the process and $f(z)$ decays faster than any power law for large $z$. Indeed, using a ``block renormalization'' argument, one can still apply the same universal result as for i.i.d. variables when $T\gg \xi$ \cite{MP20}.

Even though in many cases one is interested in the magnitude $M$ of the maximum, an equally important observable is the time $\tm$ at which the maximum is attained (see Fig.~\ref{fig:tmax_schem}). Indeed, determining the time at which a time series will reach its global maximum is relevant in many different situations, from finance \cite{DW80,BC04,RM07,MB08} to sports \cite{CK15}. For instance, the time $\tm$ at which a stock price in the financial market reaches its global maximum within a fixed time window $T$ (e.g., a trading day) is a quantity of clear practical importance. The distribution $\PT$ of the time $\tm$ of the maximum has been investigated for a wide range of processes. For instance, when the variables $x(\tau)$ for $0\leq \tau\leq T$ are i.i.d. it is easy to show that the distribution of $\tm$ is uniformly distributed in the interval $[0,T]$, i.e., that
\begin{equation}
\PT=\frac1T\,,
\label{uniform_mead_intro}
\end{equation}
for $0\leq\tm\leq T$.

When correlations are present the probability density function (PDF) $\PT$ is usually more complicated. For instance, in the paradigmatic case of an overdamped Brownian motion (BM) in one dimension the distribution of $\tm$ was first computed by L\'evy, who showed that \cite{Levy,Feller,SA53}
\begin{equation}
\PT=\frac{1}{\pi\sqrt{\tm(T-\tm)}}\,.
\label{arcsine_intro}
\end{equation} 
Since the corresponding cumulative distribution reads
\begin{equation}
P(\tm\leq t|T)=\int_{0}^{t}d\tm\PT=\frac{2}{\pi}\sin^{-1}\left(\sqrt{\frac{\pi}{T}}\right)\,,
\end{equation}
this distribution is known as L\'evy's arcsine law. More recently, the distribution $\PT$ has been studied for several generalizations of BM, including constrained BM \cite{She79,B03,RFM07,MB08,MRK08,SLD10,MY10,MMS19,MLD20}, BM with stochastic resetting starting from the origin \cite{SP21,MMSS21}, BM with drift \cite{She79,B03,MB08}, fractional BM \cite{DW16,SDW18}, Bessel process \cite{SLD10}, L\'evy flights \cite{SA53,M10}, random acceleration process \cite{MRZ10}, and heterogeneous diffusion \cite{S22}. The distribution of $\tm$ has also been studied in the case of run-and-tumble particles (RTP) \cite{SK19,MLD20a,MLD20} and for $N$ vicious walkers \cite{RS11}. Moreover, the distribution of the time of the maximum plays a central role for computing the mean area of the convex hull of a two-dimensional process \cite{RMC09,MCR10,DMR13,HMSS20,MMSS21,SKMS22} and for determining the hitting probability for anomalous diffusion processes \cite{MRZ10}. However, to the best of our knowledge, before our recent Letter \cite{MMS21}, the time of the maximum was never systematically investigated in the case of {\it stationary} stochastic processes.

Stationary processes, i.e., stochastic processes that are invariant under a time shift, can be observed at very different scales in nature, from Brownian motors inside the cell \cite{D97} to climate systems \cite{WFM20}. A fundamental step in characterizing a stationary system is to determine whether it is at equilibrium or out of equilibrium. In addition to being stationary, equilibrium processes satisfy a stronger condition, namely detailed balance, which requires all probability currents in phase space to vanish. As a consequence of the detailed balance condition, equilibrium processes are also invariant under time-reversal symmetry and their physical properties are generally well-understood within the framework of statistical physics. On the other hand, nonequilibrium processes are characterized by probability currents in the steady state. Moreover, even though in recent years several general results have been derived concerning the fluctuations in out-of-equilibrium systems \cite{jarzynski,K98,C1999,seifert05,seifert12,HG20}, it still remains challenging to characterize precisely the statistical properties of these fluctuations. For this reason, several techniques for detecting nonequilibrium fluctuations in steady states have been developed -- for a review see \cite{GMG18}.

\begin{figure}[t]
\includegraphics[scale=1]{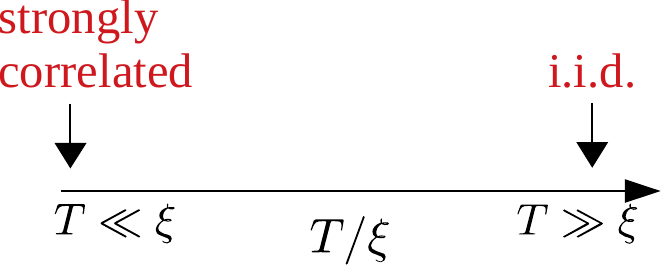} \caption{\label{fig:cross} The behavior of stationary stochastic processes of total duration $T$ and correlation time $\xi$ is controlled by the dimensionless parameter $T/\xi$. When $T\ll \xi$ the process is strongly correlated, while one can map the process into a collection of i.i.d.~variables for $T\gg\xi$. }
\end{figure}

 Notably, the case in which the autocorrelation function of the process decays over a typical timescale $\xi$ (as in Eq.~\eqref{autocorrelation_function}) has been recently investigated in the context of EVS \cite{EM_2011,MP20,MMSS21,MMSS21b}. In particular, both for the Ornstein-Uhlenbeck process \cite{MP20} and BM with resetting \cite{EM_2011,MP20,MMSS21}, it has been shown that the distribution of the maximum $M$, when properly rescaled, converges to the universal Gumbel form for $T\gg \xi$, where the correlation timescale $\xi$ depends on the details of the process. A similar late-time universality has also been observed for the record statistics of random walks with resetting \cite{MMSS21b}. The reason for these universal results is that, when $T\gg \xi$, one can apply a block renormalization argument which reduces the system to a collection of i.i.d.~variables (for the details of this argument, see \cite{MP20}). On the other hand, when $T\ll \xi$ the process is strongly correlated and one cannot apply the universal results, valid for i.i.d.~variables. Therefore changing the observation time $T$, the process interpolates between a strongly correlated state (for $T\ll\xi$) and an independent state (for $T\gg\xi$), as summarized in Fig.~\ref{fig:cross}. For this reason, stationary processes provide a natural laboratory to investigate the role of correlations in EVS.

Since the time $\tm$ is one of the central quantities in EVS, it is natural to ask whether the universality at late times also applies to the distribution $\PT$. Note that, even for short times, one could naively argue that for a stationary process the distribution $\PT$ should be given by the uniform measure in Eq.~\eqref{uniform_mead_intro}, as a consequence of the time-translational invariance. Interestingly, this is not the case due to the time-correlations of the process, as shown in \cite{MMS21}. Nevertheless, one expects the time-correlations of the process to become negligible for $T\gg \xi$, leading to the uniform distribution in Eq.~\eqref{uniform_mead_intro}. In our recent Letter \cite{MMS21}, using a path-decomposition technique, we have shown that this is the case only in the ``bulk'' of the distribution $\PT$, i.e., for $\xi\ll\tm\ll (T-\xi)$. In the ``edge regimes'' for $\tm\to 0$ and $\tm \to T$ the distribution $\PT$ strongly deviates from the uniform distribution. Moreover, we have also shown that for a large class of equilibrium processes the full distribution $\PT$, including the edge regimes, becomes universal at late times. These results were recently announced, albeit without any further details, in \cite{MMS21}. The present paper provides a detailed description of these derivations, which we believe could be useful to investigate other problems in EVS.

In this paper we study the distribution of $\PT$ for several processes, both in and out of equilibrium. It turns out that computing $\PT$ analytically is very hard, except for a 
handful of processes. In this paper, we present analytical solutions of $\PT$ for two equilibrium processes corresponding to an overdamped BM in a confining potential, 
respectively, of the form (i) $V(x)= \alpha |x|$ and (ii) $V(x)=\alpha x^2$ (the latter is the standard Ornstein-Uhlenbeck process). Similarly, we also obtain analytical solutions 
of $\PT$ for two very different out-of-equilibrium processes: (iii) a resetting Brownian motion (RBM) in one dimension and (iv) an RTP moving in the presence 
of a confining potential $V(x)= \mu |x|$. In the case of RBM, previous results for $\PT$ were known only for the case when the particle starts initially at a fixed position $x_0=0$, namely at the origin \cite{MMSS21,SP21}. In contrast, we show in this paper that when the initial position $x_0$ is sampled from the stationary distribution, 
$\PT$ is considerably different and is harder to compute. In addition to these four cases, we also study several other examples using numerical simulations and we highlight 
some universal properties of $\PT$, in particular at the edges when $\tm\to 0$ or $\tm\to T$. We then provide a block renormalization group argument to compute some of these 
universal edge scaling functions. Finally, we also provide a rather general sufficiency test to decide whether a given stationary process is out of equilibrium without having any 
a priori knowledge of its underlying dynamics. This test turns out to be incredibly simple: if $\PT$ turns out to be asymmetric around $T/2$ (either from simulations or analytical 
computations), the underlying process is surely out of equilibrium. If $\PT$ turns out to be symmetric around $T/2$, the test is inconclusive.

The rest of the paper is organized as follows. In Section \ref{sec:summary}, we provide a summary of our main results. In Section \ref{sec:eq}, we investigate the time $\tm$ of the maximum in the case of equilibrium processes. We consider the paradigmatic model of an overdamped Brownian particle in one dimension subject to an external potential $V(x)$ such that $V(x)\approx\alpha|x|^p$ for large $|x|$. Using a path-decomposition technique, we derive an exact result in the cases $V(x)=\alpha |x|$ (subsection \ref{sec:p1}) and $V(x)=\alpha x^2$ -- corresponding to the Ornstein-Uhlenbeck process (subsection \ref{sec:p2}). Moreover, in subsection \ref{sec:univ} we show that for $p>0$ the distribution of $\tm$ becomes universal at late times. In Section \ref{sec:neq}, we investigate the distribution of $\tm$ for nonequilibrium processes, including RBM (subsection \ref{sec:res_BM}) and a confined RTP (subsection \ref{sec:rtp}). In addition, in subsection \ref{sec:criterion}, we formulate a simple criterion, based on the estimation of the distribution $\PT$, to detect nonequilibrium fluctuations in steady states. Finally, in Section \ref{sec:conclusion}, we conclude with a summary and we discuss possible perspectives. Some details of the computations are presented in the appendices.

\section{Models and summary of the main results}

\label{sec:summary}

Since the paper is rather long, we provide a concise description of the models and a summary of our main results, so that the main mathematical formulae can be easily retrieved without a detailed search in the main body of the paper. We consider a one-dimensional stationary process $x(\tau)$ for $0\leq \tau\leq T$. We assume that at time $\tau=0$, the process has already reached a steady state. This is equivalent to assuming that the system is initialized at some arbitrary state at time $\tau=-\infty$ and that we start to observe it at time $\tau=0$. Our goal is to compute analytically the distribution $\PT$ of the time $\tm$ at which the process reaches its global maximum up to time $T$. Note that the domain of $\tm$ is the time interval $[0,T]$. We consider different stochastic models, both at equilibrium and out-of-equilibrium.

\subsection{Equilibrium processes}

We consider a class of equilibrium processes corresponding to an overdamped Brownian particle in a confining potential $V(x)$, such that $V(x)\approx\alpha |x|^p$, with $\alpha>0$ and $p>0$. The position $x(\tau)$ of the process evolves according to the Langevin equation
\begin{equation}
\frac{dx(\tau)}{d\tau}=-V'(x)+\sqrt{2D}\eta(\tau)\,,
\end{equation}
where $\eta(\tau)$ is a Gaussian white noise with zero mean and correlator $\langle \eta(\tau)\eta(\tau')\rangle=2D\delta(\tau-\tau')$, $D>0$ is the diffusion constant, and $V'(x)=dV(x)/dx$. The equilibrium stationary state of this process is given by the Boltzmann weight $P_{\rm st}(x_0)\propto e^{-V(x)/D}$. Computing the full distribution $\PT$ for any $p>0$ is challenging. Nevertheless, we are able to calculate this quantity in two exactly solvable cases.

\subsubsection{The case $p=1$}
In the case $V(x)=\alpha |x|$, we show that
\begin{equation}
P(\tm|T)=\frac{\alpha^2}{4D}F_1\left(\frac{\alpha^2}{4D}\tm,\frac{\alpha^2}{4D}(T-\tm)\right)\,,\label{scaling_p1_summary}
\end{equation}
where the double Laplace transform of $F_1(T_1,T_2)$ is given by
\begin{eqnarray}
\label{LT_scaling_p1_summary}
&&\int_{0}^{\infty}dT_1 e^{-s_1 T_1} \int_{0}^{\infty}dT_2 e^{-s_2 T_2} F_1(T_1,T_2)\\&=& \frac{1}{2(1+\sqrt{1+s_1})(1+\sqrt{1+s_2})}\Bigg[1+\int_{0}^{\infty}dz\,e^{-z}\frac{\left(\sqrt{1+s_1}+1-e^{-\sqrt{1+s_1}z}\right)\left(\sqrt{1+s_2}+1-e^{-\sqrt{1+s_2}z}\right)}{\left(\sqrt{1+s_1}-1+e^{-\sqrt{1+s_1}z}\right)\left(\sqrt{1+s_2}-1+e^{-\sqrt{1+s_2}z}\right)}\Bigg]\,.\nonumber
\end{eqnarray}
Inverting this double Laplace transform is highly nontrivial. Nevertheless, from this expression it is easy to check that $\PT$ is symmetric around the midpoint $\tm=T/2$, i.e., that $P(\tm|T)=P(T-\tm|T)$. This implies that the first moment of $\tm$ is simply given by $\langle\tm \rangle=T/2$. Interestingly, this property, which is a consequence of the time-reversal symmetry, is valid for any equilibrium process and is confirmed by numerical simulations (see Fig.~\ref{fig:comparis_eq-neq}{\bf a}). This observation will lead us to formulate the criterion discussed below to decide whether or not a stationary time series is at equilibrium.

In addition, from the expression in Eq.~\eqref{LT_scaling_p1_summary}, it is possible to extract the asymptotic behavior of $\PT$ for small and large $T$. When $T\ll \xi$, where $\xi=(4D)/\alpha^2$ is the correlation time of the process, we find
\begin{equation}
\PT\approx\frac{1}{\pi\sqrt{\tm(T-\tm)}}\,,
\end{equation}
which corresponds to the arcsine law, valid for free BM (see Eq.~\eqref{arcsine_intro}). Thus, for short times the process is strongly correlated and behaves as a BM. On the other hand, in the late-time regime $T\gg\xi$, we find
\begin{equation}
\PT\approx    \begin{cases}\frac1T 
G\left(\frac{\alpha^2}{4D}\tm\right)\quad &\text{ for }\quad\tm\lesssim 4D/\alpha^2\,,\\
\\
\frac1T \quad &\text{ for }\quad 4D/\alpha^2\ll \tm \ll T-4D/\alpha^2\,,\\
\\
\frac1T G\left[\frac{\alpha^2}{4D}(T-\tm)\right]\quad &\text{ for }\quad\tm\gtrsim T- 4D/\alpha^2\,,\\
\end{cases}
\label{intro_PT_asymptotics_1}
\end{equation}
where
\begin{equation}
G(z)=\frac12 \left[1+\operatorname{erf}(\sqrt{z})+\frac{1}{\sqrt{\pi z}}e^{-z}\right]\,,
\label{G_summary}
\end{equation}
and $\operatorname{erf}(z)=(2/\sqrt{\pi})\int_{0}^{z}du~e^{-u^2}$. This function $G(z)$ has asymptotic behaviors
\begin{equation}
G(z)\approx    \begin{cases}
1/(2\sqrt{\pi z})\quad &\text{ for } z\to 0\,,\\
\\
1+e^{-z}/(4\sqrt{\pi}z^{3/2})\quad &\text{ for } z\to \infty\,.\\
\end{cases}
\label{PT_asymp_p1_summary}
\end{equation}
Thus, the PDF $\PT$ becomes constant in the bulk regime where $4D/\alpha^2\ll \tm \ll T-4D/\alpha^2$. The edge regimes $\tm\to0$ and $\tm\to T$ are instead described by the function $G(z)$ in Eq.~\eqref{G_summary}. In particular, $\PT$ diverges as $\sim 1/\sqrt{\tm}$ for $\tm \to 0$ and by symmetry as $\sim 1/\sqrt{T-\tm}$ for $\tm \to T$. The width of the edge regime is in this case $\mathcal{O}(1)$.

\subsubsection{The case $p=2$}

In the case $V(x)=\alpha x^2$, corresponding to the Ornstein-Uhlenbeck process, we obtain
\begin{equation}
\PT=\alpha F_{\rm OU}(\alpha\tm,\alpha(T-\tm))\,.
\label{summary_scaling_relation_OU}
\end{equation}
where
\begin{equation}
\int_{0}^{\infty}dT_1~\int_{0}^{\infty}dT_2~e^{-s_1 T_1-s_2 T_2}F_{\rm OU}(T_1,T_2)=\frac{1}{\sqrt{8\pi}}\int_{-\infty}^{\infty}dz~e^{-z^2/2}\frac{D_{-1-s_1/2}\left(-z\right)}{D_{-s_1/2}\left(-z\right)}\frac{D_{-1-s_2/2}(-z)}{D_{-s_2/2}(-z)}\,.
\label{summary_scaling_OU}
\end{equation}
Here, $D_p(z)$ is the parabolic cylinder function \cite{DLMF}. From this expression, we find that the distribution $\PT$ is symmetric around $\tm=T/2$, implying $\langle\tm\rangle=T/2$. This is in agreement with the fact that the process is at equilibrium.

The asymptotic behaviors of $\PT$ for short and late times are qualitatively similar to the ones we obtained for $p=1$ (see Eq.~\eqref{intro_PT_asymptotics_1}). In particular, in the short-time regime $T\ll\xi$, where $\xi=1/\alpha$ for this model, we find that the distribution $\PT$ approaches the arcsine law in Eq.~\eqref{arcsine_intro}. Thus, for short times the process is strongly correlated and we find that the distribution of $\tm$ approaches that of a BM. On the other hand, in the late-time regime $T\gg\xi$, we obtain
\begin{equation}
\PT\approx    \begin{cases}\frac1T 
G\left(\alpha \ln(T)~\tm\right)\quad &\text{ for }\quad\tm\lesssim 1/(\alpha \ln(T))\,,\\
\\
\frac1T \quad &\text{ for }\quad 1/(\alpha \ln(T))\ll \tm \ll T-1/(\alpha \ln(T))\,,\\
\\
\frac1T G\left(\alpha \ln(T)~(T-\tm)\right)\quad &\text{ for }\quad\tm\lesssim 1/(\alpha \ln(T))\,,\\
\end{cases}
\label{PT_asymp_p2_summary}
\end{equation}
where the function $G(z)$ is given again in Eq.~\eqref{G_summary}. Interestingly, we find that the late-time behavior of $\PT$ is the same for $p=1$ and $p=2$. The only difference is the width of the edge regimes, which is $\mathcal{O}(1)$ for $p=1$ and $\mathcal{O}(1/\ln(T))$ for $p=2$. This result is quite unexpected and led us to ask whether this universality extends to any $p>0$.

\subsubsection{Universality at late times}

\begin{figure*}[t]
\includegraphics[scale=0.7]{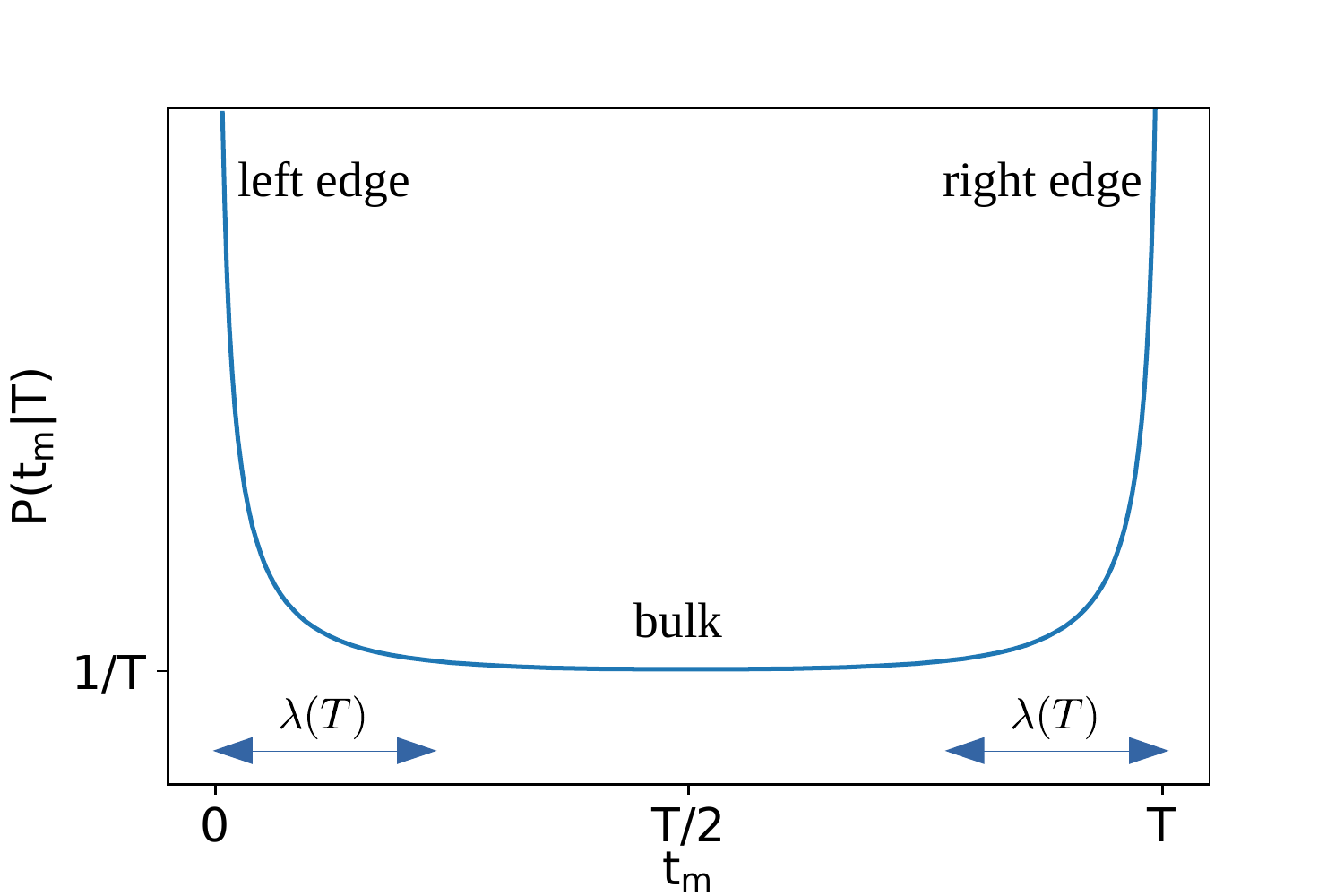} 
\caption{\label{fig:ptm_asymp} Schematic representation of the late-time distribution $\PT$ as a function of $\tm$ for Brownian motion in a potential $V(x)=\alpha |x|^p$ with diffusion constant $D$. The blue curve represents the universal result in Eq.~\eqref{universal_G_summary}. The distribution is flat in the bulk regime $\lambda(T)\ll \tm \ll (T-\lambda(T))$, while it diverges in the edge regimes $\tm\to 0$ and $\tm \to T$. The width of the edge regimes is $\lambda(T)$, given in Eq.~\eqref{lambda_summary}. Since the process is at equilibrium, the distribution is symmetric around the midpoint $\tm=T/2$, i.e., $\PT=P(T-\tm|T)$.}
\end{figure*}

Using a ``block renormalization'' argument, we indeed show that the late time behavior of $\PT$ is universal for any $p>0$. In particular,  for $T\gg\xi$, we find that
\begin{equation}
P(\tm|T)\approx
\begin{cases}
\frac{1}{T}G\left(\frac{\tm}{\lambda(T)}\right) &~~\text{  for   }~~ \tm\lesssim \lambda(T)\\
\\
\frac1T &~~\text{  for   } ~~\lambda(T)\ll \tm\ll T- \lambda(T)\\
\\
\frac{1}{T}G\left(\frac{T-\tm}{\lambda(t)}\right) &~~\text{  for   } ~~ \tm\gtrsim T- \lambda(T)
\,,
\end{cases}
\label{universal_G_summary}
\end{equation}
where $G(z)$ is given in Eq.~\eqref{G_summary} and
\begin{equation}
\lambda(T)=\frac{4D}{\alpha^2 p^2}\left(\frac{D}{\alpha}\ln(T)\right)^{-2(p-1)/p}\,.
\label{lambda_summary}
\end{equation}
Interestingly, at late times the distribution $\PT$, once appropriately scaled, becomes completely universal, i.e., independent of the specific details of the model. Note that the model parameters $\alpha$ and $p$ appear in the expression of $\PT$ only through the width $\lambda(T)$ of the edge regime. This quantity is constant with increasing $T$ (for large $T$) for $p=1$, it shrinks as $\ln(T)^{-2(p-1)/p}$ for $p>1$, while it grows as $\ln(T)^{2(1-p)/p}$ for $0<p<1$. Setting $p=1$ or $p=2$ in Eq.~\eqref{universal_G_summary}, we recover the results in Eqs.~\eqref{PT_asymp_p1_summary} and \eqref{PT_asymp_p2_summary}.

\subsection{Out-of-equilibrium processes} 

We also investigate the distribution of the time of the maximum in the case of out-of-equilibrium stationary processes. In this case, the system does not satisfy the time-reversal symmetry. Consequently, the distribution $\PT$ is generally not symmetric around the midpoint $\tm=T/2$. The two exactly solvable processes that we consider are a single RBM and a confined RTP.

\subsubsection{Resetting Brownian motion}

\begin{figure*}[t]\includegraphics[scale=0.5]{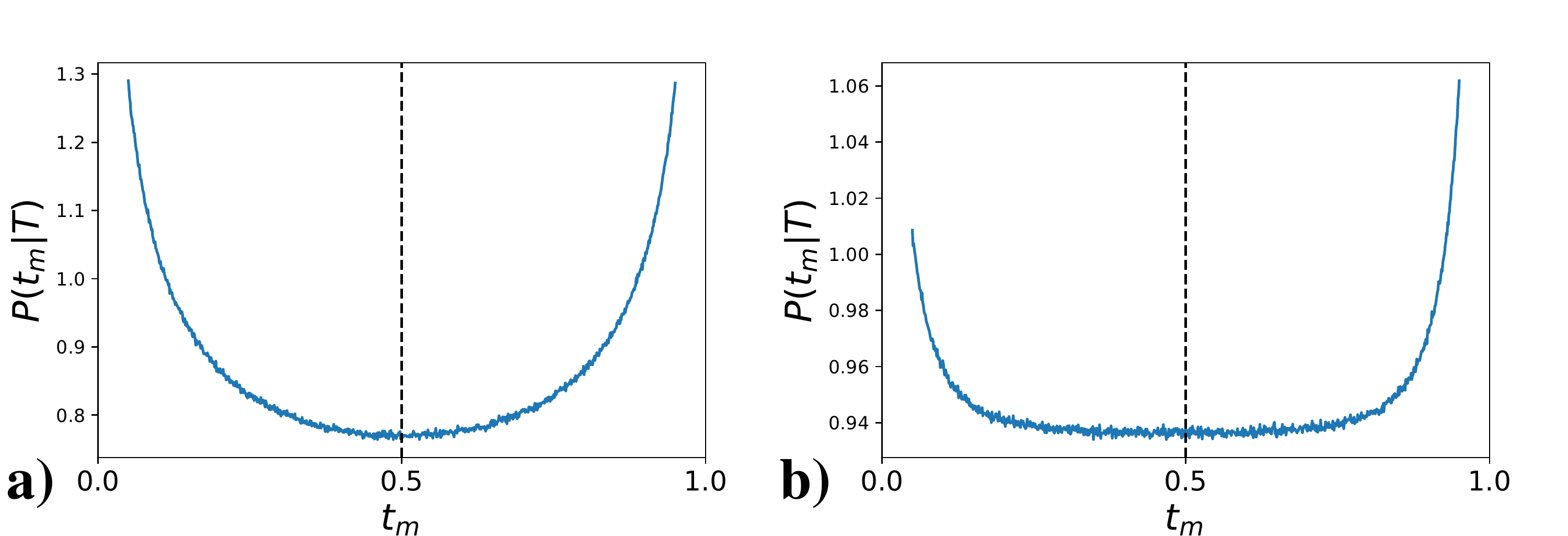} 
\caption{\label{fig:comparis_eq-neq} {\bf a)} Probability density function $\PT$ as a function of the time $\tm$ of the maximum for the Ornstein-Uhlenbeck process of duration $T=1$, with $\alpha=D=1$. The curve is symmetric around the midpoint $\tm=T/2$ (vertical dashed line). {\bf b)} Probability density function $\PT$ versus $\tm $ for Brownian motion with stochastic resetting, obtained from numerical simulations with $D=T=1$ and $r=10$. The curve is not symmetric around the midpoint $\tm=T/2$ (see also Fig.~\ref{fig:avg_tmax}). }
\end{figure*}

We consider a Brownian particle with diffusion coefficient $D$. The particle is reset to the origin $x=0$ at a constant rate $r$. In other words, in a small time interval $dt$, the position $x(\tau)$ of the particle evolves according to \cite{EM_2011,EMS20}
\begin{equation}
x(t+dt)=\begin{cases}
x(t)+\sqrt{2D}\eta(t)dt&\quad\text{ with probabilty }1-rdt\,,\\
\\
0&\quad\text{ with probabilty }rdt\,.\\
\end{cases}
\label{resetting_rule_1}
\end{equation}
The resetting process admits the following nonequilibrium steady state \cite{EM_2011}
\begin{equation}
P_{\rm st}(x_0)=\frac{1}{2}\sqrt{\frac{r}{D}}\exp\left(-\sqrt{\frac{r}{D}}|x_0|\right)\,.
\label{NEQSS_RES}
\end{equation}
Note that the detailed balance is manifestly violated by the resetting move in Eq.~\eqref{resetting_rule_1} that induces a nonzero current to the resetting point $x=0$, even
in the stationary state. Consequently, the RBM is a nonequilibrium process.

Using a path-decomposition technique, we show that $\PT$ has the scaling form
\begin{equation}
\PT=rF_R(r\tm,r(T-\tm))\,,
\label{scaling_res_summary}
\end{equation}
where 
\begin{eqnarray}
\nonumber &&\int_{0}^{\infty}dT_1~e^{-s_1 T_1}\int_{0}^{\infty}dT_2~e^{-s_2 T_2}~F_R(T_1,T_2)=\frac{1}{2}  
 \frac{1}{(1+\sqrt{1+s_1})\sqrt{1+s_2}}\\&+&\frac12 \frac{\sqrt{1+s_2}}{\sqrt{1+s_1}-1}\int_{0}^{\infty}dz~e^{-(1+\sqrt{1+s_1})z}\frac{e^{z\sqrt{1+s_1}} s_1-\sqrt{1+s_1}+1} {\left(s_1+ e^{-z\sqrt{1+s_1}}\right)\left(s_2+ e^{-z\sqrt{1+s_2}}\right)}\,.
 \label{FR_LT_summary}
 \end{eqnarray}
 Interestingly, in this case we find that $P(\tm|T)\neq P(T-\tm|T)$, as a consequence of the nonequilibrium nature of the process. This is confirmed by numerical simulations (see Fig.~\ref{fig:comparis_eq-neq}{\bf b}). Thus, the first moment of $\tm$ deviates from the equilibrium value $T/2$. In particular, we find
\begin{equation}
\langle \tm\rangle =T f(rT)\,,
\end{equation} 
where $f(t)$ is given in Eq.~\eqref{foft} and is shown in Fig.~\ref{fig:avg_tmax}. We observe that $f(t)>1/2$ for any $t>0$. This function is nonmonotonous and has a maximum at $t^*\approx 2.218$ with $f(t^*)\approx0.519$. Note that this is different from the case where the RBM starts from a fixed position
$x_0$ in space, as investigated in \cite{MMSS21,SP21}. In contrast, in our case the initial position $x_0$ is sampled from the nonequilibrium steady state in Eq.~\eqref{NEQSS_RES}.

From Eq.~\eqref{FR_LT_summary} one can also extract the asymptotic behaviors of $\PT$. In the short-time regime $T\ll\xi$, where $\xi=1/r$ in this case, we obtain once again that the distribution $\PT$ approaches the arcsine law in Eq.~\eqref{arcsine_intro}. This is because for $T\ll 1/r$ the system typically does not reset and the process therefore reduces to a standard BM. On the other hand, for $T\gg\xi$, we get
\begin{equation}
\PT\approx\begin{cases}
\frac1T G(r\tm)\quad &\text{ for }\tm\ll1/r\,\\
\\
\frac1T\quad &\text{ for }1/r\ll\tm\ll(T-1/r)\\
\\
\frac1T\left[2 G(r(T-\tm))-1\right]\quad &\text{ for }\tm\ll1/r\,,\\
\end{cases}
\end{equation}
where $G(z)$ is given in Eq.~\eqref{G_summary} (see Fig.~\ref{fig:comparis_eq-neq}b). Interestingly, the late time behavior of $\PT$ is qualitatively similar to the one of the equilibrium processes described above. Note however that in this case the distribution $\PT$ is not symmetric around $\tm=T/2$. Indeed, for $\tm\to 0$ the PDF diverges as $1/(2T\sqrt{\pi \tm})$ while it diverges as $1/(T\sqrt{\pi (T-\tm)})$ for $\tm \to T$.

\subsubsection{Run-and-tumble particle in a potential $V(x)=\mu |x|$}

We consider a single RTP moving in a one-dimensional potential $V(x)=\mu |x|$. The state $(x(\tau),\sigma(\tau))$ of the system at time $\tau$ is specified by the position $x(\tau)$ of the particle and its direction $\sigma(\tau)=\pm1$. The position of the particle evolves according to the differential equation
\begin{equation}
\frac{dx(\tau)}{d\tau}=-V'(x)+v_0\sigma(t)=-\mu\operatorname{sign}(x)+v_0\sigma(t)\,,
\end{equation}
where $\operatorname{sign}(x)$ denotes the sign of $x$ and $v_0>\mu$ is the speed of the particle. The direction $\sigma(t)$ of the particle is flipped at a constant rate $\gamma$. As explained in Section \ref{sec:neq}, the persistent motion of the particle breaks the detailed balance condition and thus the system is out-of-equilibrium. In the steady state, the probability of finding the particle at $x_0$ with a positive (negative) velocity is given by \cite{SKM_19}
\begin{equation}
P_{\rm st}^{\pm }(x_0)=\frac12 \left(1\pm \frac{\mu}{v_0}\operatorname{sign}(x)\right)\frac{\gamma~\mu}{v_0^2-\mu^2}\exp\left(-\frac{2\gamma\mu}{v_0^2-\mu^2}|x_0|\right)\,.
\label{joint_distribution_RTP}
\end{equation}
We assume that at the initial time $\tau=0$ the position $x_0$  (with a positive/negative velocity) is drawn from this distribution in Eq.~\eqref{joint_distribution_RTP}. Our goal is to compute the distribution of the time $\tm$ at which the position of the particle becomes maximal. 

We show that the distribution of $\tm$ can be written as 
\begin{equation}
\PT=P_0(T)\delta(\tm)+P_{\rm bulk}(\tm|T)+P_1(T)\delta(\tm-T)\,,
\end{equation}
where $P_0(T)$, $P_{\rm bulk}(\tm|T)$ and $P_1(T)$ are given in Eqs.~\eqref{PO_LT}, \eqref{P1_LT}, and \eqref{eq:PDF_RTP_LT_bulk}. Interestingly, the events ``$\tm=0$'' and ``$\tm=T$'' occur with finite probability, as highlighted by the two $\delta$ functions in the expression above. The function $P_{\rm bulk}(\tm|T)$ has support in $0<\tm<T$. 

Interestingly, even though the system is nonequilibrium, the bulk of the distribution is still symmetric around the midpoint $\tm=T/2$, i.e., $P_{\rm bulk}(\tm|T)=P_{\rm bulk}(T-\tm|T)$. Nevertheless, the amplitudes $P_0(T)$ and $P_1(T)$ of the $\delta$ functions are different, meaning that the distribution $\PT$ is overall not symmetric around $\tm=T/2$.

\subsubsection{Criterion to detect nonequilibrium dynamics}

From the exact computations performed for different models of stationary processes, we have observed that, when the system is at equilibrium, the distribution of $\tm$ is symmetric around the midpoint $\tm=T/2$, i.e., $\PT=P(T-\tm|T)$. On the contrary the PDF $\PT$ does not satisfy this symmetry for the nonequilibrium processes described above. This observation turns out to be quite general. Indeed, we show that if the process is at equilibrium then necessarily the distribution of $\tm$ is symmetric around $\tm=T/2$. This property is a consequence of the time-reversal symmetry of equilibrium processes. On the other hand, for nonequilibrium processes, $\PT$ is typically not symmetric. Note, however, that there exist nonequilibrium systems for which $\PT$ is symmetric.

This result leads to a simple criterion to detect nonequilibrium fluctuations in stationary time series. Imagine that one has access to a long stationary time series $x(\tau)$ (for instance, this could result from some experimental measurement). Without knowing the specific details of the dynamics of the process, how can one determine whether or not the underlying system is nonequilibrium?

Building on the observation that if the distribution of $\tm$ is not symmetric then the process is necessarily nonequilibrium, we propose the following simple method. First, divide the time series into $N$ blocks each of duration $T$ (assuming that the total duration of the time series is much larger than $T$) and measure the time $\tm^i$ at which the maximum is reached within the $i$-th block. From this $N$ values $\tm^1\,,\ldots\,,\tm^N$ one can build the empirical PDF $\PT$, with $0\leq\tm\leq T$. If this distribution is not symmetric around the midpoint $\tm=T/2$ (as in Fig.~\ref{fig:comparis_eq-neq}{\bf b}), then one can conclude that the system is nonequilibrium. However, if $\PT$ turns out to be symmetric (as in Fig.~\ref{fig:comparis_eq-neq}{\bf a}) our test is inconclusive. This test can also be applied to systems composed of many interacting degrees of freedom. Indeed, finding that the distribution of $\tm$ for one of the variables describing the system is not symmetric is sufficient to conclude that the full system is out of equilibrium.

\section{Equilibrium processes}

\label{sec:eq}

\begin{figure*}[t]\includegraphics[scale=0.7]{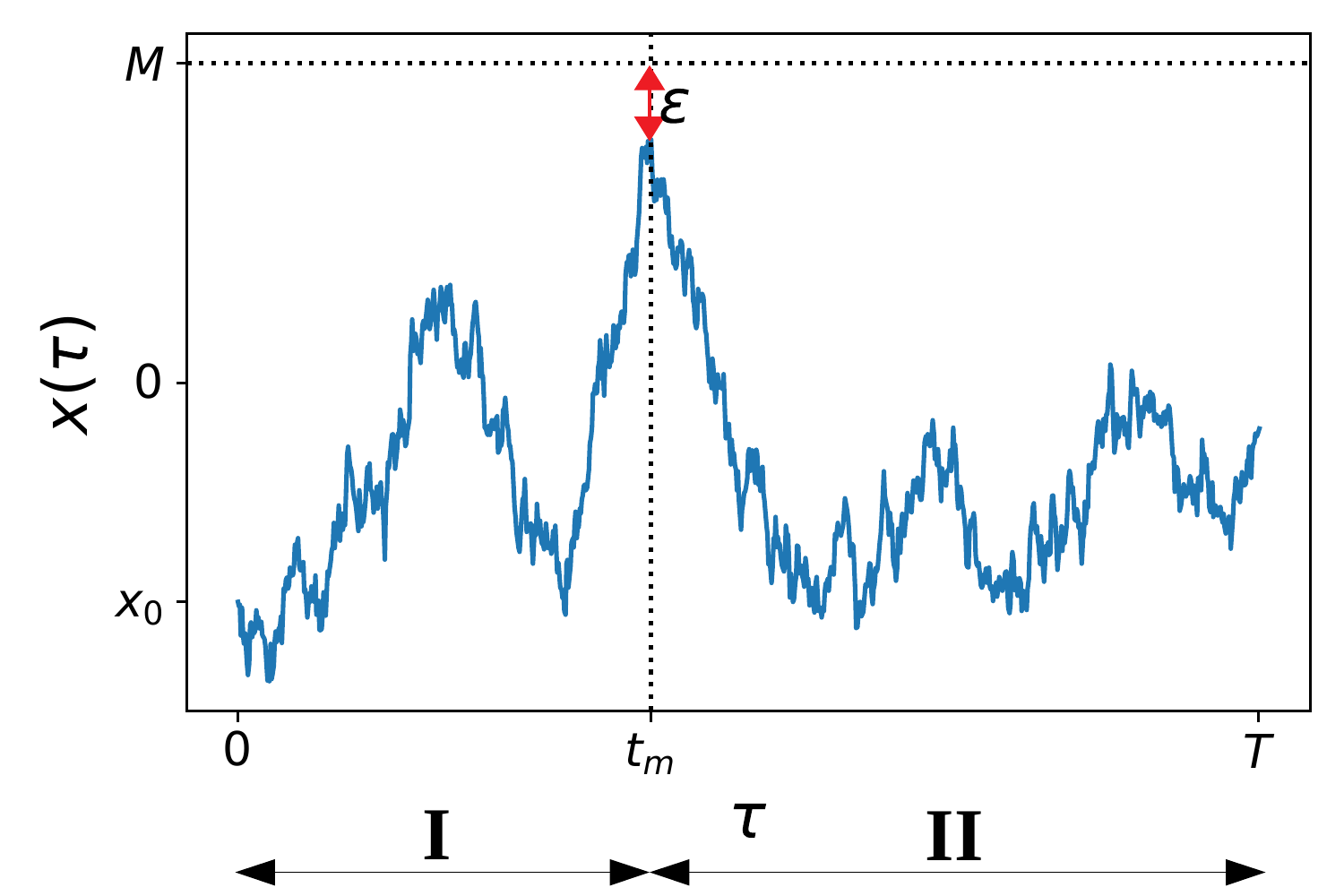} 
\caption{\label{fig:deco} Stationary process $x(\tau)$ during the time interval $[0,T]$. The value of the global maximum is $M-\epsilon$, with $\epsilon>0$, and the time of the maximum is $\tm$. The time interval $[0,T]$ is divided into the two subintervals $[0,\tm]$ (I) and $[\tm , T]$ (II). }
\end{figure*}

In this section, we investigate the distribution $\Pt$ of the time $\tm$ at which an equilibrium process reaches the global maximum. We focus on the paradigmatic case of an overdamped Brownian particle in a confining potential $V(x)$. The Langevin equation that describes the evolution of the position $x(\tau)$ of the particle reads
\begin{equation}
\frac{d x(\tau)}{d\tau}=-V'(x)+\sqrt{2D}\eta(\tau)\,,
\label{langevin}
\end{equation}
where $V'(x)=dV(x)/dx$ and $\eta(\tau)$ is a Gaussian zero-mean white noise with correlator $\langle\eta(\tau)\eta(\tau')\rangle=\delta(\tau-\tau')$ and $D>0$ is the diffusion constant. If the potential $V(x)$ grows sufficiently fast with $|x|$, the process admits the Boltzmann equilibrium state
\begin{equation}
P_{\rm st}(x)=\frac{1}{Z}e^{-V(x)/D}\,,
\label{boltzmann}
\end{equation}
where $Z$ is the normalization constant. Here we assume that $V(x)$ is sufficiently confining such that $P_{\rm st}(x)$ is normalizable. In particular, we focus on the class of potentials $V(x)$ such that $V(x)\approx  \alpha |x|^p$ for large $|x|$, where $\alpha>0$ and $p>0$ are fixed constants. We investigate the distribution of the time $\tm$ at which the position $x(\tau)$ of the particle reaches its maximal value before time $T$. We assume that at the initial time $\tau=0$ the process has already reached the equilibrium state, meaning that the initial position $x_0=x(0)$ is drawn from the PDF $P_{\rm st}(x)$ in Eq.~\eqref{boltzmann}. We will first identify two cases ($p=1$ and $p=2$) in which the distribution of $\tm$ can be exactly computed. Then, we will show that this distribution $\Pt$ becomes universal for any $p> 0$ at late times.

We use a path-decomposition technique to compute analytically the distribution of the time $\tm$ of the maximum. Doing so, we first obtain the joint distribution $P(\tm,M|T)$ of $\tm$ and of the maximum $M=x(\tm)$. Then, integrating over $M$, we find $\Pt$. This path-decomposition approach is similar to the one adopted in Refs.~\cite{MRK08,MMS19,MMS20} and can be described as follows. Using the Markov property of the process, we can write the joint probability of $\tm$ and $M$ as the product of the probabilities of two independent segments: (I) $[0,\tm]$ and (II) $[\tm, T]$ (see Fig.~\ref{fig:deco}). In the first interval (I), the process starts from position $x_0=x(0)$, which is random and distributed according to the steady state in Eq.~\eqref{boltzmann}, and it reaches the global maximum $M$ at time $\tm$. In the second interval (II), the walker starts from position $M$ at time $\tm$ and has to remain below this position $M$ up to time $T$.

To compute the probability weight of the first interval, one has to solve the Fokker-Planck equation of this process with absorbing boundary condition at $x=M$ (see details below). Moreover, one must also impose that the particle arrives exactly at $M$ at time $\tm$. However, due to the continuous-time nature of the process, one cannot constrain the trajectory to arrive at the absorbing boundary at a given time. Indeed, if the process arrives exactly at $M$ at time $\tm$, it will go above position $x=M$ infinitely many times in any time interval $[\tm-\delta,\tm]$ with $\delta>0$ \cite{Feller}. In other words, one cannot satisfy $x(\tau)<M$ for $\tau<\tm$ while imposing $x(\tm)=M$. A possible solution to this issue is to introduce a cutoff $\epsilon>0$ and to impose that at time $\tm$ the process reaches position $x(\tm)=M-\epsilon$ (see Fig.~\ref{fig:deco}). In this way, one can compute $\Pt$ for fixed $\epsilon$ and then take the limit $\epsilon\to 0$ at the very end of the computation. This approach was, for instance, used in Refs.~\cite{MRK08,RS11}.

Let us first consider the interval $[0,\tm]$. It is useful to define the constrained propagator $G^{M}(x,t|x_0)$ as the probability that the process goes from position $x_0$ at time $\tau=0$ to position $x$ at time $t$, while always remaining below position $M$. The probability weight $P_{\rm I}$ of the first interval (I) is $G^M(M-\epsilon,\tm|x_0)$. The constrained propagator satisfies the Fokker-Plank equation \cite{Feller}
\begin{equation}
\partial_t G^M(x,t|x_0)=D\partial_x^2 G^M(x,t|x_0)+\partial_x\left[V'(x) ~G^M(x,t|x_0)\right]\,,
\label{forward_FP}
\end{equation}
valid for $x\in (-\infty,M]$ with initial condition 
\begin{equation}
G^M(x,t=0|x_0)=\delta(x-x_0)\,.\label{initial_condition_fw}
\end{equation}
The first boundary condition is
\begin{equation}
G^M(M,t|x_0)=0\,,
\label{absorbing_condition_fw}
\end{equation}
corresponding to an absorbing wall at $x=M$. This boundary condition selects only those trajectories that remain below position $M$. The second boundary condition is
\begin{equation}
\lim_{x\to-\infty}G^M(x,t|x_0)=0\,,
\label{boundary_condition_fw}
\end{equation}
since the probability to find the particle infinitely far from its starting position after a finite amount of time vanishes.

In the second interval $[\tm,T]$, the process starts from position $M-\epsilon$ and remains below position $M$ up to time $T$. The corresponding probability weight can be expressed in terms of the survival probability $Q^M(x,t)$, i.e, the probability that the process starts from $x$ and remains below position $M$ up to time $t$. The weight $P_{\rm II}$ of the second interval can be written as $Q^M(M-\epsilon,T-\tm)$. The survival probability satisfies the backward Fokker-Planck equation \cite{Feller}
\begin{equation}
\partial_t Q^M(x,t)=D\partial^2_x Q^M(x,t)-V'(x)\partial_x Q^M(x,t)\,,
\label{backward_FP}
\end{equation}
with initial condition (for $x<M$)
\begin{equation}
Q^M(x,t=0)=1\,.
\label{initial_condition_bw}
\end{equation}
The boundary conditions are
\begin{equation}
Q^M(M,t)=0\,,
\label{absorbing_condition_bw}
\end{equation}
meaning that the particle at $x=M$ is immediately absorbed and 
\begin{equation}
\lim_{x\to-\infty}Q^M(x,t)=1\,,
\label{boundary_condition_bw}
\end{equation}
since a particle starting infinitely far away from the absorbing wall will never be absorbed in a finite time.

Then, the joint distribution of $M$ and $\tm$ can be obtained as the product of the probability weights of the first ($[0,\tm])$) and of the second ($[\tm ,T]$) intervals. Since the starting position $x_0$ is also random, one has also to integrate over $x_0$ with the correct probability weight given in Eq.~\eqref{boltzmann}. Therefore, for a fixed value of $\epsilon$, we get
\begin{equation}
P(\tm,M|T,\epsilon)=\mathcal{N}(\epsilon) \int_{-\infty}^{M}dx_0~P_{\rm st}(x_0)G^M(M-\epsilon,\tm|x_0)Q^M(M-\epsilon,T-\tm)\,,
\end{equation}
where $\mathcal{N}(\epsilon)$ is a normalization constant, i.e., $\mathcal{N}(\epsilon)$ is chosen to satisfy
\begin{equation}
\int_{0}^{T}d\tm~\int_{-\infty}^{\infty}dM~P(\tm,M|T,\epsilon)=1\,.
\end{equation} 
This constant $\mathcal{N}(\epsilon)$ could, in principle, depend on the total time $T$, but we will show a posteriori that it does not. Integrating over $M$, one finds
\begin{equation}
P(\tm|T,\epsilon)=\mathcal{N}(\epsilon) \int_{-\infty}^{\infty}dM~\int_{-\infty}^{M}dx_0~P_{\rm st}(x_0)G^M(M-\epsilon,\tm|x_0)Q^M(M-\epsilon,T-\tm)\,.
\end{equation}
Finally, taking the limit $\epsilon\to 0$, we obtain
\begin{equation}
P(\tm|T)=\lim_{\epsilon\to 0}\left[\mathcal{N}(\epsilon) \int_{-\infty}^{\infty}dM~\int_{-\infty}^{M}dx_0~P_{\rm st}(x_0)G^M(M-\epsilon,\tm|x_0)Q^M(M-\epsilon,T-\tm)\right]\,.
\label{Ptm_integral}
\end{equation}
For a given potential $V(x)$, one needs first to compute the constrained propagator $G^M(x,t|x_0)$ and the survival probability $Q^M(x,t)$. Then, the distribution of $\tm$ can be obtained using the formula \eqref{Ptm_integral}. As shown in the next sections, this can be done in the cases $V(x)=\alpha |x|$ (corresponding to $p=1$) and $V(x)=\alpha x^2$ (corresponding to $p=2$).

In general, it is easier to compute the propagator $G^M(x,t|x_0)$ and the survival probability $Q^M(x,t)$ in Laplace space (with respect to the time $t$). Therefore, it is useful to express the relation in Eq.~\eqref{Ptm_integral} in terms of the Laplace transforms of these quantities. To do this, we introduce the variables $t_1=\tm$, corresponding to the time of the maximum, and $t_2=T-\tm$, corresponding to the time after the maximum. Considering the double Laplace transform of Eq.~\eqref{Ptm_integral} with Laplace variables $s_1$ and $s_2$, corresponding to $t_1$ and $t_2$ respectively, we obtain
\begin{eqnarray}
\label{Ptm_integral_LT}
&&\int_{0}^{\infty}dt_1~e^{-s_1 t_1}\int_{0}^{\infty}dt_2~e^{-s_2 t_2}P(\tm=t_1|T=t_1+t_2)\\
&=&\lim_{\epsilon\to 0}\left[\mathcal{N}(\epsilon) \int_{-\infty}^{\infty}dM~\int_{-\infty}^{M}dx_0~P_{\rm st}(x_0)\tilde{G}^M(M-\epsilon,s_1|x_0)\tilde{Q}^M(M-\epsilon,s_2)\right]\,,\nonumber
\end{eqnarray}
where we have defined 
\begin{equation}
\tilde{G}^M(x,s|x_0)=\int_{0}^{\infty}dt~e^{-st}G^M(x,t|x_0)\,,
\end{equation}
and
\begin{equation}
\tilde{Q}^M(x,s)=\int_{0}^{\infty}dt~e^{-st}Q^M(x,t)\,.
\end{equation}
In the next sections, we derive an exact expression for $\Pt$ in the cases $p=1$ and $p=2$.

\subsection{The case $p=1$}
\label{sec:p1}

We first consider the case $p=1$, corresponding to the potential $V(x)=\alpha |x|$. The associated equilibrium steady state is
\begin{equation}
P_{\rm st}(x)=\frac{D}{2\alpha}e^{-\alpha |x|/D}\,.
\label{stat_p1}
\end{equation}
We first compute the forward propagator for this process. Setting $V(x)=\alpha |x|$ in Eq.~\eqref{forward_FP}, we obtain
\begin{equation}
\partial_t G^M(x,t|x_0)=D\partial_x^2 G^M(x,t|x_0)+2\alpha\delta(x)~G^M(x,t|x_0)+\alpha \operatorname{sign}(x)~\partial_x G^M(x,t|x_0)\,,
\label{forward_FP_p1}
\end{equation}
Taking a Laplace transform of this equation with respect to $t$ and using the initial condition in Eq.~\eqref{initial_condition_fw}, we find that $\tilde{G}^M(x,s|x_0)$ satisfies the equation
\begin{equation}
s\tilde{G}^M(x,s|x_0)-\delta(x-x_0)=D\partial_x^2 \tilde{G}^M(x,s|x_0)+2\alpha\delta(x)~\tilde{G}^M(x,s|x_0)+\alpha \operatorname{sign}(x)~\partial_x \tilde{G}^M(x,s|x_0)\,,
\label{forward_FP_p1_LT}
\end{equation}
The boundary conditions in Eq.~\eqref{absorbing_condition_fw} and \eqref{boundary_condition_fw} become
\begin{equation}
\tilde{G}^M(M,s|x_0)=0\,,
\label{absorbing_condition_fw_LT}
\end{equation}
and
\begin{equation}
\lim_{x\to-\infty}\tilde{G}^M(x,s|x_0)=0\,.
\label{boundary_condition_fw_LT}
\end{equation}
Solving the differential equation \eqref{forward_FP_p1_LT} (see Appendix \ref{app:G_p1}), we obtain to leading order for small $\epsilon$
\begin{equation}
\tilde{G}^M(M-\epsilon,s|x_0)\approx    
\begin{cases}
\dfrac{\epsilon}{D}e^{(\alpha-k)(M-x_0)/(2D)}\quad &\text{ if }x_0<M<0\,,\\
\\
\dfrac{\epsilon}{D}\dfrac{(k-\alpha)e^{k x_0 /D}+\alpha}{(k-\alpha)e^{k M /D}+\alpha}e^{(-\alpha+k)(M-x_0)/(2D)}\quad &\text{ if }0<x_0<M\,,\\
\\
\dfrac{k \epsilon}{D}\dfrac{e^{(k-\alpha) x_0 /(2D)}e^{(-k-\alpha) M /(2D)}}{k-\alpha+\alpha e^{-k M /D}}\quad &\text{ if }x_0<0 \text{ and }M>0\,,\\
\end{cases}
\label{G_p1_solution}
\end{equation}
where $k=\sqrt{\alpha^2+4sD}$.

We next focus on the survival probability $Q^M(x,t)$. For the potential $V(x)=\alpha |x|$, the differential equation \eqref{backward_FP} becomes
\begin{equation}
 \partial_t Q^M(x,t)=\partial^2_xQ^M(x,t)-\alpha \operatorname{sign}(x)\partial_x Q^M(x,t)\,.
\label{backward_FP_LT_p1}
\end{equation}
Taking a Laplace transform with respect to $t$ on both sides and using the initial condition in Eq.~\eqref{initial_condition_bw}, we obtain that $\tilde{Q}^M(x,s)$ satisfies the equation
\begin{equation}
s\tilde{Q}^M(x,s)-1=\partial^2_x \tilde{Q}^M(x,s)-\alpha \operatorname{sign}(x)\partial_x\tilde{Q}^M(x,s)\,,
\label{backward_FP_LT}
\end{equation}
with boundary conditions (see Eqs.~\eqref{absorbing_condition_bw} and \eqref{boundary_condition_bw})
\begin{equation}
\tilde{Q}^M(M,s)=0\,,
\label{absorbing_condition_bw_LT}
\end{equation}
and 
\begin{equation}
\lim_{x\to-\infty}\tilde{Q}^M(x,s)=\frac{1}{s}\,.
\label{boundary_condition_bw_LT}
\end{equation}
Solving Eq.~\eqref{backward_FP_LT}, we find that to leading order in $\epsilon$ (see Appendix \ref{app:G_p1})
\begin{equation}
\tilde{Q}^M(M-\epsilon,s)\approx    
\begin{cases}
\dfrac{\epsilon}{s}\dfrac{k-\alpha}{2D}\dfrac{(k+\alpha)e^{kM/D}-\alpha}{(k-\alpha)e^{kM/D}+\alpha}\quad &\text{ if }M>0\,,\\
\\
\dfrac{\epsilon}{s}\dfrac{k-\alpha}{2D}\quad &\text{ if }M<0\,.\\
\end{cases}
\label{Q_p1_solution}
\end{equation}

We now have all the ingredients to compute the distribution of $\tm$. Substituting the expressions of $P_{\rm st}(x_0)$, $\tilde{G}^M(M-\epsilon,s|x_0)$, and $\tilde{Q}^M(M-\epsilon,s)$, respectively given in Eq.~\eqref{stat_p1}, \eqref{G_p1_solution}, and \eqref{Q_p1_solution}, into Eq.~\eqref{Ptm_integral}, we obtain
\begin{eqnarray}
\nonumber
&&\int_{0}^{\infty}dt_1~e^{-s_1 t_1}\int_{0}^{\infty}dt_2~e^{-s_2 t_2}~P(\tm=t_1|T=t_1+t_2)=\lim_{\epsilon\to 0}\left[\mathcal{N}(\epsilon)\epsilon^2\right]\frac{k_2-\alpha}{4D \alpha s_2}\left\{ \int_{-\infty}^{0}dM~\int_{-\infty}^{M}dx_0~e^{\alpha x_0/D}\right. \\
&\times &\left. e^{(\alpha-k_1)(M-x_0)/(2D)}+\int_{0}^{\infty}dM~\int_{0}^{M}dx_0~e^{-\alpha x_0/D}\frac{(k_1-\alpha)e^{k_1 x_0 /D}+\alpha}{(k_1-\alpha)e^{k_1 M /D}+\alpha}\right.\left.e^{(-\alpha +k_1)(M-x_0)/(2D)}\frac{(k_2+\alpha)e^{k_2M/D}-\alpha}{(k_2-\alpha)e^{k_2 M/D}+\alpha}\right. \nonumber \\&+&\left.\int_{0}^{\infty}dM~\int_{-\infty}^{0}dx_0~e^{\alpha x_0/D}k_1\frac{e^{(k_1-\alpha) x_0 /(2D)}e^{(-k_1-\alpha) M /(2D)}}{k_1-\alpha+\alpha e^{-k_1 M /D}}\right.\left.\frac{(k_2+\alpha)e^{k_2M/D}-\alpha}{(k_2-\alpha)e^{k_2 M/D}+\alpha} \right\}\,,\label{Ptm_integral_LT_p1}
\end{eqnarray}
where we have defined $k_1=\sqrt{\alpha^2+4Ds_1}$ and $k_2=\sqrt{\alpha^2+4Ds_2}$.

The normalization constant $\mathcal{N}(\epsilon)$ can then be computed by setting $s_1=s_2=s$ on both sides of Eq.~\eqref{Ptm_integral_LT_p1}. Indeed, the left-hand side becomes
\begin{equation}
\int_{0}^{\infty}dt_1~\int_{0}^{\infty}dt_2~e^{-s (t_1+t_2)}~P(\tm=t_1|T=t_1+t_2)=\int_{0}^{\infty}dT~e^{-s T}~\int_{0}^{T}d\tm~P(\tm|T)=\int_{0}^{\infty}dT~e^{-s T}=\frac1s\,,
\label{lhs}
\end{equation}
where we have used the fact that $\PT$ is normalized to unity for $0\leq \tm \leq T$. Setting $s_1=s_2=s$ on the right-hand side of Eq.~\eqref{Ptm_integral_LT_p1} and computing the integrals over $x_0$ and $M$ with Mathematica, we find
\begin{equation}
\lim_{\epsilon\to 0}\left[\mathcal{N}(\epsilon)\epsilon^2\right]\frac{D}{\alpha^2 s}=\frac1s\,.
\end{equation}
Thus,  we get
\begin{equation}
\lim_{\epsilon\to 0}\left[\mathcal{N}(\epsilon)\epsilon^2\right]=\frac{\alpha^2  }{D}\,.
\end{equation}
Substituting this expression for the normalization constant in Eq.~\eqref{Ptm_integral_LT_p1} and computing the integrals over $x_0$, we obtain
\begin{eqnarray}\nonumber
&&\int_{0}^{\infty}dt_1~e^{-s_1 t_1}\int_{0}^{\infty}dt_2~e^{-s_2 t_2}~P(\tm=t_1|T=t_1+t_2)\\
&=&\frac{2\alpha}{(k_1+\alpha)(k_2+\alpha)}\left[\frac{D}{\alpha}+\int_{0}^{\infty}dM~e^{-\alpha M/D}\frac{(k_1+\alpha-\alpha e^{-k_1 M/D})(k_2+\alpha-\alpha e^{-k_2 M/D})}{(k_1-\alpha+\alpha e^{-k_1 M/D})(k_1-\alpha+\alpha e^{-k_2 M/D})}\right]\,,
\label{Ptm_integral_LT_p1_2}
\end{eqnarray}
where we recall that $k_1=\sqrt{\alpha^2+4Ds_1}$ and $k_2=\sqrt{\alpha^2+4Ds_2}$. Notably, from Eq.~(\ref{Ptm_integral_LT_p1_2}) we can already observe that the distribution $P(\tm|T)$ is symmetric around $\tm=T/2$, i.e., $P(\tm|T)=P(T-\tm|T)$. Indeed, it is clear from Eq.~(\ref{Ptm_integral_LT_p1_2}) that the Laplace transform of $P(\tm=t_1|T=t_1+t_2)$ is symmetric under the exchange of $s_1$ and $s_2$. This symmetry is a signature of the equilibrium nature of the process.

Interestingly, the PDF $\Pt$ can be rewritten in the scaling form
\begin{equation}
P(\tm|T)=\frac{\alpha^2}{4D}F_1\left(\frac{\alpha^2}{4D}\tm,\frac{\alpha^2}{4D}(T-\tm)\right)\,,
\label{scaling_form_F1}
\end{equation}
where $F_1(T_1,T_2)$ is the scaling function and $\xi=4D/\alpha^2$ is the natural timescale of the process. Plugging this expression into Eq.~(\ref{Ptm_integral_LT_p1_2}), we find that the double Laplace transform of $F_1(T_1,T_2)$ is given by (see also Eq.~\eqref{LT_scaling_p1_summary})
\begin{eqnarray}\label{eq:LT_F_V}
\nonumber \tilde{F}_1(s_1,s_2)&=& \frac{1}{2(1+\sqrt{1+s_1})(1+\sqrt{1+s_2})}\\
&\times & \Bigg[1+\int_{0}^{\infty}dz\,e^{-z}\frac{\left(\sqrt{1+s_1}+1-e^{-\sqrt{1+s_1}z}\right)\left(\sqrt{1+s_2}+1-e^{-\sqrt{1+s_2}z}\right)}{\left(\sqrt{1+s_1}-1+e^{-\sqrt{1+s_1}z}\right)\left(\sqrt{1+s_2}-1+e^{-\sqrt{1+s_2}z}\right)}\Bigg]\,,
\end{eqnarray}
where we have defined
\begin{equation}
\tilde{F}_1(s_1,s_2)=\int_{0}^{\infty}dT_1 e^{-s_1 T_1} \int_{0}^{\infty}dT_2 e^{-s_2 T_2} F_1(T_1,T_2)\,.
\end{equation}
This scaling function manifestly satisfies the symmetry $F_1(T_1,T_2)=F_1(T_2,T_1)$, corresponding to $\Pt=P(T-\tm|T)$. As a consequence of this symmetry, the first moment of $\tm$ is given by
\begin{equation}
\langle \tm \rangle=\frac{T}{2}\,.
\end{equation}

\subsubsection{Asymptotic behaviors}

Even though it is challenging to invert the double Laplace transform in Eq.~\eqref{eq:LT_F_V} exactly, this expression in Eq.~\eqref{eq:LT_F_V} can be used to extract the asymptotic behaviors of $\PT$ in the limits of short times ($T\ll \xi$) and late times ($T\gg \xi$). Here $\xi=\alpha^2/(4D)$ is the autocorrelation time of the process.

The short-time limit ($T\ll\xi$) corresponds in Laplace space to the limit $s_1,s_2\to \infty$. Taking this limit in Eq.~\eqref{eq:LT_F_V}, we obtain to leading order
\begin{equation}
\tilde{F}_1(s_1,s_2)\approx    \frac{1}{2\sqrt{s_1s_2}}\left(1+\int_{0}^{\infty}dz~e^{-z}\right)=\frac{1}{\sqrt{s_1s_2}}\,.
\end{equation}
This Laplace transform can now be inverted using the inversion formula
\begin{equation}
\mathcal{L}^{-1}_{s\to t}\left[\frac{1}{s^{\nu}}\right]=\frac{1}{\Gamma(\nu)~t^{1-\nu}}\,,
\label{inv_lapl_sqrt}
\end{equation}
where $\Gamma(\nu)$ is the Gamma function. Using this formula with $\nu=1/2$, we find that for small $T_1$ and $T_2$
\begin{equation}
F_1(T_1,T_2)\approx     \frac{1}{\pi\sqrt{T_1 T_2}}\,.
\end{equation}
Using Eq.~\eqref{scaling_form_F1}, we find that this corresponds to
\begin{equation}
\PT\approx    \frac{1}{\pi\sqrt{\tm (T-\tm)}}\,,
\end{equation}
which is valid for $T\ll 4D/\alpha^2$. This expression coincides with the well-known arcsine law of L\'evy \cite{Levy}, describing the distribution of the time of the maximum for a free Brownian motion. Indeed, for $T\ll 4D/\alpha^2$, the process does not have enough time to feel the confining potential.

We next focus on the late time limit $T\gg4D/\alpha^2 $. In this limit, three different regimes of distribution $\PT$ can be investigated: the central ``bulk'' regime, corresponding to $\tm,T\to \infty$ with $\tm/T$ fixed, the left ``edge'' regime, corresponding to $T\to \infty$ with $\tm\sim \mathcal{O}
(1)$, and the right edge regime, where $T\to \infty$ with $T-\tm\sim \mathcal{O}
(1)$. Note that, thanks to the symmetry $\PT=P(T-\tm|T)$, it is sufficient to study the right edge regime.

\begin{figure*}[t]\includegraphics[scale=0.7]{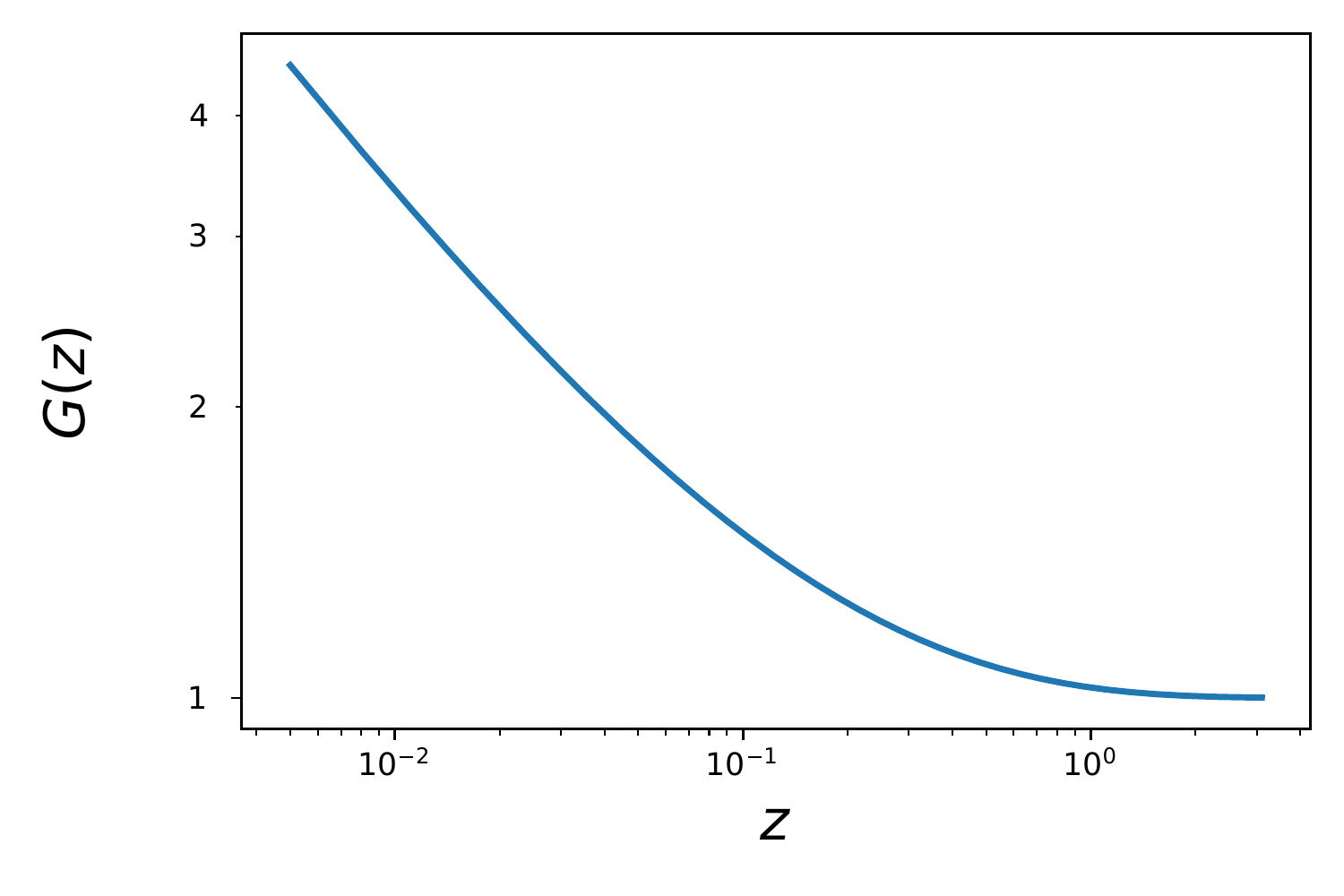} 
\caption{\label{fig:Gz} Log-log plot of the function $G(z)$, given in Eq.~\eqref{G}. For $z\ll 1$, the function diverges as $G(z)\approx1/(2\sqrt{\pi z})$ while $G(z)$ tends to the limit value $1$ as $G(z)\approx 1+e^{-z}/(4\sqrt{\pi}z^{3/2})$ for large $z$.}
\end{figure*}

To investigate the right-edge regime ($\tm\to T$), we expand Eq.~\eqref{eq:LT_F_V} to leading order for small $s_1$ while keeping $s_2\sim\mathcal{O}(1)$, yielding
\begin{eqnarray}
\tilde{F}_1(s_1,s_2)\approx     \frac{1}{2(1+\sqrt{1+s_2})}\int_{0}^{\infty}dz\,e^{-z}\frac{2-e^{-z}}{s_1+2e^{-z}}~\frac{\left[\sqrt{1+s_2}+1-e^{-\sqrt{1+s_2}z}\right]}{\left[\sqrt{1+s_2}-1+e^{-\sqrt{1+s_2}z}\right]}\,.
\end{eqnarray}
Identifying the pole $s_1=-2 e^{-z}$ and inverting the Laplace transform with respect to $T_1$, one finds
\begin{eqnarray}
\int_{0}^{\infty}dT_2~F_1(T_1,T_2)e^{-s_2 T_2}\approx     \frac{1}{2(1+\sqrt{1+s_2})}\int_{0}^{\infty}dz\,\exp\left[-2T_1e^{-z}\right]e^{-z}(2-e^{-z})~\frac{\left[\sqrt{1+s_2}+1-e^{-\sqrt{1+s_2}z}\right]}{\left[\sqrt{1+s_2}-1+e^{-\sqrt{1+s_2}z}\right]}\,.
\end{eqnarray}
For large $T_1$, the integral on the right-hand side is dominated by large values of $z$ and can thus be approximated as
\begin{eqnarray}
\int_{0}^{\infty}dT_2~F_1(T_1,T_2)e^{-s_2 T_2}\approx     \frac{1}{\sqrt{1+s_2}-1}\int_{0}^{\infty}dz\,e^{-z-2T_1\,e^{-z}}\,.
\end{eqnarray}
Performing the change of variable $z\to u=2T_1 e^{-z}$, we obtain
\begin{eqnarray}
\int_{0}^{\infty}dT_2~F_1(T_1,T_2)e^{-s_2 T_2}\approx    \frac12 \frac{1+\sqrt{1+s_2}}{s_2}\frac{1}{T_1}\int_{0}^{2T_1}du\,e^{-u}\,,
\end{eqnarray}
where we have used the relation $(1+\sqrt{1+s})(1-\sqrt{1+s})=s$. When $T_1$ is large, we can replace the upper limit of integration with infinity, yielding
\begin{eqnarray}
\int_{0}^{\infty}dT_2~F_1(T_1,T_2)e^{-s_2 T_2}\approx     \frac12 \frac{1+\sqrt{1+s_2}}{s_2}\frac{1}{T}\,,
\end{eqnarray}
where we have approximated $T_1\approx T$. The Laplace transform can be inverted by using the relation (see Appendix \ref{app:Laplace inversion})
\begin{equation}
\mathcal{L}^{-1}_{s\to t}\left[\frac{1+\sqrt{1+s}}{s}\right]=1+\operatorname{erf}(\sqrt{t})+\frac{1}{\sqrt{\pi t}}e^{-t}\,,
\label{G_LT}
\end{equation}
where $\operatorname{erf}(z)=(2/\sqrt{\pi})\int_{0}^{z}du~e^{-u^2}$. Therefore, we find that in the limit $T_1\to \infty$ with $T_2\sim O(1)$,
\begin{equation}
F(T_1,T_2)\approx    \frac{1}{T}G(T_2)\,,
\end{equation}
where
\begin{equation}
G(z)=\frac12 \left[1+\operatorname{erf}(\sqrt{z})+\frac{1}{\sqrt{\pi z}}e^{-z}\right]\,.
\label{G}
\end{equation}
This function $G(z)$ is shown in Fig.~\ref{fig:Gz} and has asymptotic behaviors
\begin{equation}
G(z)\approx    \begin{cases}
1/(2\sqrt{\pi z})\quad &\text{ for } z\to 0\,,\\
\\
1+e^{-z}/(4\sqrt{\pi}z^{3/2})\quad &\text{ for } z\to \infty\,.\\
\end{cases}
\label{G_asym}
\end{equation}
Thus, for $\tm\to T$, we find that $\PT$ diverges as $1/\sqrt{T-\tm}$. On the other hand, for $T-\tm\gg 4D/\alpha^2$ we find $\PT\approx     1/T$, smoothly connecting to the bulk regime. Similarly, using the symmetry of $F_1(T_1,T_2)$ we also find that when $T_2\to \infty$ with $T_1\sim O(1)$
\begin{equation}
F(T_1,T_2)\approx    \frac{1}{T}G(T_1)\,.
\end{equation}

In Laplace space the bulk regime is obtained in the limit $s_1,s_2\to \infty$, with $s_1/s_2$ fixed. Taking this limit in Eq.~\eqref{eq:LT_F_V}, we find
\begin{equation}
\tilde{F}_1(s_1,s_2)\approx    \frac{1}{2}\int_{0}^{\infty}dz\,e^{-z}\frac{\left(2-e^{-z}\right)^2}{\left(s_1+2e^{-z}\right)\left(s_2+2e^{-z}\right)}\,.
\end{equation}
Inverting the double Laplace transform, we get
\begin{equation}
F_1(T_1,T_2)\approx     \frac{1}{2}\int_{0}^{\infty}dz\,e^{-z}\left(2-e^{-z}\right)^2 \exp[-2(T_1+T_2)e^{-z}]\,.
\end{equation}
Finally, performing the change of variables $z\to u=2(T_1+T_2)e^{-z}$, we obtain
\begin{equation}
F_1(T_1,T_2)\approx     \frac{1}{4(T_1+T_2)}\int_{0}^{2(T_1+T_2)}du\,\left(2-\frac{u}{2(T_1+T_2)}\right)^2 e^{-u}=\frac{1}{T_1+T_2}+\mathcal{O}\left(\frac{1}{(T_1+T_2)^2}\right)\,.
\end{equation}
Thus, in the bulk regime the distribution $\PT$ can be approximated by the flat measure
\begin{equation}
\PT\approx    \frac{1}{T}\,.
\label{flat_p1_1}
\end{equation}
This uniform PDF for $\tm$ is the distribution that one would obtain when the positions of the process at different times are independent random variables. Indeed, since the observation time $T$ is much larger than the correlation time $\xi=4D/\alpha^2$, these variables are approximately independent (this argument is precisely explained in Section \ref{sec:univ}). However, note that this result in Eq.~\eqref{flat_p1_1} does not apply in the edge regimes for $\tm\to 0$ and $\tm\to T$.

To summarize, we have shown that in the late-time limit, the distribution of $\tm$ approaches the form
\begin{equation}
\PT\approx    \begin{cases}\frac1T 
G\left(\frac{\alpha^2}{4D}\tm\right)\quad &\text{ for }\quad\tm\lesssim 4D/\alpha^2\,,\\
\\
\frac1T \quad &\text{ for }\quad 4D/\alpha^2\ll \tm \ll T-4D/\alpha^2\,,\\
\\
\frac1T G\left[\frac{\alpha^2}{4D}(T-\tm)\right]\quad &\text{ for }\quad\tm\gtrsim T- 4D/\alpha^2\,.\\
\end{cases}
\label{PT_asymp_p1}
\end{equation}
Note that this expression in Eq.~\eqref{PT_asymp_p1} is asymptotically normalized to one for large $T$.

\subsection{The case $p=2$: the Ornstein-Uhlenbeck process}
\label{sec:p2}

This section focuses on BM in a harmonic potential $V(x)=\alpha x^2$, corresponding to the Ornstein-Uhlenbeck process. The equilibrium state for this process reads
\begin{equation}
P_{\rm st}(x_0)=\sqrt{\frac{\alpha}{\pi D}}\exp\left(-\frac{\alpha}{D}x_0^2\right)\,.
\label{stat_p2}
\end{equation}
As before, we first need to compute the constrained propagator and the survival probability. The propagator satisfies the forward Fokker-Plank equation \eqref{forward_FP}, which in this case reads
\begin{equation}
\partial_t G^M(x,t|x_0)=D\partial_x^2 G^M(x,t|x_0)+2\alpha G^M(x,t|x_0)+2\alpha x\partial_x G^M(x,t|x_0)\,,
\end{equation}
Taking a Laplace transform with respect to $t$ and using the initial condition in Eq. (\ref{initial_condition_fw}), we obtain
\begin{equation}\label{eq:FP_LT_ou}
D\partial_x^2 \tilde{G}^M(x,s|x_0)+2\alpha x\partial_x  \tilde{G}^M(x,s|x_0)+(2\alpha-s)   \tilde{G}^M(x,s|x_0)+\delta(x-x_0)=0\,.
\end{equation}
Eq. (\ref{eq:FP_LT_ou}) can be exactly solved (see Appendix \ref{app:G_p2}) and one obtains to leading order in $\epsilon$
\begin{eqnarray}\label{eq:G_expanded_ou}
\tilde{G}^M(M-\epsilon,s|x_0)\approx    \frac{\epsilon}{D}e^{-(M^2-x_0^2)\alpha/(2D)}~\frac{D_{-s/(2\alpha)}\left(-\sqrt{2\alpha/D}~x_0\right)}{D_{-s/(2\alpha)}\left(-\sqrt{2\alpha/D}~M\right)}\,,
\end{eqnarray}
where $D_p(z)$ is the parabolic cylinder function.

The backward Fokker-Plank equation for the survival probability, given in Eq.~\eqref{backward_FP} for a generic potential, in this case reads
\begin{equation}
\partial_t Q^M(x,t)=D\partial^2_x Q^M(x,t)-2\alpha x\partial_x Q^M(x,t)\,.
\label{backward_FP_p2}
\end{equation}
Taking a Laplace transform and using the initial condition in Eq.~\eqref{initial_condition_bw}, we find
\begin{equation}
s \tilde{Q}^M(x,s)-1=D\partial^2_x \tilde{Q}^M(x,s)-2\alpha x\partial_x \tilde{Q}^M(x,s)\,,
\label{backward_FP_LT_p2}
\end{equation}
with the boundary conditions in Eq.~\eqref{absorbing_condition_bw_LT} and \eqref{boundary_condition_bw_LT}. Solving this equation and imposing the boundary conditions (see Appendix \ref{app:G_p2}), we find, to leading order in $\epsilon$,
\begin{equation}\label{eq:Q_ou_LT_expanded}
\tilde{Q}^M(M-\epsilon,s)\approx     \frac{\epsilon}{s}\left[\frac{M2\alpha}{D}+\sqrt{\frac{2\alpha}{D}}\frac{D_{1-s/(2\alpha)}\left(-\sqrt{\frac{2\alpha}{D}}~M\right)}{D_{-s/(2\alpha)}\left(-\sqrt{\frac{2\alpha}{D}}~M\right)}\right]\,.
\end{equation}

Substituting the expressions for $P_{\rm st}(x_0)$, $\tilde{G}^M(M-\epsilon,s|x_0)$, and $\tilde{Q}^M(M-\epsilon,s)$, respectively given in Eqs.~\eqref{stat_p2}, \eqref{eq:G_expanded_ou}, and \eqref{eq:Q_ou_LT_expanded}, into the formula for $\PT$ in Eq.~\eqref{Ptm_integral_LT}, we get
\begin{eqnarray}
&&\int_{0}^{\infty}dt_1~e^{-s_1 t_1}\int_{0}^{\infty}dt_2~e^{-s_2 t_2}~P(\tm=t_1|T=t_1+t_2)=\lim_{\epsilon\to 0}\left[\mathcal{N}(\epsilon)\epsilon^2\right]\sqrt{\frac{2}{\pi}}\frac{1}{s_2}\frac{\alpha}{D^2}\int_{-\infty}^{\infty}dM~\int_{-\infty}^{M}dx_0~e^{-\alpha x_0^2/(2D)}\nonumber \\
&\times & e^{-M^2\alpha/(2D)}\frac{D_{-s_1/(2\alpha)}\left(-\sqrt{2\alpha/D}x_0\right)}{D_{-s_1/(2\alpha)}\left(-\sqrt{2\alpha/D}M\right)} ~\left[\sqrt{\frac{2\alpha}{D}}M+\frac{D_{1-s_2/(2\alpha)}\left(-\sqrt{\frac{2\alpha}{D}}M\right)}{D_{-s_2/(2\alpha)}\left(-\sqrt{\frac{2\alpha}{D}}M\right)}\right]\,.
\end{eqnarray}
To simplify this expression we first perform the change of variables $(x_0,M)\to (z=M\sqrt{2\alpha/D},w=x_0\sqrt{2\alpha/D})$, yielding
\begin{eqnarray}
&&\int_{0}^{\infty}dt_1~e^{-s_1 t_1}\int_{0}^{\infty}dt_2~e^{-s_2 t_2}~P(\tm=t_1|T=t_1+t_2)=\lim_{\epsilon\to 0}\left[\mathcal{N}(\epsilon)\epsilon^2\right]\frac{1}{\sqrt{2\pi}s_2D}\nonumber \\
&\times &\int_{-\infty}^{\infty}dz~e^{-z^2/4}\left[z+\frac{D_{1-s_2/(2\alpha)}\left(-z\right)}{D_{-s_2/(2\alpha)}~\left(-z\right)}\right]\int_{-\infty}^{z}dw~e^{-w^2/4} \frac{D_{-s_1/(2\alpha)}\left(-w\right)}{D_{-s_1/(2\alpha)}\left(-z\right)} \,.
\end{eqnarray}
Moreover, the integral over $w$ can be computed using the following identity (see Appendix \ref{app:wronskian})
\begin{equation}\label{eq:relation2}
\int_{-\infty}^{z}dw~e^{-w^2 /4}D_{-s}(-w)=\frac{1}{s}e^{-z^2/4}\left[z D_{-s}(-z)+D_{1-s}(-z)\right]\,,
\end{equation}
which yields
\begin{eqnarray}
&&\int_{0}^{\infty}dt_1~e^{-s_1 t_1}\int_{0}^{\infty}dt_2~e^{-s_2 t_2}~P(\tm=t_1|T=t_1+t_2)=\lim_{\epsilon\to 0}\left[\mathcal{N}(\epsilon)\epsilon^2\right]\frac{\sqrt{2}\alpha}{\sqrt{\pi}Ds_1 s_2}\nonumber \\
&\times &\int_{-\infty}^{\infty}dz~e^{-z^2/2}\left[z+\frac{D_{1-s_1/(2\alpha)}\left(-z\right)}{D_{-s_1/(2\alpha)}\left(-z\right)}\right]\left[z +\frac{D_{1-s_2/(2\alpha)}(-z)}{D_{-s_2/(2\alpha)}(-z)}\right] \,.
\end{eqnarray}
This expression can be rewritten in a more compact form by using the identity \cite{Gradshteyn}
\begin{equation}\label{eq:relation3}
z + \frac{D_{1-s}(-z)}{D_{-s}(-z)}= s\frac{D_{-s-1}(-z)}{D_{-s}(-z)} \,,
\end{equation}
yielding
\begin{eqnarray}\nonumber
&&\int_{0}^{\infty}dt_1~\int_{0}^{\infty}dt_2~e^{-s_1t_1-s_2 t_2}~P(\tm=t_1|T=t_1+t_2)\\ &=&\frac{\lim_{\epsilon\to 0}\left[\mathcal{N}(\epsilon)\epsilon^2\right]}{\sqrt{8\pi}D\alpha}\int_{-\infty}^{\infty}dz~e^{-z^2/2}\frac{D_{-1-s_1/(2\alpha)}\left(-z\right)}{D_{-s_1/(2\alpha)}\left(-z\right)}~\frac{D_{-1-s_2/(2\alpha)}(-z)}{D_{-s_2/(2\alpha)}(-z)}\,.
\label{N_p2}
\end{eqnarray}
Finally, to fix the constant $\mathcal{N(\epsilon})$ we impose that $\PT$ must be normalized to unity. To do so, we set $s_1=s_2=s$ on both sides of equation \eqref{N_p2}. As before (see Eq.~\eqref{lhs}) the left-hand side is simply equal to $1/s$, yielding
\begin{eqnarray}
\frac1s = \frac{A}{\sqrt{8\pi}\alpha}\int_{-\infty}^{\infty}dz~e^{-z^2/2}\left[\frac{D_{-1-s/(2\alpha)}\left(-z\right)}{D_{-s/(2\alpha)}\left(-z\right)}\right]^2\,,
\end{eqnarray}
where $A=\lim_{\epsilon\to 0}\left[\mathcal{N}(\epsilon)\epsilon^2\right]/D$ needs to be determined. Introducing the dimensionless variable $q=s/(2\alpha)$, we find that $A$ satisfies
\begin{eqnarray}
\frac1q = \frac{A}{\sqrt{2\pi}}\int_{-\infty}^{\infty}dz~e^{-z^2/2}\left[\frac{D_{-1-q}\left(-z\right)}{D_{-q}\left(-z\right)}\right]^2\,.
\end{eqnarray}
Even though the integral on the right-hand side is hard to compute analytically, we verified numerically that
\begin{equation}\label{eq:normalization_ou}
\frac{1}{\sqrt{2\pi}}\int_{-\infty}^{\infty}dz\,e^{-z^2 /2}\left[\frac{D_{-1-q}(-z)}{D_{-q}(-z)}\right]^2=\frac1q\,,
\end{equation}
implying $A=1$ and thus 
\begin{equation}
\label{mathcalN_1}
\lim_{\epsilon\to 0}[\mathcal{N}(\epsilon)\epsilon^2]=D\,.
\end{equation}
 Plugging the expression in Eq.~\eqref{mathcalN_1} into Eq.~\eqref{N_p2}, we find that the double Laplace transform of $\PT$ reads
\begin{equation}
\int_{0}^{\infty}dt_1~\int_{0}^{\infty}dt_2~e^{-s_1t_1-s_2 t_2}~P(\tm=t_1|T=t_1+t_2)=\frac{1}{\sqrt{8\pi}\alpha}\int_{-\infty}^{\infty}dz~e^{-z^2/2}\frac{D_{-1-s_1/(2\alpha)}\left(-z\right)}{D_{-s_1/(2\alpha)}\left(-z\right)}~\frac{D_{-1-s_2/(2\alpha)}(-z)}{D_{-s_2/(2\alpha)}(-z)}\,.
\label{N_p2_final}
\end{equation}
Note that even if we have verified the validity of Eq.~(\ref{eq:normalization_ou}) numerically, the mathematical proof of this relation has eluded us so far and it remains an interesting exercise to prove this identity. Interestingly, the PDF $\PT$ of the time $\tm$ of the maximum can be rewritten in the scaling form
\begin{equation}
\PT=\alpha F_{\rm OU}(\alpha\tm,\alpha(T-\tm))\,,
\label{scaling_relation_OU}
\end{equation}
where $F_{\rm OU}(T_1,T_2)$ is the scaling function and $\xi=1/\alpha$ the natural timescale of the process. Plugging this expression into Eq.~\eqref{N_p2_final}, we find that the scaling function $F_{\rm OU}(T_1,T_2)$ is given by
\begin{equation}
\int_{0}^{\infty}dT_1~\int_{0}^{\infty}dT_2~e^{-s_1 T_1-s_2 T_2}F_{\rm OU}(T_1,T_2)=\frac{1}{\sqrt{8\pi}}\int_{-\infty}^{\infty}dz~e^{-z^2/2}\frac{D_{-1-s_1/2}\left(-z\right)}{D_{-s_1/2}\left(-z\right)}~\frac{D_{-1-s_2/2}(-z)}{D_{-s_2/2}(-z)}\,.
\label{scaling_OU}
\end{equation}
From this expression, we immediately find that the PDF $\PT$ is symmetric around the midpoint $\tm=T/2$. This is a signature of the time-reversal symmetry of equilibrium processes. Consequently, the first moment of $\tm$ is simply given by $\langle \tm\rangle=T/2$. Moreover, it is interesting to notice that in this case the distribution of $\tm$ is completely independent of the diffusion coefficient $D$.

\subsubsection{Asymptotic behaviors}

We next focus on the asymptotic behaviors of $\PT$ in the limit of small and large $T$. We expect these behaviors to be qualitatively similar to the ones derived for $p=1$ in Section \ref{sec:p1}. The double Laplace transform in Eq.~\eqref{scaling_OU} can be inverted in the short-time and late-time limits, corresponding to $T\ll \xi$ and $T\gg \xi$. Here the correlation time $\xi$ is given by $\xi=1/\alpha$.

We first focus on the small $T$ behavior. This corresponds to the limit $s_1,s_2\to \infty	$ in Eq.~\eqref{scaling_OU}. We start by performing the change of variable $z\to u=z\sqrt{2/s_1}$ in Eq.~\eqref{scaling_OU}, yielding
\begin{equation}
\int_{0}^{\infty}dT_1~\int_{0}^{\infty}dT_2~e^{-s_1 T_1-s_2 T_2}F_{\rm OU}(T_1,T_2)=\frac{\sqrt{s_1}}{4\sqrt{\pi}}\int_{-\infty}^{\infty}du~e^{-s_1 u^2/4}\frac{D_{-1-s_1/2}\left(-u\sqrt{s_1/2}\right)}{D_{-s_1/2}\left(-u\sqrt{s_1/2}\right)}~\frac{D_{-1-s_2/2}(-u\sqrt{s_1/2})}{D_{-s_2/2}(-u\sqrt{s_1/2})}\,.
\end{equation}
When $s_1$ and $s_2$ are both large, we use the approximation (see Appendix \ref{app:wronskian})
\begin{equation}
\frac{D_{-(s+1)}(-\sqrt{s}u)}{D_{-s}(-\sqrt{s}u)}\approx\frac{1}{\sqrt{s}}\frac{u+\sqrt{u^2+4}}{2}\,,
\label{relation_large_p}
\end{equation} 
valid for large $s$ and fixed $u$. We obtain
\begin{equation}
\int_{0}^{\infty}dT_1~\int_{0}^{\infty}dT_2~e^{-s_1 T_1-s_2 T_2}F_{\rm OU}(T_1,T_2)\approx\frac{1}{8\sqrt{\pi s_2}}\int_{-\infty}^{\infty}du~e^{-s_1 u^2/4}(u+\sqrt{u^2+4})(\sqrt{s_1/s_2}u+\sqrt{(s_1/s_2) u^2+4})\,.
\end{equation}
To leading order, this integral becomes
\begin{equation}
\int_{0}^{\infty}dT_1~\int_{0}^{\infty}dT_2~e^{-s_1 T_1-s_2 T_2}F_{\rm OU}(T_1,T_2)\approx\frac{1}{2\sqrt{\pi s_2}}\int_{-\infty}^{\infty}du~e^{-s_1 u^2/4}=\frac{1}{\sqrt{s_1 s_2}}\,.
\end{equation}
This Laplace transform can now be inverted using Eq.~\eqref{inv_lapl_sqrt}, and we obtain, for $T_1,T_2\ll 1$,
\begin{equation}
F_{\rm OU}(T_1,T_2)\approx\frac{1}{\pi\sqrt{T_1 T_2}}\,,
\end{equation}
corresponding to
\begin{equation}
\PT\approx\frac{1}{\pi\sqrt{\tm(T-\tm)}}\,.
\end{equation}
This is precisely the same result as for the case $p=1$ and corresponds to the arcsine law of L\'evy \cite{Levy}, which describes the PDF of $\tm$ for a free BM of duration $T$. As before, this result is coherent with the fact that for very short times ($T\ll 1/\alpha$), the effect of the external potential is negligible and the process is simply diffusive.

We next focus on the limit of large $T$. As for the case $p=1$, there are three different regimes to be investigated: the left edge regime ($\tm$ small and $T\to \infty$), the central bulk regime ($1\ll \tm\ll T$), and the right edge regime ($\tm\sim T$ and $T\to \infty$). We now focus on the right-edge behavior ($\tm \to T$). Formally inverting the Laplace transform in Eq.~(\ref{scaling_OU}) with respect to $T_1$ gives
\begin{eqnarray}
\int_0^{\infty}F(T_1, T_2)\, e^{-s_2\, T_2} dT_2 =\frac{1}{\sqrt{8\pi }}\int_{-\infty}^{\infty} dz\, e^{-z^2/2} \left[
\int_{\Gamma_1} \frac{ds_1}{2\pi i}\, e^{s_1\, T_1}\, \frac{D_{-1-s_1/2}(-z)}{D_{-s_1/2}(-z)}\right]\,
\left[
\frac{D_{-1-s_2/2}(-z)}{D_{-s_2/2}(-z)}\right]\, ,
\label{F2s2.1}
\end{eqnarray}
where $\Gamma_1$ denotes a Bromwich contour in the complex $s_1$ plane. We first consider the Bromwich integral over $s_1$ in \eqref{F2s2.1}. We are interested in the large $T_1$ limit.
To evaluate this integral for large $T_1$ we should look for the pole of the integrand at some small negative $s_1$. Since $s_1$ is small, we can
replace the numerator, to leading order, by its value at $s_1=0$ 
\begin{equation}
D_{-1-s_1/2}(-z) \approx D_{-1}(-z)= \sqrt{\frac{\pi}{2}}\, e^{z^2/4}\, {\rm erfc}\left(
-\frac{z}{\sqrt{2}}\right)\, ,
\label{Dm1.1}
\end{equation}
where $\operatorname{erfc}(z)=(2/\sqrt{\pi})\int_{z}^{\infty}du~e^{-u^2}$. Furthermore, the denominator for small $s_1$ can be approximated as (simply by expanding in Taylor 
series and using $D_0(z)= e^{-z^2/4}$)
\begin{eqnarray}
D_{-s_1/2}(-z) &\approx & e^{-z^2/4}\left[ 1+ s_1\, \sqrt{\pi/2}\, \frac{1}{z}\, e^{z^2/2}\right] = \frac{\sqrt{2\pi}}{z}\, e^{z^2/4}\, \left[\frac{s_1}{2}+ \frac{z}{\sqrt{2\pi}}\, e^{-z^2/2}\right]\, .
\label{Dpsmall.1}
\end{eqnarray}
Using \eqref{Dm1.1} and \eqref{Dpsmall.1} we then obtain the following approximation for the
Bromwich integral for large $T_1$
\begin{equation}
\int_{\Gamma} \frac{ds_1}{2\pi i}\, e^{s_1\, T_1}\, \frac{D_{-1-s_1/2}(-z)}{D_{-s_1/2}(-z)}
\approx \int_{\Gamma} \frac{ds_1}{2\pi i}\, e^{s_1\, T_1}\,  
\frac{z\, {\rm erfc} (-z/\sqrt{2})}{ s_1+  \frac{\sqrt{2}z}{\sqrt{\pi}}\, e^{-z^2/2}}\, .
\label{Brom.1}
\end{equation}
Clearly the pole occurs at $s_1= - \sqrt{2/\pi}z\, e^{-z^2/2}$. Since $s_1$ is small,
the variable $z$ is necessarily large and positive. In particular, since for large $T_1$ we expect that $s_1$ scales as $1/T_1$, we can anticipate that $z\sim \sqrt{\ln T_1}$. Evaluating the residue and using ${\rm erfc}(-\infty)=2$, we get
\begin{equation}
\int_{\Gamma} \frac{ds_1}{2\pi i}\, e^{s_1\, T_1}\,\frac{D_{-1-s_1/2}(-z)}{D_{-s_1/2}(-z)}
\approx 2 z\, \exp\left[- \frac{\sqrt{2} z}{\sqrt{\pi}}\, e^{-z^2/2}\, T_1\right]\, .
\label{Brom.2}
\end{equation}
Substituting this result back in \eqref{F2s2.1} gives
\begin{eqnarray}
\int_0^{\infty}F(T_1, T_2)\, e^{-s_2\, T_2} dT_2  \approx \frac{1}{\sqrt{2\pi}}\int_{-\infty}^{\infty} dz\, z\, e^{-z^2/2} 
\exp\left[- \frac{\sqrt{2} z}{\sqrt{\pi}}\, e^{-z^2/2}\, T_1\right]\,
\frac{D_{-1-s_2/2}(-z)}{D_{-s_2/2}(-z)}\, .
\label{F2s2.2}
\end{eqnarray}
To make further progress, we perform the change of variable $z\to y=(z-a_{T_1})/b_{T_1}$, where $a_{T_1}$ and $b_{T_1}$ are positive constants that depend on $T_1$ and will be chosen appropriately, yielding
\begin{eqnarray}
\nonumber&&\int_0^{\infty}F(T_1, T_2)\, e^{-s_2\, T_2} dT_2  \approx \frac{b_{T_1}}{\sqrt{2\pi}}\int_{-\infty}^{\infty} dy\, (a_{T_1}+b_{T_1}y)\, e^{-(a_{T_1}+b_{T_1}y)^2/2} \\
 &\times &
\exp\left\{- \exp\left[\ln\left(\sqrt{\frac{2}{\pi}}\right)+\ln(a_{T_1}+b_{T_1}y)-(a_{T_1}+b_{T_1}y)^2/2+\ln T_1\right]\, \right\}
\frac{D_{-1-s_2/2}(-(a_{T_1}+b_{T_1}y))}{D_{-s_2/2}(-(a_{T_1}+b_{T_1}y))}\, .
\label{F2s2.2.1}
\end{eqnarray}
To get rid of the dependence on $T_1$ in the exponent, we choose $a_{T_1}$ such that
\begin{equation}
\ln\left(\sqrt{\frac{2}{\pi}}\right)+\ln(a_{T_1}+b_{T_1}y)-a_{T_1}^2/2+\ln T_1=0\,,
\end{equation}
meaning that to leading order 
\begin{equation}
\label{Oumax.1}
a_{T_1}\approx \sqrt{2\ln T_1}\,.
\end{equation}
Similarly, we choose $b_{T_1}=1/a_{T_1}\approx 1/\sqrt{2\ln(T_1)}$. Plugging these expressions into Eq.~\eqref{F2s2.2.1}, we find that to leading order in $T_1$
\begin{equation}
\int_0^{\infty}F(T_1, T_2)\, e^{-s_2\, T_2} dT_2  \approx \frac{1}{2T_1 a_{T_1}}\int_{-\infty}^{\infty}dy ~e^{-y-e^{-y}}\frac{D_{-1-s_2/2}(-(a_{T_1}+b_{T_1}y))}{D_{-s_2/2}(-(a_{T_1}+b_{T_1}y))}\,.
\end{equation}
Moreover, since $a_{T_1}\gg b_{T_1}$, we obtain
\begin{equation}
\int_0^{\infty}F(T_1, T_2)\, e^{-s_2\, T_2} dT_2  \approx \frac{1}{2T_1 a_{T_1}}\frac{D_{-1-s_2/2}(-a_{T_1})}{D_{-s_2/2}(-a_{T_1})}\,,
\label{F2s2.4}
\end{equation}
where $a_{T_1}\approx\sqrt{2\ln(T_1)}$ and we have used
\begin{equation}
\int_{-\infty}^{\infty}dy ~e^{-y-e^{-y}}=1\,.
\label{integrand_gumbel_1}
\end{equation}
Note that the result in \eqref{F2s2.4} is valid for large $T_1$, but for arbitrary $s_2$.
We then need to invert this Laplace transform with respect to $s_2$ in order to derive the dependence on $T_2$.

To make further progress, let us see what we anticipate and then work backwards.
We anticipate a scaling form for the edge behavior, $F(T_1,T_2)\sim (1/T_1) G\left( a\, 
(2 \ln T_1)\, T_2\right)$ for large $T_1$, where $a$ is some constant scale factor. 
This would mean that the width of the right edge would shrink very slowly with increasing $T_1$. Hence, if $T_1$ is large,
we should scale $T_2$ as $T_2 \sim t/(2 \ln T_1)$ where $t\sim O(1)$.
This would mean that in the conjugate Laplace space $s_2$ should scale as $s_2\sim (2 \ln T_1)$.
Since $a_{T_1}\approx \sqrt{2 \ln T_1}$, this means that $s_2\sim (a_{T_1})^2$ for large $a_{T_1}$.
Hence, driven by this observation, in order to derive a nontrivial scaling function at the right edge, we should analyse \eqref{F2s2.4} in the scaling limit 
by setting 
\begin{equation}
a_{T_1}= \sqrt{ \frac{s_2}{2}}\, u
\label{u_def.1}
\end{equation}
where $u\sim O(1)$, while both $s_2$ and $a_{T_1}$ are large. With this setting, i.e.,
replacing $a_{T_1}$ by $\sqrt{s_2/2}\, u$ in \eqref{F2s2.4} we get
\begin{equation}
\int_0^{\infty}F(T_1, T_2)\, e^{-s_2\, T_2} dT_2  \approx \frac{1}{2T_1}\,   \frac{\sqrt{2}}{\sqrt{s_2}\, u} 
\, \frac{D_{-1-s_2/2}(-\sqrt{s_2/2}\, u)}{D_{-s_2/2}(-\sqrt{s_2/2}\, u)} \,.
\label{F2s2.5}
\end{equation}
This leads us to analyse the asymptotic behavior of $D_{-p}(-\sqrt{p}\, u)$ in the limit of large $p$ (somewhat akin to
the Plancheral-Rotah type asymptotic limits for Hermite polynomials). To do this, we use the expansion in Eq.~\eqref{relation_large_p}, which gives the limiting large $s_2$ behavior
\begin{equation}
\int_0^{\infty}F(T_1, T_2)\, e^{-s_2\, T_2} dT_2   \approx \frac{1}{2 T_1}\, \frac{1}{s_2}\, \left[ 1+ \sqrt{1+ \frac{4}{u^2}} \right]= 
\frac{1}{2T_1}\, \frac{1}{s_2}\, \left[ 1+ \sqrt{1+ \frac{2 s_2}{a_{T_1}^2}} \right]\, ,
\label{F2s2.6}
\end{equation}
where we used $u=\sqrt{2} a_{T_1}/\sqrt{s_2}$ from \eqref{u_def.1}. Inverting formally the Laplace transform
with respect to $s_2$ gives
\begin{equation}
F(T_1,T_2)\approx 
\int_{{\Gamma}_2} \frac{ds_2}{2\pi i}\, e^{s_2\, T_2}\,
\frac{1}{2T_1}\, \frac{1}{\,s_2}\, \left[ 1+ \sqrt{1+ \frac{2 s_2}{a_{T_1}^2}}\right]\,  ,
\label{FT1T2.1}
\end{equation}
where ${\Gamma}_2$ denotes a Bromwich contour in the complex $s_2$ plane. Rescaling
$s_2= s\, a_{T_1}^2/2$ gives
\begin{equation}
F(T_1,T_2)\approx
\int_{{\Gamma}_2} \frac{ds}{2\pi i}\, e^{s\,  a_{T_1}^2 T_2/2} \, 
\frac{1}{T_1}\, \frac{1}{2s}\, \left[ 1+ \sqrt{1+ s}\right]\,  ,
\label{FT1T2.2}
\end{equation}
Using the Laplace-inversion formula in Eq.~\eqref{G_LT}, we get our main scaling result at the right edge
\begin{equation}
F(T_1,T_2) \approx \frac{1}{T_1}\, G\left( \frac{a_{T_1}^2}{2}\, T_2\right) 
\label{edge_scaling.1}
\end{equation}
where the scaling function $G(t)$ is given exactly in \eqref{G}. Note that
$a_{T_1}= \sqrt{2\, \ln T_1}$ from \eqref{Oumax.1}. The result in \eqref{edge_scaling.1} is
valid in the scaling limit when $T_1$ is large and $T_2\sim 1/{\ln T_1}$ is small.
In this limit, we can replace $T_1\approx  T=T_1+T_2$ and hence \eqref{edge_scaling.1} reads
\begin{equation}
F(T_1,T_2) \approx \frac{1}{T}\, G\left((\ln T)\, T_2\right) \, .
\label{edge_scaling.2}
\end{equation}
Using the scaling relation in Eq.~\eqref{scaling_relation_OU}, we obtain 
\begin{equation}
\PT\approx \frac1T G\left(\alpha(\ln T) (T-\tm)\right)\,,
\end{equation}
valid for large $T$ and $(T-\tm)\sim 1/\ln(T)$. Using the $\tm\to T-\tm$ symmetry of the process, we obtain the left edge regime
\begin{equation}
\PT\approx \frac1T G\left(\alpha(\ln T) \tm\right)\,,
\end{equation}
valid for large $T$ and $\tm \sim 1/\ln(T)$.

\begin{figure}[t]
\begin{center}
\includegraphics[scale=0.5]{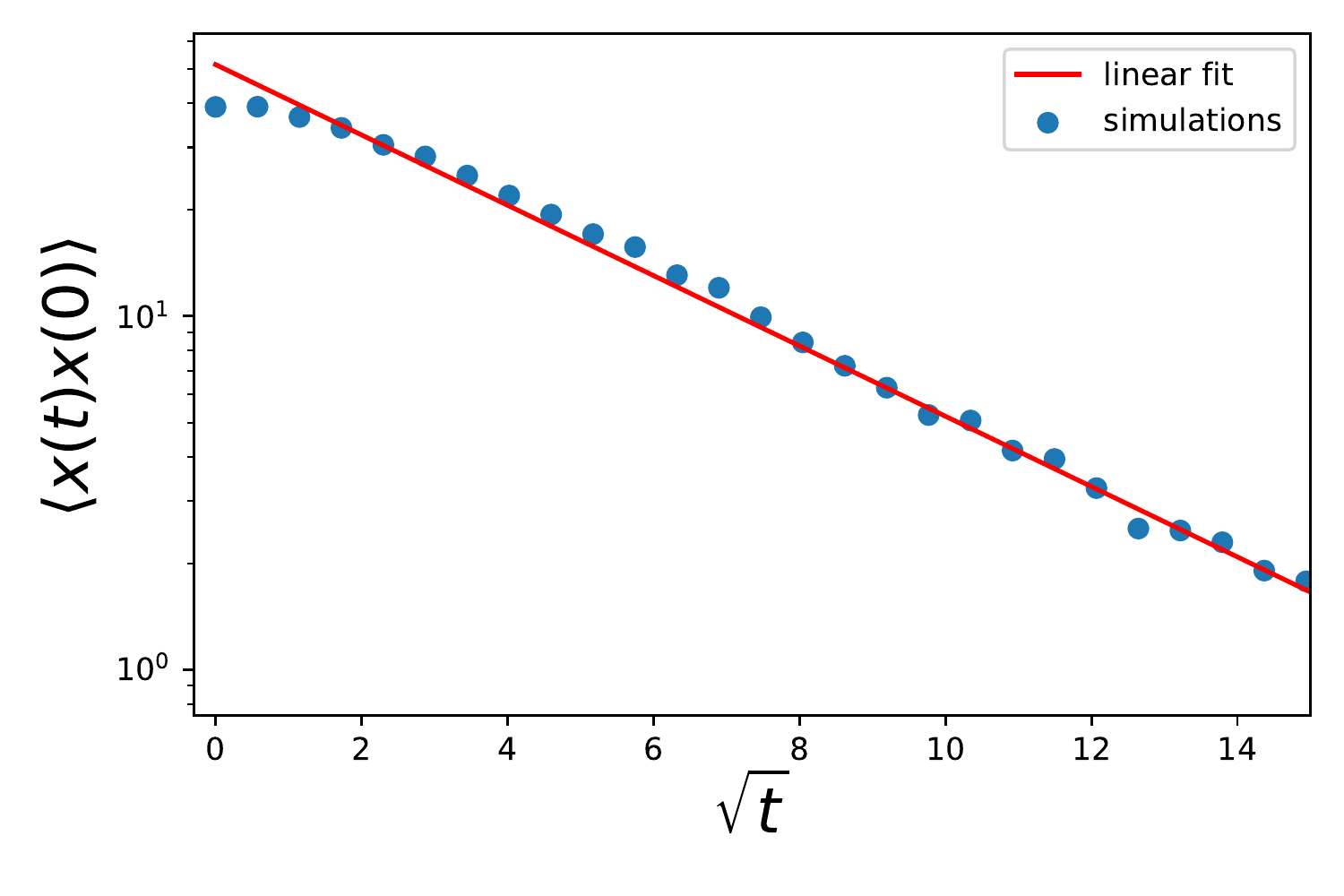}
\caption{ Semi-logarithmic plot of the correlation function $\langle x(t)x(0)\rangle$ as a function of $\sqrt{t}$ for Brownian motion with diffusion constant $D=1$ in the potential $V(x)=\sqrt{|x|}$. The initial position $x(0)$ is drawn from the equilibrium state of the system. The continuous red line shows a stretched-exponential decay of the type $\langle x(t)x(0)\rangle\sim e^{-\sqrt{t/\xi}}$.
\label{fig:correlations}}
\end{center}
\end{figure}

Finally, to compute the bulk regime ($\tm, T\to\infty$ with $\tm/T$ fixed), we first formally invert the double Laplace transform in Eq.~\eqref{scaling_OU} to obtain
\begin{equation}
F_{\rm OU}(T_1,T_2)=\frac{1}{\sqrt{8\pi}}\int_{-\infty}^{\infty}dz~e^{-z^2/2}\left[\int_{\Gamma_1}\frac{ds_1}{2\pi i}e^{s_1 T_1}\frac{D_{-1-s_1/2}\left(-z\right)}{D_{-s_1/2}\left(-z\right)}\right]\left[\int_{\Gamma_2}\frac{ds_2}{2\pi i}e^{s_2 T_2}\frac{D_{-1-s_2/2}(-z)}{D_{-s_2/2}(-z)}\right]\,,
\end{equation}
where $\Gamma_1$ and $\Gamma_2$ denote Bromwich contours in the complex $s_1$ and $s_2$ planes. The bulk regime corresponds to the limit $s_1\,,s_2\to 0$. Thus, using the small-$s$ expansion in Eq.~\eqref{Brom.2}, we find
\begin{equation}
F_{\rm OU}(T_1,T_2)\approx\sqrt{\frac{2}{\pi}}\int_{-\infty}^{\infty}dz~e^{-z^2/2}z^2\, \exp\left[- \frac{\sqrt{2}z}{\sqrt{\pi}}\, e^{-z^2/2} (T_1+T_2)\right]\,.
\end{equation}
We next perform the change of variable $z\to u=\frac{\sqrt{2}z}{\sqrt{\pi}}\, e^{-z^2/2} (T_1+T_2)$ with Jacobian
\begin{equation}
|du/dz|=\frac{\sqrt{2}z}{\sqrt{\pi}}\, e^{-z^2/2} (T_1+T_2)|1-z^2|\,.
\end{equation}
Using the fact that when $(T_1+T_2)$ is large the integral is dominated by large values of $z$, we can approximate
\begin{equation}
|du/dz|\approx\frac{\sqrt{2}z}{\sqrt{\pi}}\, e^{-z^2/2} (T_1+T_2)z^2\,.
\end{equation}
Thus, we finally obtain
\begin{equation}
F_{\rm OU}(T_1,T_2)\approx \frac{1}{T_1+T_2}\int_{0}^{\infty}du~e^{-u}=\frac{1}{T_1+T_2}\,.
\end{equation}
Therefore, in the bulk regime where $1\ll\tm\ll T$, we obtain
\begin{equation}
\PT\approx\frac{1}{T}\,.
\end{equation}
Once again, we find that the distribution of $\tm$ becomes flat in the bulk regime for late times. This is because the random variables describing the positions of the process at different times become approximately independent when $T\gg \xi$ (where $\xi$ is the correlation time).

To summarize, for $T\gg \xi=1/\alpha$, the distribution $\PT$ of the time $\tm$ of the maximum behaves as
\begin{equation}
\PT\approx    \begin{cases}\frac1T 
G\left(\alpha \ln(T)~\tm\right)\quad &\text{ for }\quad\tm\lesssim 1/(\alpha \ln(T))\,,\\
\\
\frac1T \quad &\text{ for }\quad 1/(\alpha \ln(T))\ll \tm \ll T-1/(\alpha \ln(T))\,,\\
\\
\frac1T G\left(\alpha \ln(T)~(T-\tm)\right)\quad &\text{ for }\quad\tm\lesssim 1/(\alpha \ln(T))\,.\\
\end{cases}
\label{PT_asymp_p2}
\end{equation}
Remarkably, the late-time shape of $\PT$, obtained for the harmonic potential ($p=2$) is the same as the one obtained for the potential $V(x)=\alpha|x|$ ($p=1$, see Eq.~\eqref{PT_asymp_p1}). The only difference is the scale of the edge regime, which is $\sim 4D/\alpha^2$ for $p=1$ and $\sim 2/(\alpha \ln(T))$ for $p=2$. The shape of the edge regime is described by the function $G(z)= \left[1+\operatorname{erf}(\sqrt{z})+e^{-z}/\sqrt{\pi z}\right]/2$, which is the same for both $p=1$ and $p=2$. This universality is unexpected and lead us to two natural questions: (i) What is the origin of this function $G(z)$? and (ii) Is the universality of the edge behavior valid for any $p>0$?

\subsection{Universality at late times}
\label{sec:univ}

\begin{figure*}[t]\includegraphics[scale=0.9]{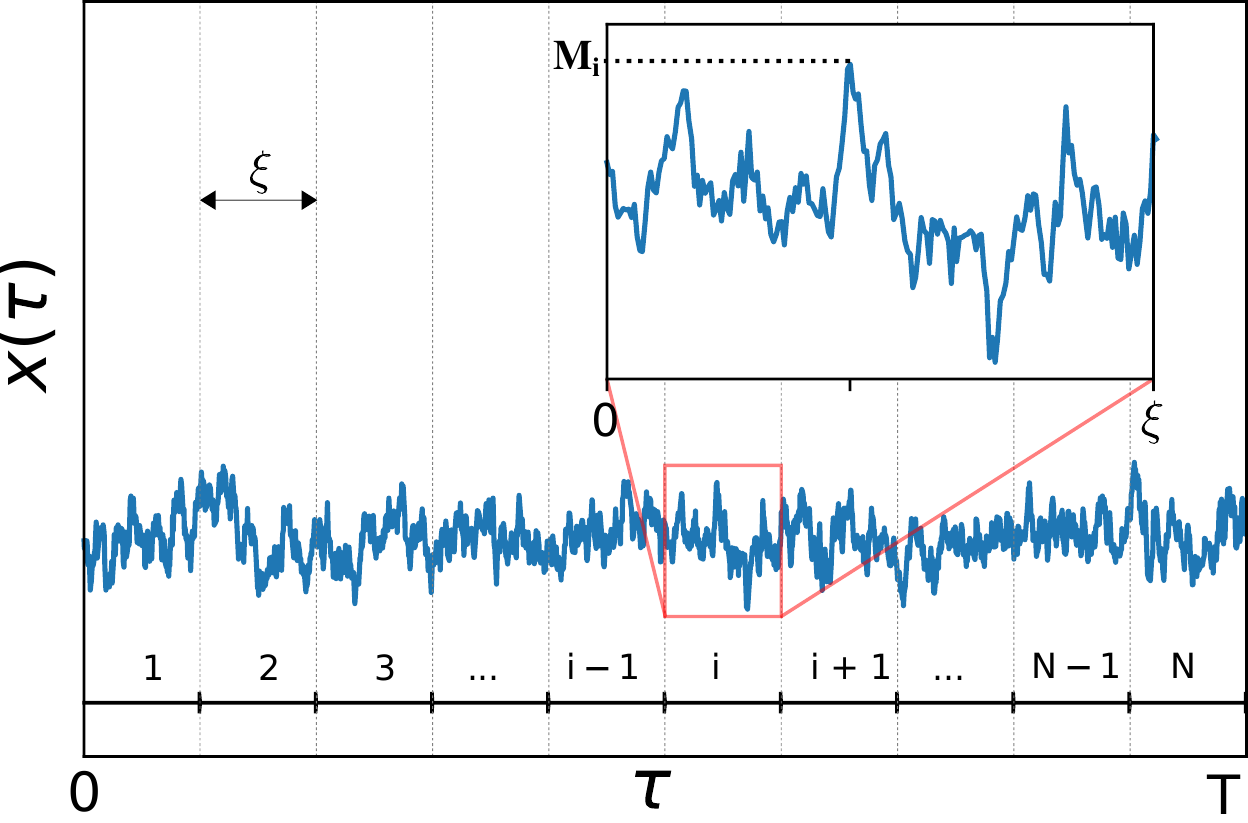} 
\caption{\label{fig:block_argument} Typical trajectory of a stationary process of duration T. The time interval $[0,T]$ is divided into the $N$ subintervals of duration $\xi$, where $\xi$ is the correlation time of the process. The maximum of the process during the $i$-th time window is denoted by $M_i$. The variables $M_1\,,\ldots\,, M_N$ are approximately independent.}
\end{figure*}

In this section, we show that the late-time universality of the distribution of $\tm$ that we have observed for $p=1$ and $p=2$ is actually valid for any $p>0$. This is based on a real-space “blocking argument”, which we describe below. For $p\geq 1$, i.e., if the potential $V(x)$ grows faster than $|x|$ for large $|x|$, one can show that the correlation function decays exponentially in time as \cite{SM20}
\begin{equation}
\langle x(\tau)x(\tau')\rangle-\langle x(\tau)\rangle\langle x(\tau')\rangle\sim e^{-|\tau-\tau'|/\xi}\,,
\end{equation}
where $\xi$ is the correlation time. For $0<p<1$, we have verified numerically that the autocorrelation function has a stretched-exponential decay in time. For instance, for $p=1/2$, we verified numerically that (see Fig.~\ref{fig:correlations})
\begin{equation}
\langle x(t)x(t')\rangle\sim e^{-\sqrt{|t-t'|/\xi}}\,,
\end{equation}
for some timescale $\xi>0$. Thus, also for $0<p<1$ one has a typical timescale over which correlations decay and one can still apply the blocking argument. Note that in our recent Letter \cite{MMS21}, we used an heuristic argument to show that the distribution of $\tm$ becomes universal at late times for $p\geq 1$. Here, we present a more precise version of this argument and we show that this universality is actually valid for $p>0$.

For large $T$ we can thus divide the time interval $[0,T]$ into $N$ blocks of duration $\xi=T/N$ (see Fig.~\ref{fig:block_argument}). Let $M_i$ be the maximal position reached in the $i$-th block. It is clear that the variables $M_1\,,\ldots\,,M_N$ are identically distributed, since the process is in the steady state. Moreover, since we assume that $\xi$ is of the order of the correlation time, they can also be considered independent. Thus the maximum will be reached in a given box with probability $1/N=\xi/T$ and the late-time probability distribution of the $\tm$ is approximately given by the uniform measure
\begin{equation}
P(\tm|T)\approx \frac1N \frac{1}{\xi}=\frac1T\,.
\label{uniform2}
\end{equation}
Note, however, that this argument is only valid when $\xi\ll \tm\ll T-\xi$, i.e., in the bulk of the distribution $P(\tm|T)$. As observed in the exactly solvable cases $p=1$ and $p=2$, in the edge regions $0<\tm<\xi$ and $(T-\xi)<\tm<T$, a more precise analysis is required.

To proceed, let us analyze the behavior of the global maximum $M$ in the limit of large $T$. The global maximum $M$ of the process can be written as the maximum of the local i.i.d. variables $M_1\,,\ldots\,,M_N$
\begin{equation}
M=\max_{1\leq i\leq N}\left(M_i\right)\,.
\end{equation}
Even though we do not know the PDF $P(M_i)$ of the local maximum $M_i$, we can guess that it will have the same large-$M_i$ tail as the equilibrium distribution in Eq.~\eqref{boltzmann}, i.e., that for large $M_i$ one has
\begin{equation}
P(M_i)\sim \exp\left(-\frac{\alpha }{D}~ M_i^p\right)\,.
\label{tail_behavior_1}
\end{equation}
Thus, one can apply the standard extreme value theory for i.i.d. random variables (see, e.g., \cite{MP20}). In particular, to find the leading-order behavior of the global maximum $M$, we need to solve the equation
\begin{equation}
\int_{M}^{\infty}dM'~P(M')=\frac1N\,.
\label{1Nfrac}
\end{equation}
Indeed, we expect the probability of observing a value larger than the global maximum to be $1/N$, since $M_1\,,\ldots\,,M_N$ are i.i.d. variables (this argument can be made more precise, see \cite{MP20}). Plugging the expression for $P(M)$, given in Eq.~\eqref{tail_behavior_1}, into Eq.~\eqref{1Nfrac} and integrating by parts, we find that to leading order
\begin{equation}
	M \approx \left(\frac{D}{\alpha}\ln(N)\right)^{1/p}\,.
\end{equation} 
Moreover, since $N=T/\xi$, we obtain
\begin{equation}
	M \approx \left(\frac{D}{\alpha}\ln(T)\right)^{1/p}\,.
\label{M_leading_order}
\end{equation} 
Interestingly, one can also show that the global maximum concentrates around this deterministic value in Eq.~\eqref{M_leading_order} for large $T$ \cite{MP20}, meaning that the fluctuations around this value are subleading in $T$. Indeed, one can show that the global maximum $M$ of $N$ i.i.d.~variables, each drawn from the PDF in Eq.~\eqref{tail_behavior_1}, grows as
\begin{equation}
M\approx \left(\frac{D}{\alpha}\ln N\right)^{1/p}+\frac{D}{\alpha p}\left(\frac{D}{\alpha}\ln N\right)^{(1-p)/p}z\,,
\end{equation}
where $z$ is Gumbel distributed. Note that the relative ratio of  the fluctuations and the mean decays as $1/\ln N$. Therefore, we can consider the value of the global maximum to be fixed and given by Eq.~\eqref{M_leading_order}.

\begin{figure*}[t]\includegraphics[scale=0.9]{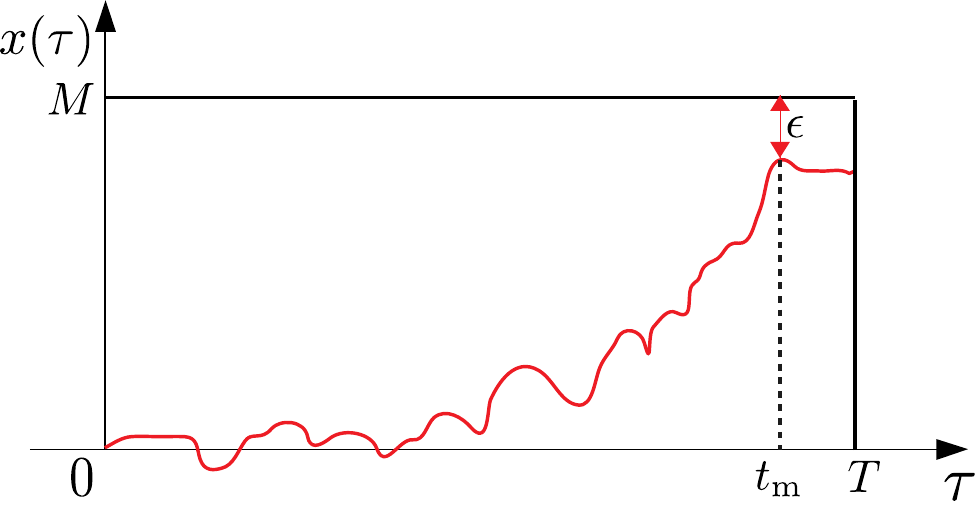} 
\caption{\label{fig:last_block} Schematic representation of a stochastic process where the global maximum $M-\epsilon$ is reached within the right edge, i.e., at time $\tm$ with $T-\tm\ll T$. }
\end{figure*}

Then, we can apply the path-decomposition technique derived at the beginning of this section (see Eq.~\eqref{Ptm_integral}), with the only difference that $M$ is now fixed. Thus, Eq.~\eqref{Ptm_integral} gets modified as follows
\begin{equation}
P(\tm|T)=\lim_{\epsilon\to 0}\left[\mathcal{N}(\epsilon) \int_{-\infty}^{M}dx_0~P_{\rm st}(x_0)G^M(M-\epsilon,\tm|x_0)Q^M(M-\epsilon,T-\tm)\right]\,,
\label{Ptm_integral_2}
\end{equation}
where now $M$ depends on $T$ and is given in Eq.~\eqref{M_leading_order}. We recall that $P_{\rm st}(x_0)$, $G^M(x,t|x_0)$, and $Q^M(x,t)$ respectively indicate the equilibrium distribution, the constrained propagator, and the survival probability of the process. We focus on the right-edge regime, corresponding to configurations in which the global maximum is reached at the end of the time interval $[0,T]$, i.e., for $T-\tm\ll T$ (see Fig.~\ref{fig:last_block}). In this region we can approximate $\tm \approx T$ and therefore Eq.~\eqref{Ptm_integral_2} becomes
\begin{equation}
P(\tm|T)\approx\lim_{\epsilon\to 0}\left[\mathcal{N}(\epsilon) \int_{-\infty}^{M}dx_0~P_{\rm st}(x_0)G^M(M-\epsilon,T|x_0)Q^M(M-\epsilon,T-\tm)\right]\,.
\end{equation}
Absorbing the constant terms, i.e., the terms that are independent of $\tm$, into $\mathcal{N}(\epsilon)$, we obtain
\begin{equation}
P(\tm|T)\approx\lim_{\epsilon\to 0}\left[\mathcal{N}'(\epsilon) Q^M(M-\epsilon,T-\tm)\right]\,,
\label{Ptm_integral_2_new}
\end{equation}
where the constant $\mathcal{N}'(\epsilon)$ can be determined by matching this edge expression in Eq.~\eqref{Ptm_integral_2_new} with the bulk result $\PT\approx 1/T$.

\begin{figure*}[t]
\includegraphics[scale=0.6]{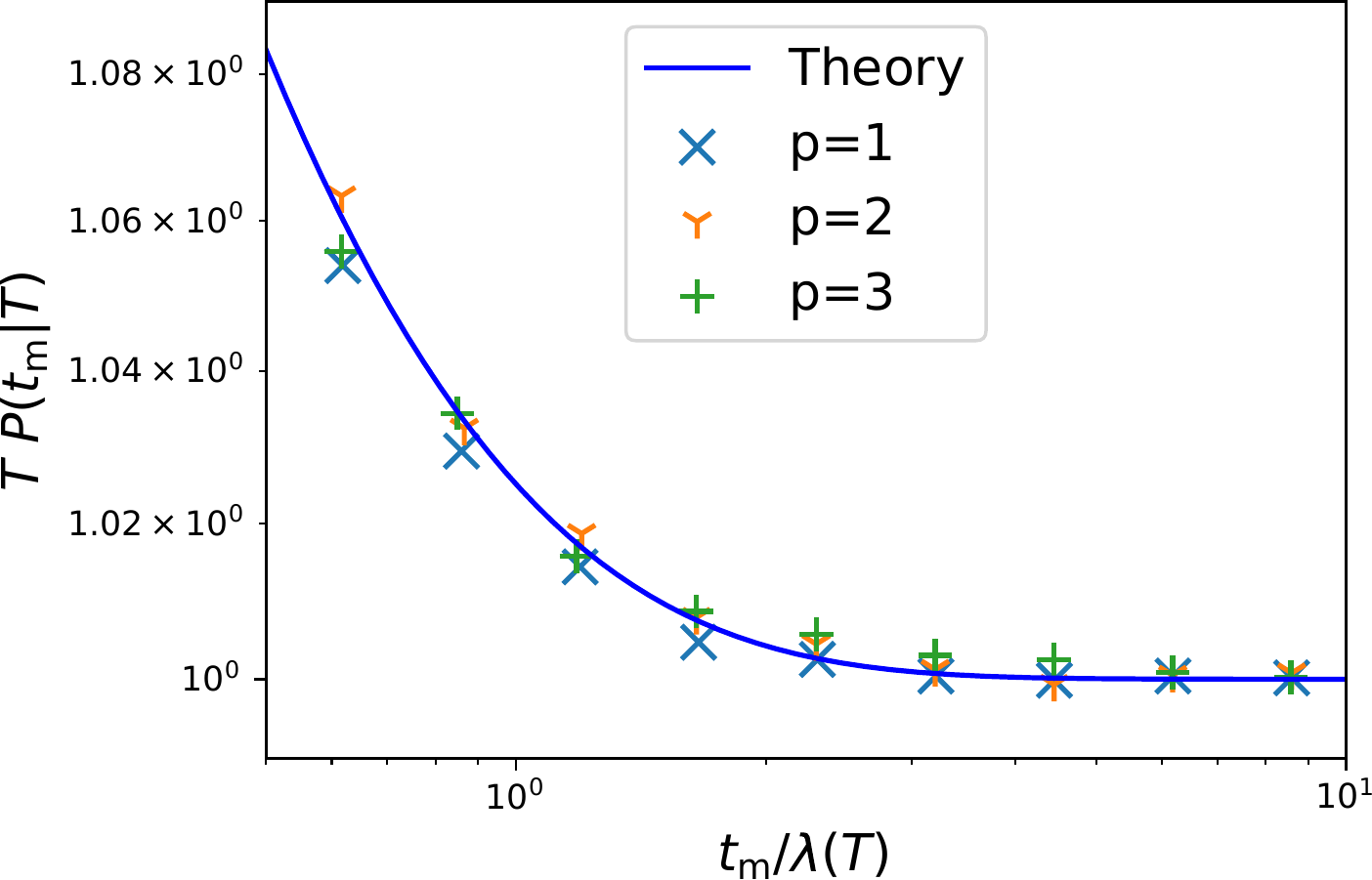} 
\caption{\label{fig:comparison} The scaled probability density function $T\PT$ as a function of the scaled time of the maximum $\tm/\lambda(T)$. The symbols correspond to numerical simulations of Brownian motion in a potential $V(x)=|x|^p$, for different values of $p$ and $T$ large ($T=6400$ for $p=1$ and $T=800$ for $p=2$ and $p=3$). The continuous blue curve corresponds to the exact scaling function $G(z)$, given in Eq.~\eqref{G_summary} and valid for large $T$. For further evidence of the convergence to this scaling function, see Fig.~\ref{fig:conv}.}
\end{figure*}

Thus, we need to compute the survival probability $Q^M(M-\epsilon,T-\tm)$, defined as the probability that the process remains below position $M$ for a time $T-\tm$, having started from position $M-\epsilon$. As previously explained, this is, in general, hard (the only two solvable models are $p=1$ and $p=2$). Nevertheless, since the time interval $[T-\tm,T]$ is short, we expect the position of the particle within this time interval to remain close to the global maximum $M$. Thus, we linearize the potential $V(x)$ around $x=M$ and we obtain
\begin{equation}
V(x)\approx V(M)+(x-M)V'(M)\,.
\end{equation}
As a consequence, to leading order, the effective Langevin equation of the process (see Eq.~\eqref{langevin}) becomes
\begin{equation}
\frac{dx(\tau)}{d\tau}=-V'(M)+\eta(\tau)\,,
\label{largevin2}
\end{equation}
meaning that the particle is subject, in first approximation, to a constant negative drift $\mu=-V'(M)$. Using $V'(x)=\alpha p x^{p-1}$ for $x>0$ and the expression for $M$ in Eq.~\eqref{M_leading_order}, we find that the constant drift $\mu$ is given by
\begin{equation}
\mu=-V'(M)\approx-\alpha~ p\left(\frac{D}{\alpha}\ln(T)\right)^{(p-1)/p}\,.
\label{mu}
\end{equation}
Here we use the definition that the drift $\mu$ is positive when it is pointing towards increasing values of $x$ and negative otherwise (in our case the drift is negative). Crucially, the survival probability of a BM subject to a constant drift $\mu$ can be computed exactly (see Appendix \ref{app:surv_drift}) and reads
\begin{equation}
Q^M(x_0,t)=\frac{1}{2}\left[\operatorname{erfc}\left(-\frac{M-x_0-\mu t}{\sqrt{4Dt}}\right)-e^{\mu (M-x_0)/D}\operatorname{erfc}\left(\frac{M-x_0+\mu t}{\sqrt{4Dt}}\right)\right]\,,
\end{equation}
for $x<M$, where $\operatorname{erfc}(z)=(2/\sqrt{\pi})\int_{z}^{\infty}du~e^{-u^2}$. We recall that the survival probability $Q^M(x,t)$ is defined as the probability that the process remains below position $M$ for a total time $t$, having started from position $x$. Evaluating this expression for $x=M-\epsilon$ and expanding to leading order in $\epsilon$, we find
\begin{equation}
Q^M(M-\epsilon,t)\approx \frac{\epsilon |\mu|}{D}G\left(\frac{\mu^2 t}{4D}\right)\,,
\label{QM_1_exp}
\end{equation}
where $G(z)= \left[1+\operatorname{erf}(\sqrt{z})+e^{-z}/\sqrt{\pi z}\right]/2$. Plugging the expression for $Q^M(M-\epsilon,t)$ in Eq.~\eqref{QM_1_exp} into Eq.~\eqref{Ptm_integral_2} and using the expression for $\mu$ in Eq.~\eqref{mu}, we obtain
\begin{equation}
P(\tm|T)\approx\lim_{\epsilon\to 0}\left[\mathcal{N}'(\epsilon) \frac{\epsilon|\mu|}{D}G\left(\frac{T-\tm}{\lambda(T)}\right)\right]\,,
\label{Ptm_integral_3}
\end{equation}
where
\begin{equation}
\lambda(T)=\frac{4D}{\mu^2}= \frac{4D}{\alpha^2 p^2}\left(\frac{D}{\alpha}\ln(T)\right)^{-2(p-1)/p}\,,
\label{lambda}
\end{equation}
where we have used the expression for $\mu$ in Eq.~\eqref{mu}. To determine the constant $\mathcal{N}'(\epsilon)$ we impose that this edge-regime expression matches the bulk expression in Eq.~\eqref{uniform2}, i.e., that
\begin{equation}
\lim_{\epsilon\to 0}\left[\mathcal{N}'(\epsilon) \frac{\epsilon |\mu|}{D}G\left(\frac{T-\tm}{\lambda(T)}\right)\right]\approx \frac1T\,,
\end{equation}
for $T-\tm\gg \lambda(T)$. Using the fact that $G(z)\approx 1 $ for large $z$, we obtain
\begin{equation}
\lim_{\epsilon\to 0}\left(\mathcal{N}'(\epsilon) \epsilon\right)=\frac{D}{|\mu| T}\,.
\end{equation}
Finally, we get
\begin{equation}
P(\tm|T)\approx \frac{1}{T}G\left(\frac{T-\tm}{\lambda(T)}\right)\,,
\label{eq:edge2_drift_final}
\end{equation}
for $T-\tm \lesssim \lambda(T)$.

\begin{figure*}[t]
\includegraphics[scale=0.4]{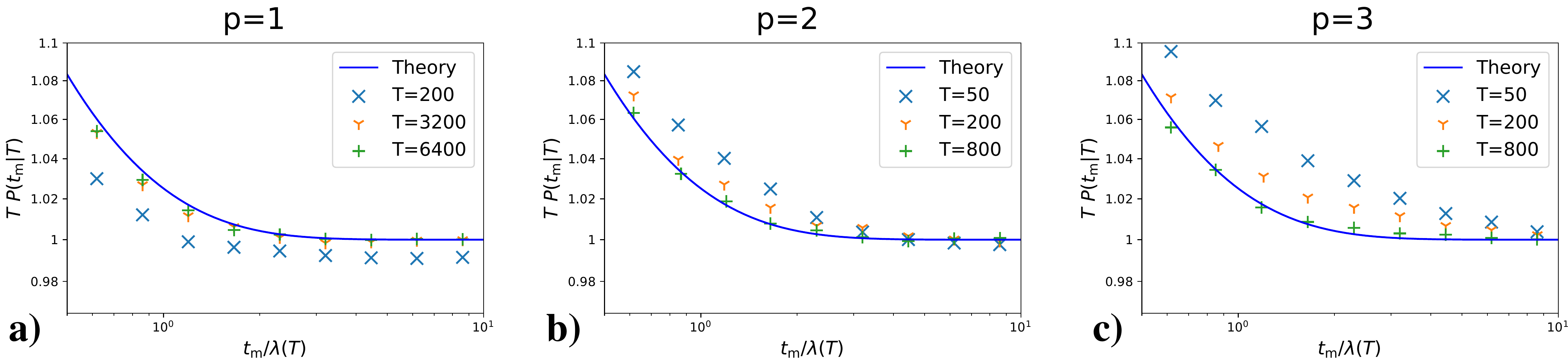} 
\caption{\label{fig:conv} The scaled probability density function $T\PT$ of the time $\tm$ of the maximum for Brownian motion in a confining potential $V(x)=\alpha|x|^p$ as a function of the scaled time of the maximum $\tm/\lambda(T)$ for $p=1$ (a), $p=2$ (b), and $p=3$ (c). The timescale $\lambda(T)$, given in Eq.~\eqref{lambda}, is a function of $\alpha$ and $p$. The continuous blue curve corresponds to the theoretical scaling function $G(z)$, given in Eq.~\eqref{G_summary} and valid for large $T$. This universal curve is valid for any value of $\alpha>0$ and $p\geq 1$. As $T$ increases, the numerical scaling functions for different values of $p$, shown by symbols, approaches the same theoretical scaling function, shown by a solid line.}
\end{figure*}

Even if the width $\lambda(T)$ of the edge region depends on the details of the potential, the shape of $P(\tm|T)$, encoded in the scaling function $G(z)$, becomes completely universal, i.e. independent of $V(x)$, for large $T$. Note that in the special case $p=1$, i.e. when $V(x)=\alpha~|x|$, one finds that $\lambda(T)=4D/\alpha^2$, coinciding with the results of Section \ref{sec:p1}. Thus, for $p=1$ the width of the edge region is independent of $T$. On the other hand, for $p>1$ we observe that $\lambda(T)$ decreases very slowly with $T$ and thus for $T\to \infty$ the edge region slowly disappears. In particular, for $p=2$, we find $\lambda(T)=1/(\alpha\ln(T))$, in agreement with Eq.~\eqref{PT_asymp_p2}. The universal scaling function $G(z)$ has asymptotic behaviors given in Eq.~\eqref{G_asym} and is plotted in Fig.~\ref{fig:Gz}.

\begin{figure}[t]
\begin{center}
\includegraphics[scale=0.5]{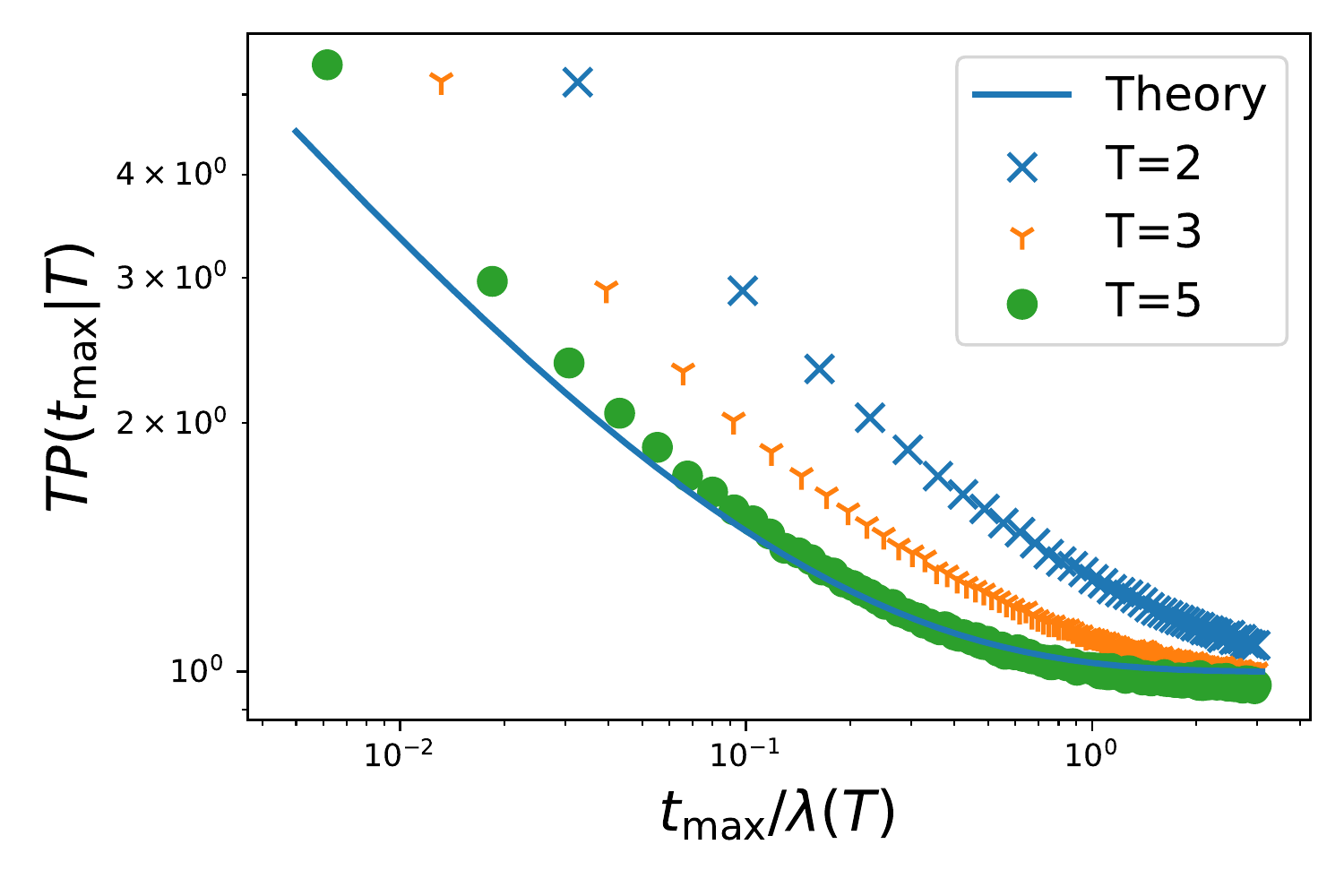}
\caption{Left edge of the scaled distribution $T\PT$ as a function of the scaled time of the maximum $\tm/\lambda(T)$ for $p=1/2$. Note that here the width $\lambda(T)=(16D/\alpha^2)(D\ln(T)/\alpha)^2$ is increasing with $T$. The continuous blue line corresponds to the universal result in Eq.~\eqref{eq:edge1_drift_final}. The symbols are the results of numerical simulations of Brownian motion in the potential $V(x)=4x^2$ for $|x|<1$ and $V(x)=4\sqrt{|x|}$ for $|x|>1$ with different total times $T$. We choose the quadratic part for small $x$ to avoid the divergence in the first derivative $V'(x)$. We observe that already at $T=5$ the numerical results are in excellent agreement the analytical prediction.
\label{fig:convergence_p05}}
\end{center}
\end{figure}

Since the process is at equilibrium, the PDF of $\tm$ satisfies the symmetry $\PT=P(T-\tm |T)$ (see Subsection \ref{sec:criterion}). Therefore, in the left-edge regime, i.e., for $\tm\lesssim \lambda(T)$, we have
\begin{equation}
P(\tm|T)\approx \frac{1}{T}G\left(\frac{\tm}{\lambda(T)}\right)\,.
\label{eq:edge1_drift_final}
\end{equation}
Thus, the late-time behavior of the distribution of $\tm$ can be summarized as
\begin{equation}
P(\tm|T)\approx
\begin{cases}
\frac{1}{T}G\left(\frac{\tm}{\lambda(T)}\right) &~~\text{  for   }~~ \tm\lesssim \lambda(T)\\
\\
\frac1T &~~\text{  for   } ~~\lambda(T)\ll \tm\ll T- \lambda(T)\\
\\
\frac{1}{T}G\left(\frac{T-\tm}{\lambda(t)}\right) &~~\text{  for   } ~~ \tm\gtrsim T- \lambda(T)
\,.
\end{cases}
\label{universal_G}
\end{equation}
Remarkably, the shape of the distribution $\PT$ is completely independent of the details of the potential and valid for any $V(x)$ such that $V(x)\approx\alpha |x|^p$ for large $|x|$, with $\alpha>0$ and $p\geq 1$. The details of the potential, i.e., the parameters $\alpha$ and $p$, only appear in $\PT$ through the width $\lambda(T)$ of the edge region, which slowly shrinks as $\ln(T)^{-2(p-1)/p}$ for $p>1$, is of order one for $p=1$ and expands as $\ln(T)^{2(1-p)/p}$ for $0<p<1$. In Figs.~\ref{fig:comparison}, \ref{fig:conv}, and \ref{fig:convergence_p05}), we compare this universal result in Eq.~\eqref{universal_G} with numerical simulations performed for different values of $p$. We observe that for large $T$ the results of numerical simulations performed with different values of $p$ and appropriately rescaled with the corresponding value of $\lambda(T)$ collapse into the same universal function $G(z)$, given in Eq.~\eqref{G}.

\section{Out-of-equilibrium processes}

\label{sec:neq}

We next focus on the case of nonequilibrium steady states (NESS). This class of stochastic processes is characterized by the violation of the detailed balance condition and by the presence of steady-state probability currents. In the last decades, there has been a surge of interest in characterizing the properties of NESS, especially in the context of living systems. As a consequence of the violation of the detailed balance condition, NESS do not satisfy time-reversal symmetry. To better understand the properties of $\tm$ for nonequilibrium processes, we first investigate two canonical models for which $\PT$ can be computed exactly: resetting Brownian motion (RBM) and a confined run-and-tumble particle (RTP).

\subsection{Resetting Brownian motion}

\label{sec:res_BM}

\begin{figure*}[t]
\includegraphics[scale=0.7]{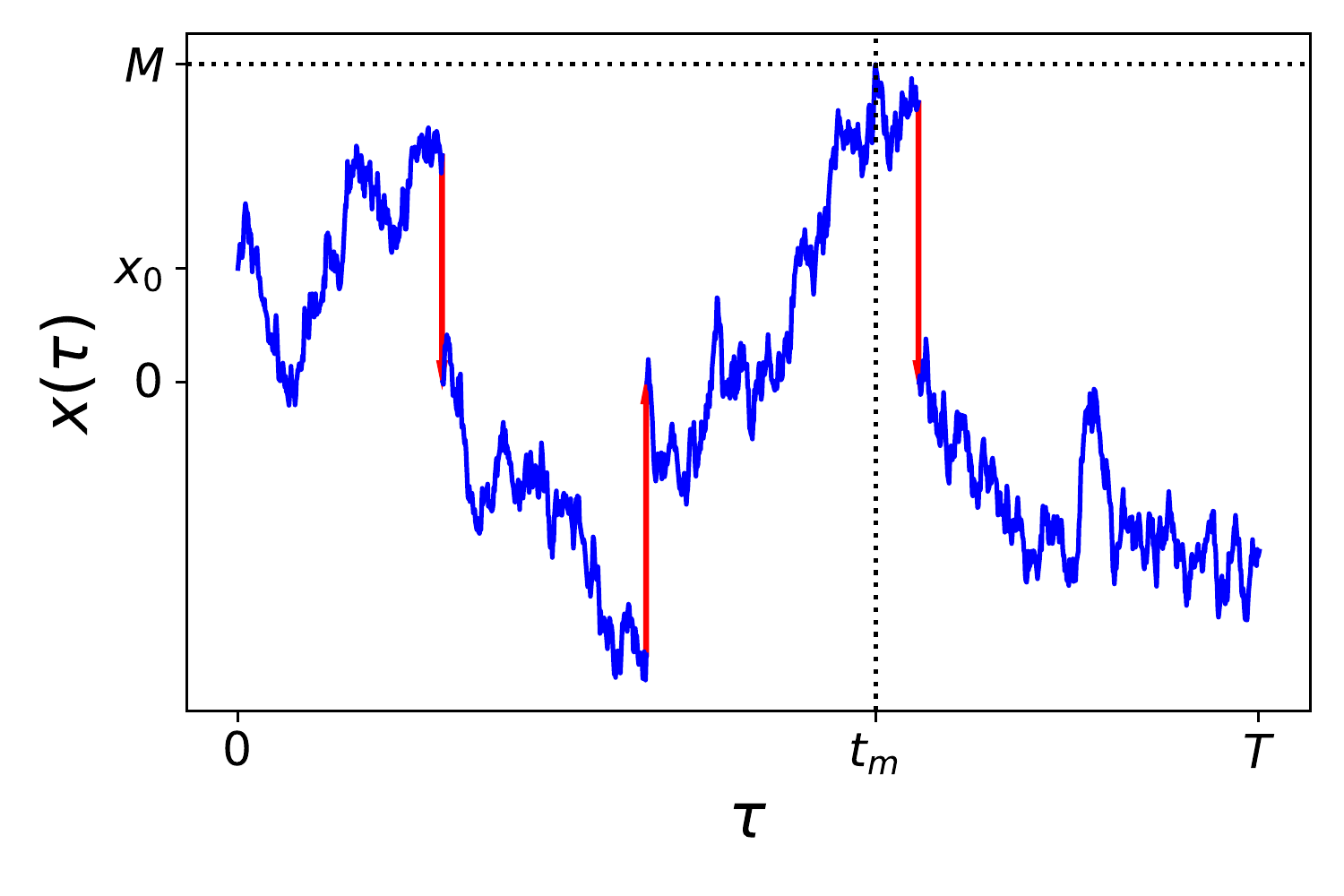} 
\caption{\label{fig:res} Typical trajectory $x(\tau)$ of a Brownian motion with stochastic resetting as a function of time $\tau$ in the interval $[0,T]$. The red segments indicate the resetting events. The particle starts from position $x_0$, drawn from the steady state \eqref{eq:stationary_resetting} and reaches the global maximum $M$ at time $\tm$.}
\end{figure*}

A nonequilibrium version of BM which has been widely investigated recently is BM with stochastic resetting \cite{EM_2011,EMS20}. Stochastic resetting describes dynamical processes that are restarted from some fixed state at random times. Processes with stochastic restarts appear in different disciplines, from computer science \cite{montanari2002optimizing} to chemistry \cite{reuveni2014role}. The restarting dynamics drives the system to a nonequilibrium steady state \cite{EM_2011,EMS20} and induces many interesting phenomena, including dynamical phase transitions \cite{MSS15,BBPM20,FBPC21,MVB20,MVB22}. Besides Brownian motion, resetting has been investigated for several other random processes, including L\'evy flights \cite{kusmierz2014first,kusmierz2015optimal,campos2015phase}, active particles \cite{evans2018run,masoliver2019telegraphic}, fluctuating interfaces \cite{GMS14}, and the Ising model \cite{magoni2020ising}. Moreover, many theoretical predictions for stochastic resetting have been recently verified in experiments \cite{TFPS20,BBPM20,FBPC21,MVB20,MVB22}.

In this section, we investigate the time $\tm$ of the maximum for a Brownian particle $x(\tau)$, evolving with diffusion coefficient $D$ up to time $T$. In addition to the usual diffusive motion, we assume that the particle undergoes stochastic resetting to the origin with constant rate $r$. In a small time interval $dt$, the position of the particle evolves according to
\begin{equation}
x(t+dt)=\begin{cases}
x(t)+\sqrt{2D}\eta(t)dt&\quad\text{ with probabilty }1-rdt\,,\\
\\
0&\quad\text{ with probabilty }rdt\,,\\
\end{cases}
\end{equation}
where $\eta(t)$ is a Gaussian white noise. A typical trajectory is shown in Fig.~\ref{fig:res}. The resetting dynamics drives the system to the stationary state \cite{EM_2011}
\begin{equation}\label{eq:stationary_resetting}
P_{\rm st}(x_0)=\frac{1}{2}\sqrt{\frac{r}{D}}\exp\left(-\sqrt{\frac{r}{D}}|x_0|\right)\,.
\end{equation}
Interestingly, the resetting events produce a net probability current towards the resetting location $x=0$, driving the system out of equilibrium.

Note that the distribution of the time $\tm$ of the maximum for RBM has been also investigated in \cite{SP21}, where the authors considered the case where the initial position of the particle is fixed to $x_0=0$. Here, we assume instead that at the initial time the particle has already reached the steady state, meaning that $x_0=x(0)$ is drawn from $P_{\rm st}(x_0)$ given in Eq.~\eqref{eq:stationary_resetting}.  To compute the distribution of $\tm$, we will use the path-decomposition technique described in Section \ref{sec:eq}. Indeed, it is easy to show that the result in Eq.~\eqref{Ptm_integral_LT} remains valid in the case of RBM. Note that also in this case one has to consider a cutoff $\epsilon$, as explained at the beginning of Section \ref{sec:eq}. In the case of RBM, the result in Eq.~\eqref{Ptm_integral_LT} becomes
\begin{eqnarray}
\nonumber &&\int_{0}^{\infty}dt_1~e^{-s_1 t_1}\int_{0}^{\infty}dt_2~e^{-s_2 t_2}~P(\tm=t_1|T=t_1+t_2)\\
&=&\lim_{\epsilon\to 0}\left[\mathcal{N}(\epsilon) \int_{-\infty}^{\infty}dM~\int_{-\infty}^{M}dx_0~P_{\rm st}(x_0)\tilde{G}_r^M(M-\epsilon,s_1|x_0)\tilde{Q}_r^M(M-\epsilon,s_2)\right]\,,
\label{Ptm_integral_LT_res}
\end{eqnarray}
where we now use the subscript $r$ in $\tilde{G}_r^M(x,s|x_0)$ and $\tilde{Q}_r^M(x,s)$ to stress the dependence on the resetting rate $r$. We recall that $\tilde{G}_r^M(x,s|x_0)$ is the Laplace transform with respect to $t$ of the constrained propagator $G_r^M(x,t|x_0)$, defined as the probability that the process arrives at position $x$ at time $t$, while always remaining below position $M$. Similarly, $\tilde{Q}_r^M(x,s)$ is the Laplace transform with respect to $t$ of the survival probability $Q_r^M(x,t)$, defined as the probability that the process remains below position $M$ for a total time $t$, having started from position $x$. Note that the constant $\mathcal{N}(\epsilon)$ has to be fixed using the normalization condition of $\PT$. To exploit this relation in Eq.~\eqref{Ptm_integral_LT_res}, we first have to derive the constrained propagator $\tilde{G}_r^M(M-\epsilon,s_1|x_0)$ and the survival probability $\tilde{Q}_r^M(M-\epsilon,s_2)$.

We start by computing the survival probability $Q_r^M(x,t)$. It is useful to consider the cases $M<0$ and $M>0$ separately. If the global maximum is negative, then no resetting event has occurred. Indeed, a resetting event would bring the particle to the origin, implying $M\geq 0$. As a consequence, for $M<0$, the survival probability is simply given by
\begin{equation}
Q_r^M(x,t)=e^{-rt}Q_0^M(x,t)\,,
\label{relation_Qr_M<0}
\end{equation}
where the term $e^{-rt}$ is the probability that no resetting event occurs up to time $t$ and the term $Q_0^M(x,t)$ is the survival probability of BM without resetting. The latter quantity can be easily computed by solving the diffusion equation \cite{Redner_book,M05,BMS13} and is given by
\begin{equation}
Q_0^M(x,t)=\operatorname{erf}\left(\frac{M-x}{\sqrt{4Dt}}\right)\theta(M-x)\,,
\end{equation}
where $\theta(z)$ is the Heaviside theta function, i.e., $\theta(z)=0$ for $z<0$ and $\theta(z)=1$ for $z>0$. Considering the Laplace transform of this quantity, we obtain
\begin{equation}
\tilde{Q}_0^M(x,s)= \frac{1}{s}\left[1-e^{-\sqrt{s/D}(M-x)}\right]\theta(M-x)\,.
\label{Q0_LT}
\end{equation}
Setting $x=M-\epsilon$ and expanding for small $\epsilon>0$, we get
\begin{equation}
\tilde{Q}_0^M(M-\epsilon,s)\approx \frac{\epsilon}{\sqrt{ Ds}}\,.
\label{Q0_LT_expanded}
\end{equation}
Taking a Laplace transform of the relation in Eq.~\eqref{relation_Qr_M<0} and using this expansion in Eq.~\eqref{Q0_LT_expanded}, we obtain 
\begin{equation}
\tilde{Q}_r^M(M-\epsilon,s)=\tilde{Q}_0^M(M-\epsilon,s+r)\approx \frac{\epsilon}{\sqrt{ D(s+r)}}\,,
\label{Q0_LT_M<0}
\end{equation}
valid for $M<0$.

When $M>0$, resetting events are possible. Thus, the survival probability $Q_r^M(x,t)$ can be computed by using the following renewal equation
\begin{equation}
Q_r^M(x,t)=e^{-rt}Q_0^M(x,t)+r\int_{0}^{t}d\tau~e^{-r\tau}Q_0^M(x,\tau)Q_r^M(0,t-\tau)\,.
\label{renew_1}
\end{equation}
The first term on the right-hand side of Eq.~\eqref{renew_1} corresponds to the survival of the process with no resetting up to time $t$. The second term describes the case where the first resetting occurs at time $0<\tau<t$. The factor $Q_0^M(x,\tau)$ is the survival probability in the interval $[0,\tau]$, while the factor $Q_r^M(0,t-\tau)$ is the survival probability in the remaining interval $[\tau,t]$. Taking a Laplace transform of Eq.~\eqref{renew_1} and using the convolution theorem, we get
\begin{equation}
\tilde{Q}_r^M(x,s)=\tilde{Q}_0^M(x,r+s)+r\tilde{Q}_0^M(x,r+s)\tilde{Q}_r^M(0,s)\,.
\label{renew_1_LT_}
\end{equation}
Setting $x=0$, we find
\begin{equation}
\tilde{Q}_r^M(0,s)=\frac{\tilde{Q}_0^M(0,r+s)}{1-r~\tilde{Q}_0^M(0,r+s)}\,.
\end{equation}
Substituting this last expression back into Eq.~\eqref{renew_1_LT_}, we get
\begin{equation}
\tilde{Q}_r^M(x,s)=\frac{\tilde{Q}_0^M(x,s+r)}{1-r\tilde{Q}_0^M(0,s+r)}\,,
\end{equation}
which is valid for $M>0$. Using the expression for $\tilde{Q}_0^M(x,s+r)$, given in Eq.~\eqref{Q0_LT}, we have
\begin{equation}
\tilde{Q}_r^M(x,s)=\frac{1-e^{-\sqrt{(s+r)/D}(M-x)}}{s+r e^{-\sqrt{(s+r)/D}M}}\theta(M-x)\,.
\label{LT_Qxs}
\end{equation}
Setting $x=M-\epsilon$, we obtain
\begin{equation}
\tilde{Q}_r^M(M-\epsilon,s)\approx\frac{\epsilon}{\sqrt{D}}\frac{\sqrt{s+r}}{s+r e^{-\sqrt{(s+r)/D}M}}\,.
\end{equation}
To summarize, so far we have shown that to leading order in $\epsilon$
\begin{equation}
\tilde{Q}_r^M(M-\epsilon,s)\approx\begin{cases}
\dfrac{\epsilon}{\sqrt{D}}\dfrac{\sqrt{s+r}}{s+r e^{-\sqrt{(s+r)/D}M}}\quad &\text{ for }M>0\,,\\
\\
\dfrac{\epsilon}{\sqrt{D(s+r)}}\quad &\text{ for }M<0\,.
\end{cases}
\label{Q_final_expanded}
\end{equation}

We next focus on the constrained propagator $G_r^M(x,t|x_0)$. As before, it is useful to consider the cases $M<0$ and $M>0$ separately. For $M<0$, no resetting can occur and the constrained propagator is simply given by
\begin{equation}
G_r^M(x,t|x_0)=e^{-rt} G_0^M(x,t|x_0)\,,
\label{G_M<0}
\end{equation}
where $e^{-rt}$ is the probability that no resetting occurs up to time $t$ and $G_0^M(x,t|x_0)$ is the constrained propagator of BM without resetting. The latter quantity can be computed using the method of images \cite{Redner_book} and reads
\begin{equation}\label{eq:propagtor_r0}
G^M_0(x,t|x_0)=\frac{1}{\sqrt{2\pi Dt}}\left(e^{-(x-x_0)^2 /(4Dt)}-e^{-(2M-x+x_0)^2 /(4Dt)}\right)\theta(M-x)\,.
\end{equation}
Setting $x=M-\epsilon$ and expanding to leading order for small $\epsilon$, we find
\begin{equation}
G^M_0(M-\epsilon,t|x_0)\simeq\frac{(M-x_0)\epsilon}{D t\sqrt{4\pi D t}}e^{-(M-x_0)^2/(4Dt)}\,.
\end{equation}
Taking a Laplace transform with respect to $t$ gives, to leading order in $\epsilon>0$
\begin{equation}\label{eq:tilde_G_0_expanded}
\tilde{G}^M_0(M-\epsilon,s|x_0)\simeq\frac{\epsilon}{D}e^{-\sqrt{s/D}(M-x_0)}\,.
\end{equation}
Considering the Laplace transform of Eq.~\eqref{G_M<0} and using the expansion in Eq.~\eqref{eq:tilde_G_0_expanded}, we obtain
\begin{equation}\label{eq:tilde_G_r_expanded_M<0}
\tilde{G}^M_r(M-\epsilon,s|x_0)\simeq\frac{\epsilon}{D}e^{-\sqrt{(s+r)/D}(M-x_0)}\,,
\end{equation}
valid for $M<0$.

On the other hand, in the case $M>0$, resetting events can occur and the propagator satisfies the renewal equation
\begin{equation}
G^M_r(x,t|x_0)=e^{-rt}G^M_0(x,t|x_0)+r \int_{0}^{t}d\tau\,e^{-r\tau}Q^M_r(x_0,t-\tau)G^M_0(x,\tau|0)\,.
\end{equation}
The first term on the right-hand side corresponds to the case where no resetting occurs. The second term corresponds to the case where the last resetting event occurs at time $t-\tau$ and the factor $Q^M_r(x_0,t-\tau)$ is the probability that the particle remains below position $M$ up to time $t-\tau$. Taking a Laplace transform with respect to $t$ yields
\begin{equation}
\tilde{G}^M_r(x,s|x_0)=\tilde{G}^M_0(x,s+r|x_0)+r \tilde{Q}_r^M(x_0,s)\tilde{G}^M_0(x,s+r|0)\,.
\end{equation}
Finally, setting $x=M-\epsilon$, using Eq.~\eqref{LT_Qxs}, and expanding to leading order in $\epsilon$, we find
\begin{equation}\label{eq:G_r_lt_final}
\tilde{G}_r(M-\epsilon,s|x_0)\simeq\frac{\epsilon}{D}~\frac{\left[r+s\, e^{\sqrt{(s+r)/D}x_0} \right]}{\left[r+s\, e^{\sqrt{(s+r)/D}M}\right]}\,,
\end{equation}
which is valid for $M>0$. To summarize, we have shown that to leading order in $\epsilon$
\begin{equation}
\tilde{G}_r(M-\epsilon,s|x_0)\approx\begin{cases}
\dfrac{\epsilon}{D}~\dfrac{\left[r+s\, e^{\sqrt{(s+r)/D}x_0} \right]}{\left[r+s\, e^{\sqrt{(s+r)/D}M}\right]}\quad &\text{ for }M>0\,,\\
\\
\dfrac{\epsilon}{D}e^{-\sqrt{(s+r)/D}(M-x_0)}\quad &\text{ for }M<0\,.
\end{cases}
\label{G_final_expanded}
\end{equation}

\begin{figure*}[t]\includegraphics[scale=0.7]{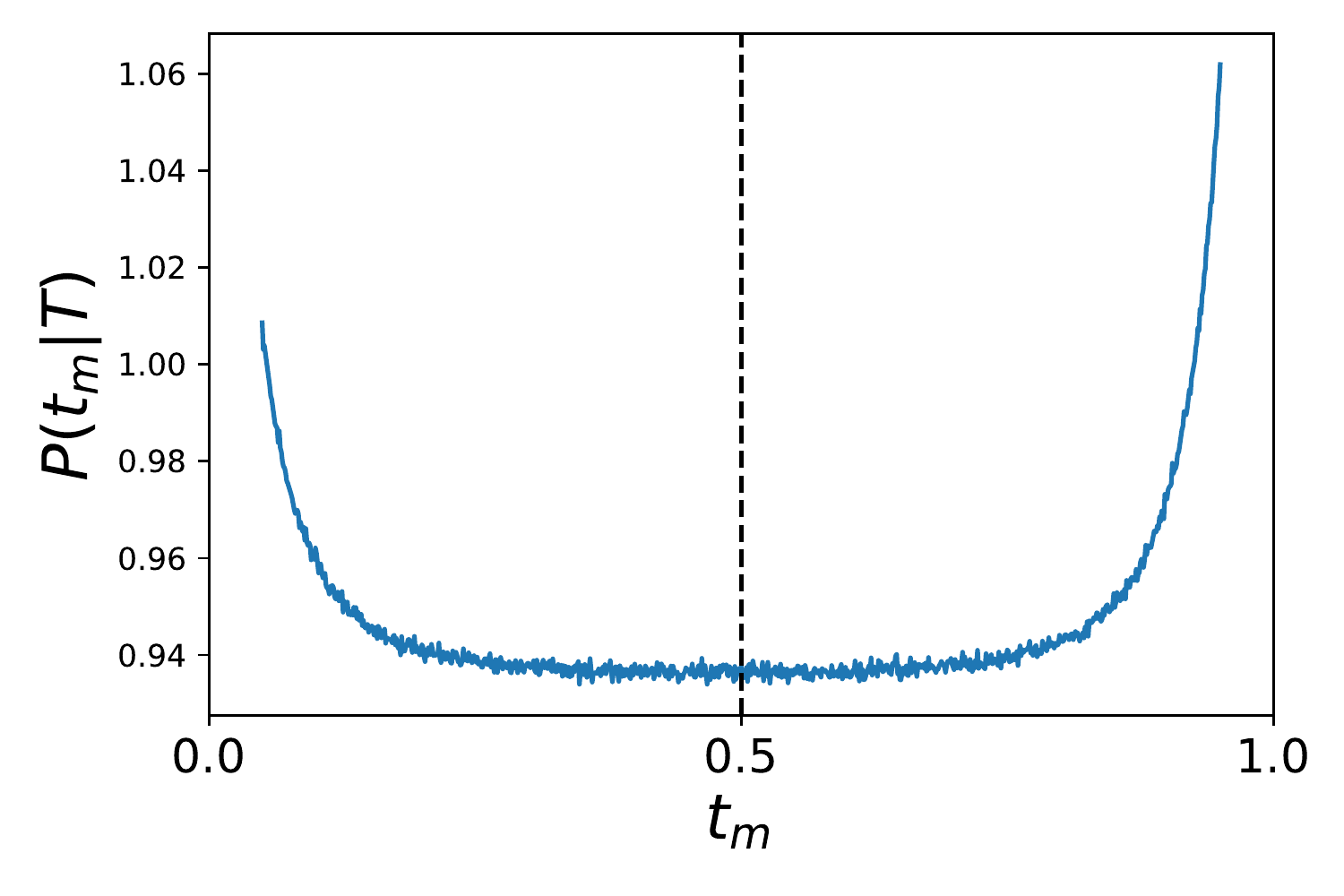} 
\caption{\label{fig:res_ptm} Probability density function $\PT$ as a function of the time $\tm$ of the maximum, obtained from numerical simulations of resetting Brownian motion with $D=T=1$ and $r=10$. The vertical dashed line indicates the midpoint $\tm=T/2$. As a consequence of the nonequilibrium nature of the process, the distribution $\PT$ is not symmetric around $\tm=T/2$. }
\end{figure*}

We now have all the ingredients to compute the PDF $\PT$. Substituting the expressions for $P_{\rm st}(x_0)$, $\tilde{Q}_r^M(M-\epsilon,s)$, and $\tilde{G}_r^M(M-\epsilon,s|x_0)$, respectively given in Eqs.~\eqref{eq:stationary_resetting}, \eqref{Q_final_expanded}, and \eqref{G_final_expanded}, into Eq.~\eqref{Ptm_integral_LT_res} we obtain
\begin{eqnarray}
&&\int_{0}^{\infty}dt_1~e^{-s_1 t_1}\int_{0}^{\infty}dt_2~e^{-s_2 t_2}~P(\tm=t_1|T=t_1+t_2)=\lim_{\epsilon\to 0}\left[\mathcal{N}(\epsilon)\epsilon^2\right]\frac{1}{2}\frac{\sqrt{r}}{D^2} \left\{\int_{-\infty}^{0}dM~\int_{-\infty}^{M}dx_0~e^{\sqrt{r/D} x_0}\right.\nonumber \\ &\times & \left. 
\frac{e^{-\sqrt{(s_1+r)/D}(M-x_0)}}{\sqrt{s_2+r}}+\int_{0}^{\infty}dM~\int_{-\infty}^{M}dx_0~e^{-\sqrt{r/D}|x_0|} \frac{\left[r+s_1\, e^{\sqrt{(s_1+r)/D}x_0} \right]}{\left[r+s_1\, e^{\sqrt{(s_1+r)/D}M}\right]}~\frac{\sqrt{s_2+r}}{\left[s_2+r e^{-\sqrt{(s_2+r)/D}M}\right]}\right\}\,,
\end{eqnarray}
where we recall that we integrate the initial position $x_0$ over the interval $(-\infty,M)$ because by definition the variable $M$ is the global maximum and hence $M>x_0$.
This expression can be simplified by performing the change of variables $(x_0,M)\to(w=x_0\sqrt{r/D},z=M\sqrt{r/D})$, which gives
\begin{eqnarray}
&&\int_{0}^{\infty}dt_1~e^{-s_1 t_1}\int_{0}^{\infty}dt_2~e^{-s_2 t_2}~P(\tm=t_1|T=t_1+t_2)=\lim_{\epsilon\to 0}\left[\mathcal{N}(\epsilon)\epsilon^2\right]\frac{1}{2}\frac{1}{D\sqrt{r}} \left\{\int_{-\infty}^{0}dz~\int_{-\infty}^{z}dw~e^{w}\right.\nonumber \\ &\times & \left. 
\frac{e^{-(z-w)\sqrt{1+s_1/r}}}{\sqrt{s_2+r}}+\int_{0}^{\infty}dz~\int_{-\infty}^{z}dw~e^{-|w|} \frac{\left[r+s_1\, e^{w\sqrt{1+s_1/r}} \right]}{\left[r+s_1\, e^{z\sqrt{1+s_1/r}}\right]}~\frac{\sqrt{s_2+r}}{\left[s_2+r e^{-z\sqrt{1+s_2/r}}\right]}\right\}\,.
\end{eqnarray}
Computing the integrals over $w$, we get
\begin{eqnarray}
&&\int_{0}^{\infty}dt_1~e^{-s_1 t_1}\int_{0}^{\infty}dt_2~e^{-s_2 t_2}~P(\tm=t_1|T=t_1+t_2)=\lim_{\epsilon\to 0}\left[\mathcal{N}(\epsilon)\epsilon^2\right]\frac{1}{2}\frac{1}{D r}   \Bigg\{ \frac{1}{(1+\sqrt{1+s_1/r})\sqrt{1+s_2/r}}\nonumber \\ &+& \frac{\sqrt{1+s_2/r}}{\sqrt{1+s_1/r}-1}\int_{0}^{\infty}dz~e^{-(1+\sqrt{1+s_1/r})z}~\frac{e^{z\sqrt{1+s_1/r}} s_1/r-\sqrt{1+s_1/r}+1} {\left(s_1/r+ e^{-z\sqrt{1+s_1/r}}\right)\left(s_2/r+ e^{-z\sqrt{1+s_2/r}}\right)}\Bigg\}\,.
 \label{eq:int_res}
\end{eqnarray}
In order to fix the normalization constant $\mathcal{N}(\epsilon)$, we set $s_1=s_2=s$ on both sides of Eq.~\eqref{eq:int_res}. The left-hand side can be evaluated by using the fact that the PDF $\PT$ is normalized to unity over $\tm$ (see Eq.~\eqref{lhs}) and is equal to $1/s$. Evaluating the integrals on the right-hand side with Mathematica, we find
\begin{equation}
\frac1s=\lim_{\epsilon\to 0}\left[\mathcal{N}(\epsilon)\epsilon^2\right]\frac{1}{D s}\,,
\end{equation}
and hence
\begin{equation}
\lim_{\epsilon\to 0}\left[\mathcal{N}(\epsilon)\epsilon^2\right]=D\,.
\end{equation}
Using this expression, we can finally write the PDF $\PT$ in the scaling form
\begin{equation}
\PT=rF_R(r\tm,r(T-\tm))\,,
\label{scaling_res}
\end{equation}
where 
\begin{eqnarray}
\int_{0}^{\infty}dT_1~e^{-s_1 T_1}\int_{0}^{\infty}dT_2~e^{-s_2 T_2}~F_R(T_1,T_2)&=&\frac{1}{2}  
 \frac{1}{(1+\sqrt{1+s_1})\sqrt{1+s_2}}\\&+&\frac12 \frac{\sqrt{1+s_2}}{\sqrt{1+s_1}-1}\int_{0}^{\infty}dz~e^{-(1+\sqrt{1+s_1})z}\frac{e^{z\sqrt{1+s_1}} s_1-\sqrt{1+s_1}+1} {\left(s_1+ e^{-z\sqrt{1+s_1}}\right)\left(s_2+ e^{-z\sqrt{1+s_2}}\right)}\,.\nonumber
 \label{FR_LT}
\end{eqnarray}
Interestingly, this expression is not invariant under exchange of $s_1$ and $s_2$. As a consequence $P(T_1,T_2)\neq P(T_2,T_1)$ and thus the PDF $\PT$ is not symmetric around the midpoint $\tm=T/2$. This is confirmed by numerical simulations (see Fig. ~\ref{fig:res_ptm}).

\subsubsection{Expected time of the maximum}

As a consequence of the asymmetry of the time $\tm$ of the maximum, the average value $\langle \tm \rangle$ is different from $T/2$. Therefore, it is interesting to investigate the behavior of $\langle \tm \rangle$ as a function of the total time $T$. The deviations of this quantity $\langle \tm \rangle$ from the equilibrium value $T/2$ quantify the degree of asymmetry of the distribution and consequently the nonequilibrium nature of the process.

To study this average value, we differentiate both sides of Eq.~\eqref{FR_LT_der} with respect to $s_1$ and then we set $s_1=s_2=s$, yielding
\begin{eqnarray}\nonumber
&&\int_{0}^{\infty}dT_1~T_1 e^{-s T_1}\int_{0}^{\infty}dT_2~e^{-s T_2}~F_R(T_1,T_2)=  
 \frac{1}{4(1+s)(1+\sqrt{1+s})^2}\\&+& \frac{1}{4\sqrt{1+s}(\sqrt{1+s}-1)^2}\int_{0}^{1}du~u^{1/\sqrt{1+s}}\frac{s\left[3+s/u-2\sqrt{1+s}+(\sqrt{1+s}-1)\ln(u)\right]-2(\sqrt{1+s}-1)}{(s+u)^3}\,,
 \label{FR_LT_der}
\end{eqnarray}
where we have performed the change of variable $z\to u=-\ln(z)/\sqrt{1+s}$. Let us first consider the left-hand side of Eq.~\eqref{FR_LT_der}, which can be rewritten as
\begin{equation}
\int_{0}^{\infty}dT_1~T_1 e^{-s T_1}\int_{0}^{\infty}dT_2~e^{-s T_2}~F_R(T_1,T_2)=\int_{0}^{\infty}d\tilde{T}~e^{-s \tilde{T}}\int_{0}^{\tilde{T}}dT_1~T_1~F_R(T_1,T_2)=\int_{0}^{\infty}d\tilde{T}~e^{-s \tilde{T}}\langle T_1 (\tilde{T})\rangle
\end{equation} 
where we have made the change of variable $(T_1,T_2)\to (T_1,\tilde{T}=T_1+T_2)$ and we have defined 
\begin{equation}
\langle T_1 (\tilde{T})\rangle=\int_{0}^{\tilde{T}}dT_1~T_1~F_R(T_1,\tilde{T}-T_1)\,.
\end{equation}
Note that $T_1$ and $\tilde{T}=T_1+T_2$ respectively correspond to the rescaled time of the maximum $T_1=r\tm$ and the rescaled total time $\tilde{T}=rT$ (see Eq.~\eqref{scaling_res}).

The Laplace transform in Eq.~\eqref{FR_LT_der} can be inverted (see Appendix \ref{app:LI_2}), yielding
\begin{equation}
\langle T_1(\tilde{T})\rangle=\tilde{T} f(\tilde{T})\,.
\end{equation}
Reintroducing dimensions, this corresponds to
\begin{equation}
\label{avg}
\langle \tm(T)\rangle=T f(rT)\,.
\end{equation}
where the scaling function $f(t)$ is given by
\begin{eqnarray} \label{foft}
f(t)&=&\frac{1}{96}\left[-4(2t^2+3t-18)+\frac{2}{\sqrt{\pi}}\frac{1}{\sqrt{t}}(3+16t+4t^2)e^{-t}+(-3-30t+36t^2+8t^3)\frac1t \operatorname{erf}(\sqrt{t})\right]\nonumber\\ &+&\frac{1}{2t}\left[e^{-t}-\frac{2}{\sqrt{\pi}}\Gamma\left(\frac32,t\right)\right]+\sum_{k=1}^{\infty} \frac1t g_k(t)\,,
\end{eqnarray}
and $\Gamma(a,t)=\int_{t}^{\infty}x^{a-1}e^{-x}$ is the upper incomplete Gamma function. The function $g_k(t)$ reads 
\begin{equation}
g_k(t)=(-1)^k\frac12 (k+1)(k+2)\int_{0}^{t}d\tau\,h_k(t-\tau)\tau^{k+1}\left(\frac{1}{(k+1)!}+\frac{\tau}{(k+2)!}\right)\,,
\end{equation}
where
\begin{eqnarray}
\label{hk_1_}
h_k(t)&=&\frac{1}{k^2}\left\{-e^{-t+t/k^2}k(1-k)^2+e^{-t}\frac{k\left[k(1+k)^3-2k^3 t\right]}{\sqrt{\pi t}(1+k)^3}\right\}+\frac{1}{k^2}\left[\operatorname{erf}\left(\frac{\sqrt{t}}{k}\right)e^{-t+t/k^2}(1-k)^2\right]\nonumber \\
&\times &\frac{1}{(1+k)^4}e^{-t+t/(1+k)^2}\left[(1+k)^2 (k^2-2)+2kt\right]\left[1-\operatorname{erf}\left(\frac{\sqrt{t}}{(k+1)}\right)\right]\,.
\end{eqnarray}

\begin{figure}[t]\includegraphics[scale=0.7]{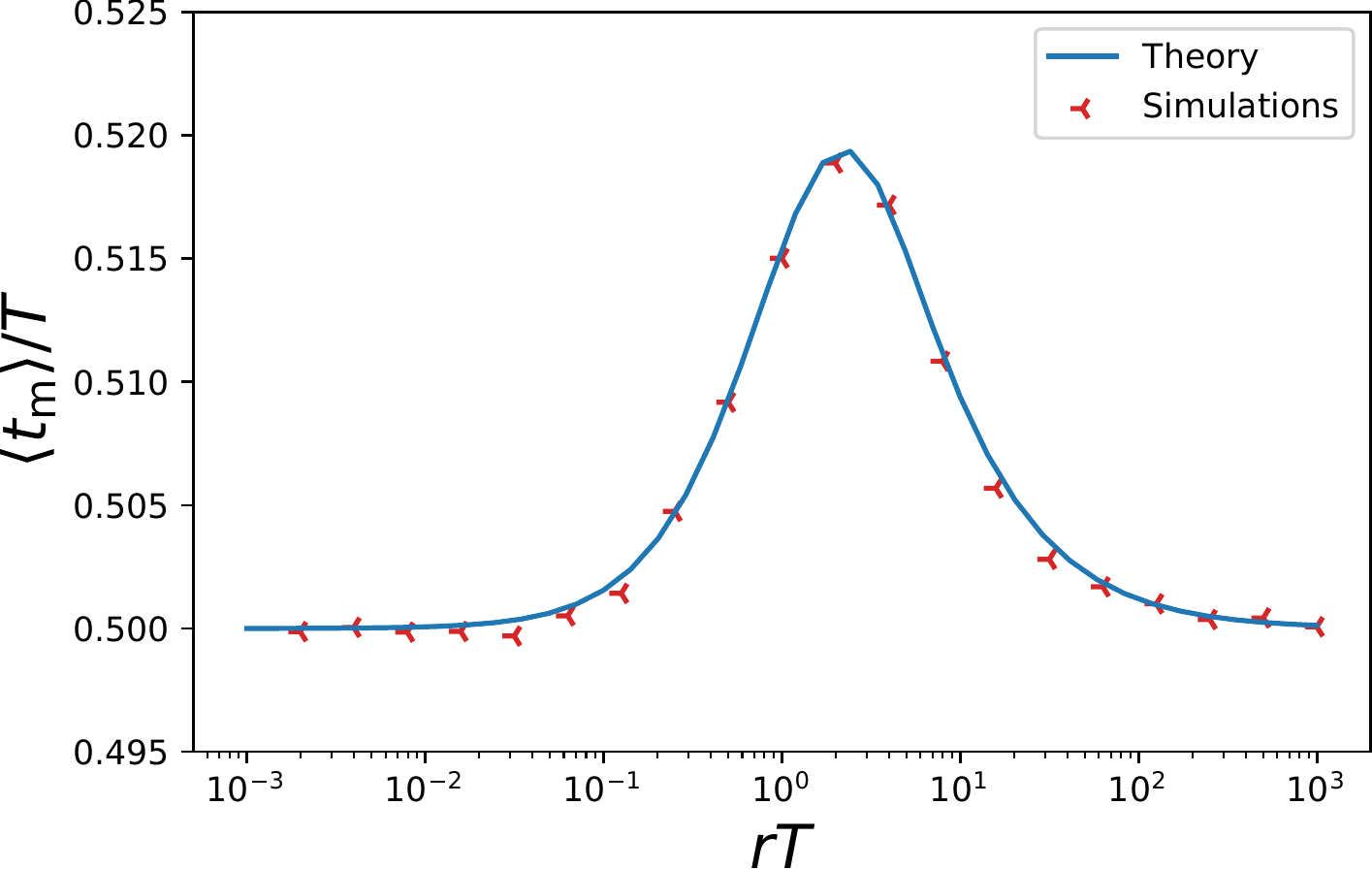} 
\caption{\label{fig:avg_tmax} The scaled average $\langle t_{\rm m}\rangle/T$ as a function of $rT$ for Brownian motion with resetting rate $r$. The symbols depict the results of numerical simulations (performed with $r=1$) while the continuous line corresponds to the analytical results in Eqs.~(\ref{avg}-\ref{hk_1_}). In the case of an equilibrium process, one expects $\langle  t_{\rm m}\rangle/T=1/2$ for any $T$.}
\end{figure}

The exact result in Eqs.~\eqref{avg} and \eqref{foft} is shown in Fig. \ref{fig:avg_tmax} and is in perfect agreement with numerical simulations. We observe that the ratio $\langle t_{\rm m}\rangle/T$ is manifestly different from the constant value $1/2$, signaling that the process violates detailed balance. Note also that the function $f(t)$ has a maximum at $t^*\approx 2.218$ with $f(t^*)\approx 0.519$. Thus, keeping $T$ fixed, there exists a value of the resetting rate $r$ that maximizes the deviation from the equilibrium result.

\subsubsection{Time asymptotics}

Although it is quite challenging to invert exactly the double Laplace transform in Eq.~\eqref{FR_LT}, this expression can be used to extract the asymptotic behavior of the distribution $\PT$ in the limit of short times ($T\ll \xi$) and late times ($T\gg \xi$). Here, the correlation time $\xi$ of the process is $\xi=1/r$, where $r$ is the resetting rate. This quantity $\xi$ represents the typical time between two consecutive resetting events. When $T\ll 1/r$, hardly any resetting event has occurred and we expect to recover the results obtained for Brownian motion without resetting. On the other hand, for $T\gg 1/r$, the positions of the process at different times become uncorrelated and we expect the distribution $\PT$ to become uniform in the interval $[0,T]$, with corrections for $\tm\to 0$ and $\tm \to T$.

To investigate the short time regime $T\ll 1/r$, we take the limit $s_1,s_2\to \infty$ on the right-hand side of Eq.~\eqref{FR_LT}
\begin{eqnarray}
\int_{0}^{\infty}dT_1~e^{-s_1 T_1}\int_{0}^{\infty}dT_2~e^{-s_2 T_2}~F_R(T_1,T_2)\approx\frac{1}{2}  
 \frac{1}{\sqrt{s_1}\sqrt{s_2}}+\frac12 \frac{1}{\sqrt{s_1}\sqrt{s_2}}\int_{0}^{\infty}dz~e^{-z}\,,
 \label{FR_LT_short_time}
\end{eqnarray}
and hence
\begin{equation}
\int_{0}^{\infty}dT_1~e^{-s_1 T_1}\int_{0}^{\infty}dT_2~e^{-s_2 T_2}~F_R(T_1,T_2)\approx \frac{1}{\sqrt{s_1}\sqrt{s_2}}\,.
\end{equation}
Using the Laplace inversion in Eq.~\eqref{inv_lapl_sqrt}, we get
\begin{equation}
F_R(T_1,T_2)\approx\frac{1}{\pi\sqrt{T_1T_2}}\,.
\end{equation}
Hence, for $T\ll1/r$, we obtain
\begin{equation}
\PT\approx\frac{1}{\pi\sqrt{\tm(T-\tm)}}\,,
\end{equation}
which is once again L\'evy arcsine law \cite{Levy}, i.e., the distribution of the time of the maximum for a free BM, as expected.

It turns out that the late time regime $T\gg 1/r$ is somewhat similar to the one described in Section \ref{sec:eq} for equilibrium processes. Indeed, there are three distinct regimes, depending on $\tm$: the left edge regime, where $\tm\sim 1/r$, the bulk regime, where $1/r\ll\tm\ll (T-1/r)$, and the right edge, where $T-\tm\sim 1/r$. We first focus on the left edge, corresponding to $T_1=r \tm\sim O(1)$ and $T_2=r(T-\tm)\to \infty$. Performing the change of variable $z\to u=\exp(-\sqrt{1+s_2}z)$ in Eq.~(\ref{FR_LT}), we obtain
\begin{eqnarray}\label{eq:LT_F_edge1_res}
&&\tilde{F}_R(s_1,s_2)\equiv \int_{0}^{\infty}dT_1\,\int_{0}^{\infty}dT_2\,e^{-s_1 \,T_1-s_2 \,T_2} F_R(T_1,T_2)=\frac{1}{2} \frac{1}{\sqrt{1+s_2}\left(\sqrt{1}+\sqrt{1+s_1}\right)}\\
&+&\frac12\frac{1}{\sqrt{s_1+1}-1}\int_{0}^{1}du\,\frac{u^{(1+\sqrt{1+s_1})/\sqrt{1+s_2}-1}\left(s_1 u^{-\sqrt{(1+s_1)/(1+s_2)}}-\sqrt{s_1+1}+1 \right)}{\left(s_1+u^{\sqrt{(1+s_1)/(1+s_2)}}\right)\left(s_2+u\right)}\,.\nonumber
\end{eqnarray}
To investigate the limit of large $T_2$, we expand the right-hand side of Eq. (\ref{eq:LT_F_edge1_res}) for small $s_2$ and we obtain
\begin{eqnarray}\label{eq:LT_F_edge1_res2}
\tilde{F}_R(s_1,s_2)\approx\frac12\frac{1}{\sqrt{s_1+1}-1}\int_{0}^{1}du\,\frac{s_1 +(1-\sqrt{s_1+1})u^{\sqrt{1+s_1}}}{\left(s_1+u^{\sqrt{1+s_1}}\right)\left(s_2+u\right)}\,.
\end{eqnarray}
We observe that the expression on the right-hand side of Eq. (\ref{eq:LT_F_edge1_res2}) has a pole at $s_2=-u$, thus the integral over $u$ will be dominated by small values of $u$. Thus, we neglect the term $u^{\sqrt{1+s_1}}$ and we obtain
\begin{eqnarray}\label{eq:LT_F_edge1_res3}
\tilde{F}_R(s_1,s_2)=\frac12\frac{1}{\sqrt{s_1+1}-1}\int_{0}^{1}du\,\frac{1}{s_2+u}\,.
\end{eqnarray}
Using the Laplace-inversion formula in Eq.~\eqref{G_LT} to invert the double Laplace transform, we obtain 
\begin{equation}
F_R(T_1,T_2)\simeq G(T_1)\int_{0}^{1}du~e^{-u T_2}\,,
\end{equation}
where
\begin{equation}\label{eq:G}
G(z)= \frac{1}{2}\left[ 1+ {\rm erf}\left(\sqrt{z}\right)+ \frac{1}{\sqrt{\pi z}}\, e^{-z}\right]\, .
\end{equation}
Finally, computing the integral over $u$, we find that, for $T_1\sim \mathcal{O}(1)$ and $T_2\to \infty$, 
\begin{equation}\label{eq:edge1_res}
F_R(T_1,T_2)\simeq \frac1T G(T_1)\,.
\end{equation}
Interestingly, we find that the same scaling function $G(z)$, that describes the left-edge behavior for equilibrium processes in Section \ref{sec:eq} (see Eq.~\eqref{universal_G}), also describes the left-edge behavior for this out-of-equilibrium Brownian resetting process.

Let us now consider the right edge, i.e., the limit $T_1\to \infty$ with $T_2=T-T_1\sim O(1)$. Performing the change of variable $z\to u=\exp(-\sqrt{1+s_1}z)$ in Eq. (\ref{FR_LT}), we obtain
\begin{eqnarray}\label{eq:LT_F_edge2_res}
\tilde{F}_R(s_1,s_2)=\frac{1}{2} \frac{1}{\sqrt{1+s_2}\left(\sqrt{1}+\sqrt{1+s_1}\right)}+\frac12\frac{1}{\sqrt{s_1+1}-1}\sqrt{\frac{1+s_2}{1+s_1}}\int_{0}^{1}du\,\frac{u^{1/\sqrt{1+s_1}}\left(\frac{s_1}{u}-\sqrt{1+s_1}+1 \right)}{\left(s_1+u\right)\left(s_2+u^{\sqrt{(1+s_2)/(1+s_1)}}\right)}\,.
\end{eqnarray}
For small values of $s_1$ the integral on the right-hand side of Eq. (\ref{eq:LT_F_edge2_res}) is dominated by small values of $u$. Thus, expanding for small $s_1$ and small $u$, we get
\begin{eqnarray}\label{eq:LT_F_edge2_res2}
\tilde{F}_R(s_1,s_2)\approx\frac{\sqrt{s_2+1}}{s_2}\int_{0}^{1}du\,\frac{1}{s_1+u}\,.
\end{eqnarray}
Inverting the double Laplace transform with Eq.~\eqref{G_LT}, we find that when $T_1\to\infty$ with $T_2\sim O(1)$
\begin{equation}\label{eq:edge2_res}
F_R(T_1,T_2)\approx \frac{1}{T}\left(2 G(T_2)-1\right)\,,
\end{equation}
where $G(T_2)$ is given in Eq. (\ref{eq:G}).

The bulk regime corresponds instead to the limit $s_1,s_2\to 0$ in Eq.~\eqref{FR_LT}, yielding
\begin{equation}
\tilde{F}_R(s_1,s_2)\approx \int_{0}^{\infty}dz~e^{-z}\frac{1}{(s_1+e^{-z})(s_2+e^{-z})}\,.
\end{equation}
Inverting the double Laplace transform, we obtain
\begin{equation}
F_R(T_1,T_2)\approx \int_{0}^{\infty}dz~e^{-z}e^{(T_1+T_2)e^{-z}}=\frac{1}{T_1+T_2}\,,
\end{equation}
corresponding to the flat distribution
\begin{equation}
\PT\approx\frac{1}{T}\,.
\end{equation}
To summarize, we have shown that for $T\gg 1/r$
\begin{equation}
\PT\approx\begin{cases}
\frac1T G(r\tm)\quad &\text{ for }\tm\ll1/r\,\\
\\
\frac1T\quad &\text{ for }1/r\ll\tm\ll(T-1/r)\\
\\
\frac1T\left[2 G(r(T-\tm))-1\right]\quad &\text{ for }\tm\ll1/r\,,\\
\end{cases}
\end{equation}
where the function $G(z)$ is given in Eq.~\eqref{eq:G}. The late-time shape of the distribution $\PT$ is remarkably similar to the one of confined Brownian motion (see Eq.~\eqref{universal_G}). However, due to the nonequilibrium nature of the process, this distribution in not symmetric around the midpoint $\tm=T/2$. In particular, using the asymptotics of the function $G(z)$, given in Eq.~\eqref{G_asym}, we find that for $\tm\to 0$ and $T\gg 1/r$
\begin{equation}
\PT\approx\frac{1}{T\sqrt{2\pi r\tm}}\,.
\end{equation}
On the other hand, for $\tm \to T$ and $T\gg 1/r$ we find
\begin{equation}
\PT\approx\frac{\sqrt{2}}{T\sqrt{\pi r\tm}}\,.
\end{equation}

\subsection{Run-and-tumble particle in a confining potential}

\label{sec:rtp}

\begin{figure}[t]
\includegraphics[scale=0.55]{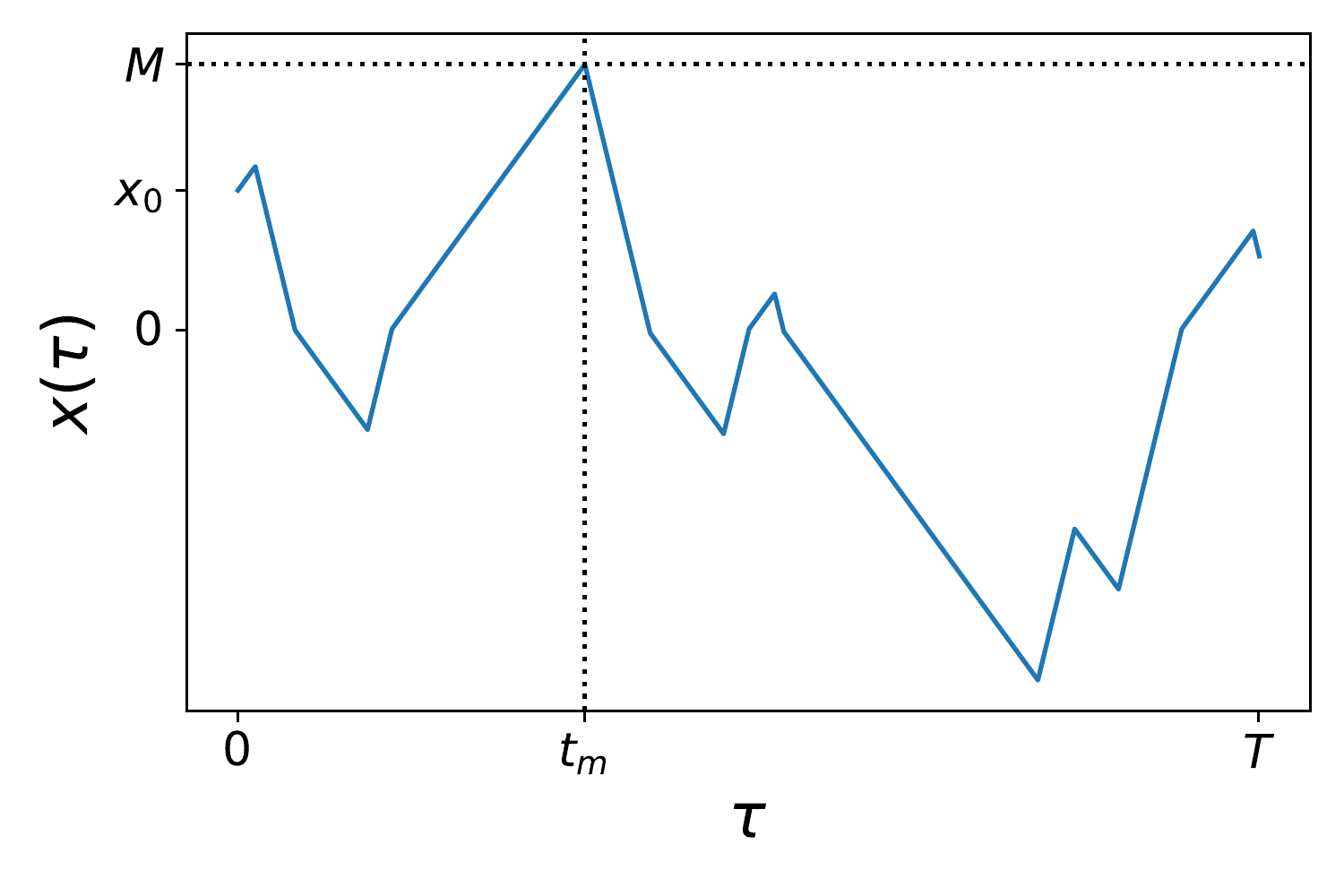} 
\includegraphics[scale=0.55]{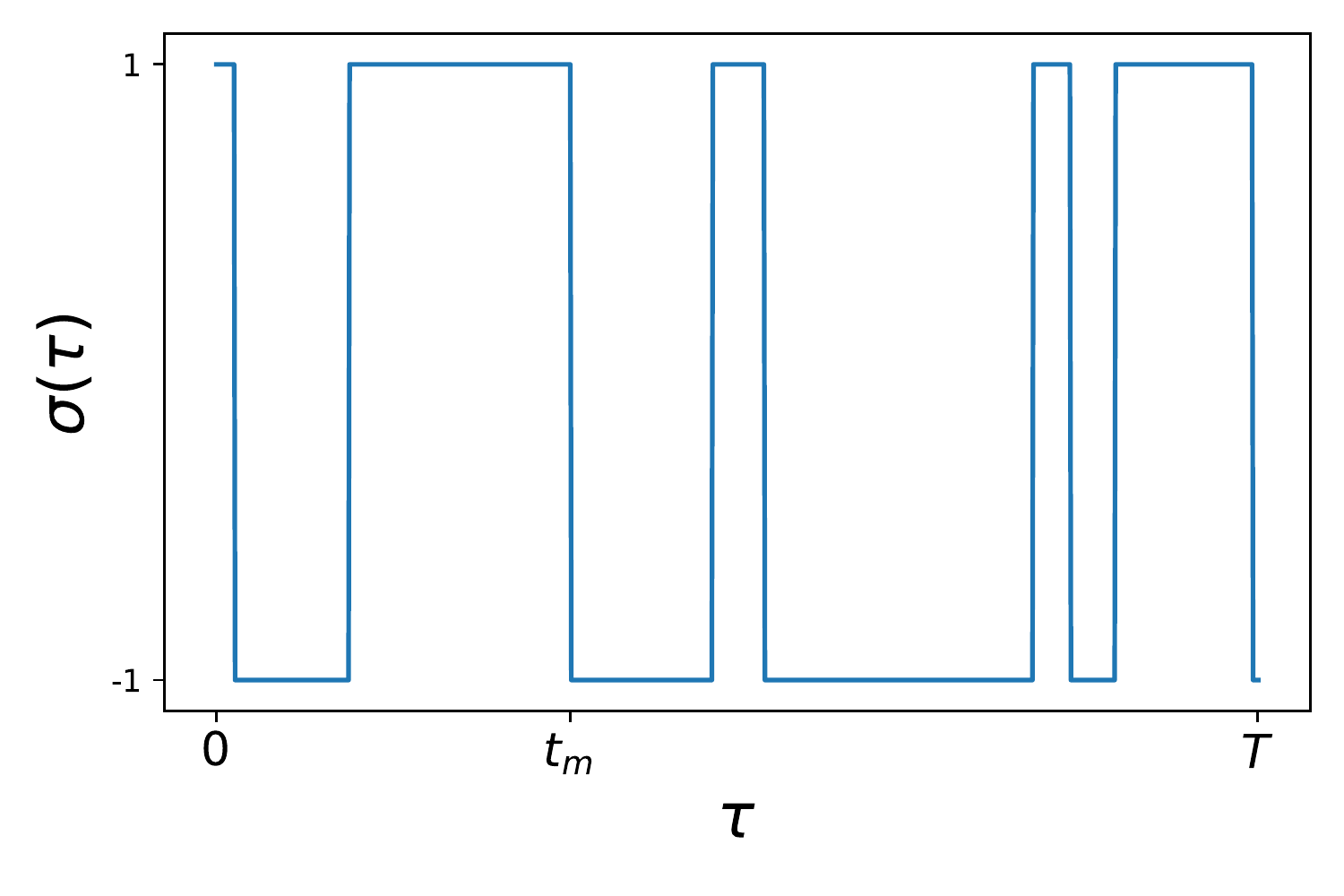} 
\caption{\label{fig:rtp_realiz} {\bf Left panel}: Typical trajectory of the position $x(\tau)$ of an RTP in a confining potential $V(x)=\mu |x|$ as a function of time $\tau$. The position of the particle reaches the maximal value $M$ at time $\tm$. {\bf Right panel}: Realization of the telegraphic noise $\sigma(\tau)$, switching sign with rate $\gamma$.}
\end{figure}

In this section, we investigate the time $\tm$ of the maximum for the RTP process. This model was first known in the literature of stochastic processes as ``persistent random walk'' \cite{kac1974stochastic,stadje1987exact,orsingher1990probability,weiss2002some}. More recently, this process was exploited to describe the persistent motion of a class of bacteria, including \emph{E. coli} \cite{berg2004coli}, which move along a  fixed direction (they ``run''), randomizing their orientation (they ``tumble'') at random times. Quite remarkably, such a simple model displays several nontrivial features, including clustering at the boundaries in a confining domain \cite{bechinger2016active}, non-Boltzmann steady-state distributions \cite{sevilla2019stationary,dhar2019run}, and jamming \cite{slowman2016jamming,metson2020jamming}.

The distribution of the time of the maximum for a free RTP has been investigated in \cite{SK19,MLD20a,MLD20}. Here we focus instead on the case of a RTP in a one-dimensional potential $V(x)$, since we want to study a stationary version of this active process. Here we consider a single RTP moving on a line and subject to a confining potential $V(x)=\mu |x|$. The evolution of the position $x$ of the particle can be described by the following Langevin equation 
\begin{equation}
\frac{dx}{dt}=f(x)+v_0~\sigma(t)\,,
\label{eq:langevin_RTP}
\end{equation}
where $v_0>0$ is the velocity of the particle, $f(x)=-V'(x)$ is the external force. The term $\sigma(t)=\pm 1$ is a telegraphic noise, describing the direction of the particle. We assume that $\sigma(t)$ flips its sign with constant rate $\gamma$. A typical realization of this process is shown in Fig.~\ref{fig:rtp_realiz}. The persistent nature of the motion of the particle drives the system out of equilibrium. Indeed, in a small time interval $dt$, the system can go from the state $(x,+1)$, i.e., position $x$ and positive direction, to the state $(x+dt(f(x)+v_0),+1)$. However, the inverse transition from $(x+dt(f(x)+v_0),+1)$ to $(x,+1)$ is not possible, inducing probability currents in phase space.

Computing analytically the distribution of $\tm$ for arbitrary $V(x)$ appears to be challenging. For this reason, we focus on the case $V(x)=\mu |x|$, which can be solved exactly. As we will show, even though it is possible to compute exactly the double Laplace transform of $\PT$ with respect to $\tm$ and $T-\tm$, the resulting expression is quite cumbersome, and even extracting asymptotics is quite hard. Nevertheless, from this exact computation one can easily check whether or not the distribution of $\tm$ satisfies the symmetry $\PT=P(T-\tm|T)$. This is precisely the goal of this section.  As we have observed that this symmetry is always present in the case of equilibrium processes while it is not satisfied by RBM, it is natural to ask whether or not this property is present in the case of nonequilibrium active particles.

We also assume that $v_0>\mu$, since in the opposite case $v_0\leq \mu$ no steady state exists \cite{SKM_19}. For $v_0>\mu$, the stationary distribution of the position is given by \cite{SKM_19}
\begin{equation}
P_{\rm st}(x_0)=\frac{\gamma~\mu}{v_0^2-\mu^2}\exp\left(-\frac{2\gamma\mu}{v_0^2-\mu^2}|x_0|\right)\,.
\label{eq:stationary_RTP_x}
\end{equation}
It is useful to define also the joint stationary distribution $P_{\rm st}^{\sigma}(x_0)$ of the position $x_0$ and of the direction $\sigma=\pm $ of the particle, which is given by \cite{SKM_19}
\begin{equation}
P_{\rm st}^{\pm }(x_0)=\frac12 \left(1\pm \frac{\mu}{v_0}\operatorname{sign}(x)\right)\frac{\gamma~\mu}{v_0^2-\mu^2}\exp\left(-\frac{2\gamma\mu}{v_0^2-\mu^2}|x_0|\right)\,.
\label{eq:stationary_RTP_x_s}
\end{equation}

We assume that the particle starts at the initial time from position $x_0$ with a positive (negative) direction $\sigma$ is drawn from the distribution in Eq.~\eqref{eq:stationary_RTP_x_s} with a positive (negative) sign, and that it evolves according to Eq.~\eqref{eq:langevin_RTP} up to time $T$. We are interested in the distribution $P(\tm|T)$ of the time $\tm$ at which the maximum of the position is reached. To compute this quantity, we will exploit a path-decomposition technique similar to the one described in the previous sections. Note that the events $\tm=0$ and $\tm=T$ happen with a finite probability and have to be considered separately.

Let us first consider the case $0<\tm<T$. As a consequence of the persistent motion of the particle, the time of the maximum coincides with a tumbling event if $0<\tm<T$ (see Fig.~\ref{fig:rtp_realiz}). Thus, we can divide the time interval $[0,T]$ into the three subintervals $[0,\tm]$ (I), $[\tm,\tm+\delta]$ (II), where $\delta$ is assumed to be small, and $[\tm+\delta,T]$ (III). In the interval (I), the particle starts from position $x_0$ with direction $\sigma$, it stays below the maximal value $M$ and reaches $M$ for the first time at time $\tm$. Note that, since we are constraining the particle to remain below position $M$, it can only arrive at position $M$ with positive velocity. Thus, the probability weight of the first interval can be written as $G^{+}_M(M,\tm|x_0,\sigma)$, where the constrained propagator $G^{\pm}_M(x,t|x_0,\sigma)$ is defined as the probability that the particle reaches position $x$ with direction $\pm$ at time $t$ while always remaining below position $M$, having started from position $x_0$ with direction $\sigma$. In the short time interval (II), the particle has to tumble, i.e., to change its direction from positive to negative. Since we assume that the tumbling events happen with a constant rate $\gamma$ and that $\delta$ is small, the probability weight of this interval is $\gamma \delta$. Finally, in the interval $[\tm+\delta,T]$ the particle starts from position $M$ with negative velocity and remains below position $M$ up to time $T$. Thus, the weight of this last interval is $Q^{-}_M(M,T-\tm)$, where the survival probability $Q^{\sigma}_M(x,t)$ is defined as the probability that the particle remains below position $M$ up to time $t$, starting from position $x$ with direction $\sigma$. Since the joint process $(x,\sigma)$ is Markov, the distribution of $\tm$ can be written as the product of the three probability weights corresponding to the three time intervals. Thus, integrating over the initial position $x_0$, the maximal value $M>x_0$ and summing over the initial direction $\sigma$, we obtain
\begin{equation}
P(\tm|T)=A~\gamma~\sum_{\sigma=\pm }\int_{-\infty}^{\infty}dx_0~P_{\rm st}^{\sigma}(x_0)~\int_{x_0}^{\infty}dM~G^{+}_M(M,\tm|x_0,\sigma)~Q_M^{-}(M,T-\tm)\,,
\label{eq:PDF_RTP_bulk}
\end{equation}
where $A$ is a normalization constant. Note that the small time interval $\delta$ is included in the normalization constant $A$.

As anticipated, the events $\tm=0$ and $\tm=T$ happen with non-zero probability. In particular, the event $\tm=0$ will happen when the particle starts from position $x_0$ with a negative velocity and remains below its starting position $x_0$ up to time $T$. Thus, since the starting position and direction are drawn from the stationary distribution $P_{\rm st}^{\sigma}(x_0)$, integrating over $x_0$ we obtain
\begin{equation}
{\rm Prob}.(\tm=0|T)=\int_{-\infty}^{\infty}dx_0~P_{\rm st}^{-}(x_0)Q_{x_0}^{-}(x_0,T)\,.
\label{eq:edge0}
\end{equation}
On the other hand, the event ``$\tm=T$'' happens when the particle reaches the maximum $M$ at the final time $T$. Since the particle is constrained to stay below position $M$, the particle can only reach the maximum coming from below, with a positive direction. Thus, summing over the initial position $x_0$ and direction $\sigma$, we find
\begin{equation}
{\rm Prob}.(\tm=T|T)=\sum_{\sigma=\pm}\int_{-\infty}^{\infty}dx_0~\int_{x_0}^{\infty}dM~P_{\rm st}^{\sigma}(x_0)~G^{+}_M(M,T|x_0,\sigma)\,.
\label{eq:edge1}
\end{equation}

Thus, using the Eqs.~\eqref{eq:PDF_RTP_bulk}, \eqref{eq:edge0}, and \eqref{eq:edge1}, we find that for $0\leq \tm\leq T$
\begin{eqnarray}
\nonumber
 && P(\tm|T)=A~\gamma~\sum_{\sigma=\pm }\int_{-\infty}^{\infty}dx_0~P_{\rm st}^{\sigma}(x_0)~\int_{x_0}^{\infty}dM~G^{+}_M(M,\tm|x_0,\sigma)~Q_M^{-}(M,T-\tm)\\
&+ & \delta\left(\tm\right)\int_{-\infty}^{\infty}dx_0~P_{\rm st}^{-}(x_0)Q_{x_0}^{-}(x_0,T)+\delta\left(\tm-T\right)\sum_{\sigma=\pm}\int_{-\infty}^{\infty}dx_0~\int_{x_0}^{\infty}dM~P_{\rm st}^{\sigma}(x_0)~G^{+}_M(M,T|x_0,\sigma)\,,\label{eq:PDF_RTP}
\end{eqnarray}
where the constant $A$ can be computed from the normalization condition
\begin{equation}
\int_{0}^{T}d\tm~P(\tm|T)=1\,.
\end{equation}
We now need to compute the constrained propagator $G_M^{\pm}(x,t|x_0,\sigma)$ and the survival probability $Q^{\pm}_M(x,t)$. These quantities can be obtained by solving the Fokker-Planck equation associated with the system.

We first consider the constrained propagator $G_M^{\pm}(x,t|x_0,\sigma)$. It is possible to show that $G_M^{+}(x,t|x_0,\sigma)$ and $G_M^{-}(x,t|x_0,\sigma)$ evolve according to the following coupled forward Fokker-Planck equations \cite{SKM_19}
\begin{equation}
\begin{cases}
\partial_tG_M^{+}(x,t|x_0,\sigma)=- \partial_x\left[\left(-\mu \operatorname{sign}(x)+v_0\right)G_M^{+}(x,t|x_0,\sigma)\right]-\gamma ~G_M^{+}(x,t|x_0,\sigma)+\gamma ~G_M^{-}(x,t|x_0,\sigma)\\
\\
\partial_tG_M^{-}(x,t|x_0,\sigma)=- \partial_x\left[\left(-\mu \operatorname{sign}(x)-v_0\right)G_M^{-}(x,t|x_0,\sigma)\right]-\gamma~ G_M^{-}(x,t|x_0,\sigma)+\gamma~ G_M^{+}(x,t|x_0,\sigma)
\end{cases}
\label{FP_RTP_G}
\end{equation}
with initial condition
\begin{equation}
G_M^{\pm}(x,t=0|x_0,\sigma)=\delta\left(x-x_0\right)\delta_{\sigma,\pm}\,,
\label{initial_condition_RTP}
\end{equation}
and boundary conditions
\begin{equation}
\begin{cases}
G_M^{\pm}(-\infty,t|x_0,\sigma)=0\\
\\
G_M^{-}(M,t|x_0,\sigma)=0\,.
\end{cases}
\label{bc_RTP}
\end{equation}
The boundary condition on the second line of Eq.~\eqref{bc_RTP} can be understood as follows. If a particle arrives at position $M$ with a negative velocity, it must have already visited the region $x>M$. However, we are constraining the particle to remain below the position $M$. Thus, $G_M^{-}(M,t|x_0,\sigma)$ has to be zero. Note that $G_M^{+}(M,t|x_0,\sigma)$ remains instead unspecified.

We limit our discussion to the case $\sigma=+$, i.e., we assume that the particle starts with a positive velocity. The complementary case $\sigma=-$ can be treated similarly. Taking a Laplace transform with respect to $t$ on both sides of the Eqs.~\eqref{FP_RTP_G} and using the initial condition in Eq.~\eqref{initial_condition_RTP}, we obtain
\begin{equation}
\begin{cases}
s\tilde{G}_M^{+}(x,s|x_0,+)-\delta(x-x_0)=- \partial_x\left[\left(-\mu \operatorname{sign}(x)+v_0\right)\tilde{G}_M^{+}(x,s|x_0,+)\right]-\gamma ~\tilde{G}_M^{+}(x,s|x_0,+)+\gamma ~\tilde{G}_M^{-}(x,s|x_0,+)\\
\\
s\tilde{G}_M^{-}(x,s|x_0,+)=- \partial_x\left[\left(-\mu \operatorname{sign}(x)-v_0\right)\tilde{G}_M^{-}(x,s|x_0,+)\right]-\gamma~ \tilde{G}_M^{-}(x,s|x_0,+)+\gamma~ \tilde{G}_M^{+}(x,s|x_0,+)
\end{cases}\,,
\label{LT_FP_RTP_G}
\end{equation}
where we have defined the Laplace transform
\begin{equation}
{\tilde G}_M^{\pm}(x,s|x_0,\sigma)=\int_{0}^{\infty}dt~e^{-st} G_M^{\pm}(x,t|x_0,\sigma)\,.
\end{equation}
The boundary conditions of the differential equations \eqref{LT_FP_RTP_G} can be obtained from Eq.~\eqref{boundary_condition_RTP} and are given by
\begin{equation}
\begin{cases}
\tilde{G}_M^{\pm}(-\infty,s|x_0,\sigma)=0\,,\\
\\
\tilde{G}_M^{-}(M,t|x_0,\sigma)=0\,.
\label{boundary_condition_RTP}
\end{cases}
\end{equation}
The solution of the coupled ordinary differential equations \eqref{LT_FP_RTP_G} is presented in Appendix \ref{app:prop}, where we show that
\begin{equation}
{\tilde G}_M^{+}(M,s|x_0,+)=\begin{cases}
\displaystyle  \dfrac{1}{v_0+\mu}e^{-(k-(s+\gamma)\mu)(M-x_0)/(v_0^2-\mu^2)}&\;{\rm for}\;x_0<0~,M<0\,,\\
\\
\displaystyle k \dfrac{e^{-(\mu(s+\gamma)+k)M/(v_0^2-\mu^2)}~e^{(-\mu(s+\gamma)+k)x_0/(v_0^2-\mu^2)}}{v_0(k-\mu(\gamma+s))+\mu(v_0(\gamma+s)-k)e^{-2kM/(v_0^2-\mu^2)}}&\;{\rm for}\;x_0<0~,M>0\,,\\
\\
\displaystyle \dfrac{1}{v_0-\mu} ~\dfrac{(k-v_0(s+\gamma))\mu+e^{2kx_0/(v_0^2-\mu^2)}v_0((s+\gamma)\mu-k)}{(k-v_0(s+\gamma))\mu+e^{2kM/(v_0^2-\mu^2)}v_0((s+\gamma)\mu-k)}\\ \times e^{(k-\mu(s+\gamma))(M-x_0)/(v_0^2-\mu^2)}&\;{\rm for}\;x_0>0~,M>0\,,\\
\end{cases}
\label{G_RTP_A}
\end{equation}
where we have defined
\begin{equation}
k=\sqrt{s^2v_0^2+2sv_0^2\gamma+\gamma^2\mu^2}\,.
\label{eq:k}
\end{equation}
Following the same steps in the case $\sigma=-$, we obtain
\begin{equation}
{\tilde G}_M^{+}(M,s|x_0,-)=\begin{cases}
\displaystyle  \dfrac{v_0(\gamma+s)-k}{\gamma(v_0^2-\mu^2)}e^{-(k-(s+\gamma)\mu)(M-x_0)/(v_0^2-\mu^2)}&\;{\rm for}\;x_0<0~,M<0\,,\\
\\
\displaystyle \dfrac{k(v_0(\gamma+s)-k)}{\gamma(v_0-\mu)} \dfrac{e^{-(\mu(s+\gamma)+k)M/(v_0^2-\mu^2)}~e^{(-\mu(s+\gamma)+k)x_0/(v_0^2-\mu^2)}}{v_0(k-\mu(\gamma+s))+\mu(v_0(\gamma+s)-k)e^{-2kM/(v_0^2-\mu^2)}}&\;{\rm for}\;x_0<0~,M>0\,,\\
\\
\displaystyle \dfrac{v_0(s+\gamma)-k}{\gamma(v_0^2-\mu^2)} ~\dfrac{(k-v_0(s+\gamma))\mu+e^{2kx_0/(v_0^2-\mu^2)}v_0((s+\gamma)\mu-k)}{(k-v_0(s+\gamma))\mu+e^{2kM/(v_0^2-\mu^2)}v_0((s+\gamma)\mu-k)}\\ \times e^{(k-\mu(s+\gamma))(M-x_0)/(v_0^2-\mu^2)}&\;{\rm for}\;x_0>0~,M>0\,,\\
\end{cases}
\label{G_RTP_B}
\end{equation}

We now want to compute the survival probability $Q_M^{\pm}(x,t)$, defined as the probability to remain below position $M$ up to time $t$, starting from position $x$ with initial direction $\pm$. It is possible to show that $Q_M^{+}(x,t)$ and $Q_M^{-}(x,t)$ satisfy the following backward Fokker-Planck equations \cite{SKM_19}
\begin{equation}
\begin{cases}
\partial_t~Q_M^{+}(x,t)= \left(-\mu \operatorname{sign}(x)+v_0\right)\partial_x Q_M^{+}(x,t)+\gamma ~Q_M^{+}(x,t)-\gamma ~Q_M^{-}(x,t)\,,\\
\\
\partial_t~Q_M^{-}(x,t)=\left(-\mu \operatorname{sign}(x)-v_0\right)\partial_x Q_M^{-}(x,t)+\gamma ~Q_M^{-}(x,t)-\gamma ~Q_M^{+}(x,t)\end{cases}
\label{FP_RTP_Q}
\end{equation}
with initial condition 
\begin{equation}
Q_M^{\pm}(x,t=0)=1\,,
\label{intial_condition_Q_RTP}
\end{equation}
for any $x<M$. The boundary conditions in this case are given by
\begin{equation}
\begin{cases}
Q_M^{\pm}(-\infty,t)=1\,,\\
Q_M^{+}(M,t)=0\,.
\end{cases}
\label{boundary_condition_RTP_2}
\end{equation}
The first boundary condition means that if the particle starts infinitely far from the absorbing barrier at $x=M$, it will never go above position $M$ in a finite time. The second boundary condition encodes that if the particle starts at $M$ with a positive velocity, it will immediately go above $M$. Note that in this case the boundary condition for $Q_M^{-}(M,t)$ remains unspecified.

It is useful to perform a Laplace transform with respect to $t$ of the equations \eqref{FP_RTP_Q}. Using the initial condition in Eq.~\eqref{intial_condition_Q_RTP}, we obtain 
\begin{equation}
\begin{cases}
s\tilde{Q}_M^{+}(x,s)-1= \left(-\mu \operatorname{sign}(x)+v_0\right)\partial_x\tilde{Q}_M^{+}(x,s)+\gamma ~\tilde{Q}_M^{-}(x,s)-\gamma ~\tilde{Q}_M^{+}(x,s)\,,\\
\\
s~\tilde{Q}_M^{-}(x,s)-1= \left(-\mu \operatorname{sign}(x)-v_0\right)\partial_x\tilde{Q}_M^{-}(x,s)+\gamma ~\tilde{Q}_M^{+}(x,s)-\gamma ~\tilde{Q}_M^{-}(x,s)\,,\end{cases}
\label{FP_RTP_Q_LT}
\end{equation}
where we have defined
\begin{equation}
\tilde{Q}^{\pm}_M(x,s)=\int_{0}^{\infty}dt ~e^{-st}Q^{\pm}_M(x,t)\,.
\end{equation}
In Laplace space, the boundary conditions in Eq.~\eqref{boundary_condition_RTP_2} become
\begin{equation}
\begin{cases}
\tilde{Q}_M^{\pm}(-\infty,s)=1/s\,,\\
\tilde{Q}_M^{+}(M,s)=0\,.
\end{cases}
\label{boundary_condition_RTP_LT}
\end{equation}
The solution of Eq.~\eqref{FP_RTP_Q_LT} is presented in Appendix \eqref{app:surv}, where we show that
\begin{equation}
{\tilde Q}_M^{-}(M,s)=\begin{cases}
\displaystyle  \dfrac{1}{s}\dfrac{k+v_0s-\gamma\mu}{k+v_0(s+\gamma)}&\;{\rm for}\;M<0\,,\\
\\
\displaystyle \dfrac{1}{s}\dfrac{1}{k+v_0(s+\gamma)}\left[k+v_0s+\mu\gamma-\dfrac{2k\gamma\mu(v_0-\mu)}{(v_0(s+\gamma)-k)\mu+v_0(k-(s+\gamma)\mu)e^{2kM/(v_0^2-\mu^2)}}\right]&\;{\rm for}\;M>0\,.\\
\\
\end{cases}
\label{Q_RTP}
\end{equation}
Note that we will not need the expression of ${\tilde Q}_M^{+}$ to compute $\PT$.

We can now use the formula in Eq.~\eqref{eq:PDF_RTP} to compute $\PT$. To proceed, we need to write this relation in Eq.~\eqref{eq:PDF_RTP} in Laplace space. Thus, we consider the double Laplace transform of Eq.~\eqref{eq:PDF_RTP} with respect to $T_1=\tm$ and $T_2=T-\tm$, yielding
\begin{eqnarray}
\nonumber &&\int_{0}^{\infty}dT_1~\int_{0}^{\infty}dt_2~e^{-s_1 T_1-s_2 T_2} P(\tm =T_1|T=T_1+T_2)= \int_{-\infty}^{\infty}dx_0~P_{\rm st}^{-}(x_0)\tilde{Q}_{x_0}^{-}(x_0,s_2) \\\nonumber
& &+\! A\gamma\sum_{\sigma=\pm }\int_{-\infty}^{\infty}dx_0~P_{\rm st}^{\sigma}(x_0)\int_{x_0}^{\infty}dM\tilde{G}^{+}_M(M,s_1|x_0,\sigma)~\tilde{Q}_M^{-}(M,s_2)\\&+&\sum_{\sigma=\pm}\int_{-\infty}^{\infty}dx_0~\int_{x_0}^{\infty}dM~P_{\rm st}^{\sigma}(x_0)~\tilde{G}^{+}_M(M,s_1|x_0,\sigma) \,.\label{eq:PDF_RTP_LT}
\end{eqnarray}

In order to determine the constant $A$, we impose that the PDF $\PT$ is correctly normalized to unity. To do this, we set $s_1=s_2=s$ on both sides of Eq.~\eqref{eq:PDF_RTP_LT}, yielding
\begin{eqnarray}
\nonumber \frac1s &=& \int_{-\infty}^{\infty}dx_0~P_{\rm st}^{-}(x_0)\tilde{Q}_{x_0}^{-}(x_0,s)+ A~\gamma~\sum_{\sigma=\pm }\int_{-\infty}^{\infty}dx_0~P_{\rm st}^{\sigma}(x_0)~\int_{x_0}^{\infty}dM~\tilde{G}^{+}_M(M,s|x_0,\sigma)~\tilde{Q}_M^{-}(M,s)\\
&+& \sum_{\sigma=\pm}\int_{-\infty}^{\infty}dx_0~\int_{x_0}^{\infty}dM~P_{\rm st}^{\sigma}(x_0)~\tilde{G}^{+}_M(M,s|x_0,\sigma) \,,
\end{eqnarray}
where we have simplified the left-hand side using Eq.~\eqref{lhs}. Computing the integrals on the right-hand side turns out to be rather nontrivial even after setting $s_1=s_2=s$. Nevertheless, using the expressions obtained above for $P^{\sigma}_{\rm st}$, $\tilde{G}^+_M$, and $\tilde{Q}^{-}_M$ and evaluating these integrals numerically with Mathematica we have verified that the correct normalization constant is $A=1$.

\begin{figure}[t]
\begin{center}
\includegraphics[scale=0.7]{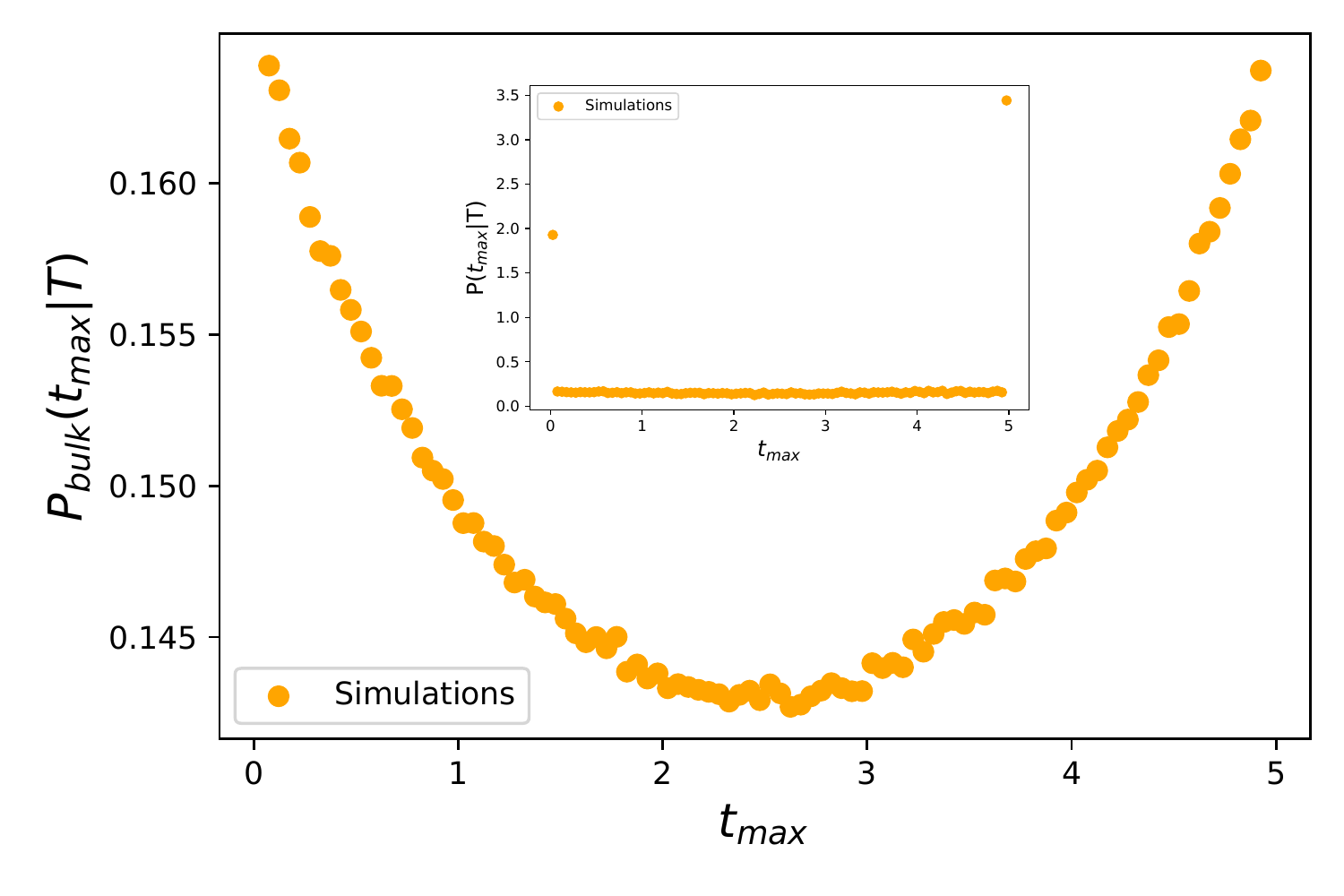}
\caption{Probability density function $P_{\rm bulk}(\tm|T)$ as a function of the time $\tm$ of the maximum for a run-and-tumble particle in a confining potential $V(x)=|x|$, for $0<\tm<T$. Note that since the events ``$\tm=0$'' and ``$\tm=T$'' occur with finite probability, the distribution is not normalized to unity for $0<\tm<T$. The curve is obtained by numerical simulations with $\gamma=1$, $T=5$, and $v_0=2$. The distribution $P_{\rm bulk}(\tm|T)$ appears to be symmetric around the midpoint $\tm=T/2$. We find numerically that $P_0(T)={\rm Prob}.(\tm=0)\approx 0.087$ and $P_1(T)={\rm Prob}.(\tm=T)\approx 0.165$. The inset shows the full distribution $\PT$, including the two asymmetric $\delta$-functions in $\tm=0$ and $\tm=T$.
\label{fig:RTP_V_tmax}}
\end{center}
\end{figure}

We then rewrite $\PT$ as
\begin{equation}
\PT=P_0(T)\delta(\tm)+P_{\rm bulk}(\tm|T)+P_1(T)\delta(\tm-T)\,,
\end{equation}
where
\begin{equation}
\int_{0}^{\infty}dt_1~\int_{0}^{\infty}dt_2~e^{-s_1 t_1-s_2 t_2} P_{\rm bulk}(\tm =t_1|T=t_1+t_2)=\gamma~\sum_{\sigma=\pm }\int_{-\infty}^{\infty}dx_0~P_{\rm st}^{\sigma}(x_0)~\int_{x_0}^{\infty}dM~\tilde{G}^{+}_M(M,s_1|x_0,\sigma)~\tilde{Q}_M^{-}(M,s_2) \,,
\label{eq:PDF_RTP_LT_bulk}
\end{equation}
\begin{equation}
\int_{0}^{\infty}dT~e^{-sT}P_0(T)=\int_{-\infty}^{\infty}dx_0~P_{\rm st}^{-}(x_0)\tilde{Q}_{x_0}^{-}(x_0,s)\,,
\label{PO_LT}
\end{equation}
and
\begin{equation}
\int_{0}^{\infty}dT~e^{-sT}P_1(T)=\sum_{\sigma=\pm}\int_{-\infty}^{\infty}dx_0~P_{\rm st}^{\sigma}(x_0)~\int_{x_0}^{\infty}dM~\tilde{G}^{+}_M(M,s|x_0,\sigma) \,.
\label{P1_LT}
\end{equation}

Exactly inverting these Laplace transforms turns out to be quite nontrivial. Nevertheless, it is possible to check, for instance using Mathematica, that the Laplace transform of $P_{\rm bulk}(\tm|T)$ in Eq.~\eqref{eq:PDF_RTP_LT_bulk} is invariant under exchange of $s_1$ and $s_2$. This implies that $P_{\rm bulk}(\tm|T)=P_{\rm bulk}(T-\tm|T)$, i.e., that the central part of the distribution of $\tm$ is symmetric around the midpoint $\tm=T/2$. This is confirmed by numerical simulations (see Fig.~\ref{fig:RTP_V_tmax}). However, it is easy to show that the amplitudes $P_0(T)$ and $P_1(T)$ of the delta functions in $\tm=0$ and $\tm=T$ are in general different. Thus, the full distribution $P(\tm|T)$, for $0\leq \tm\leq T$ is not symmetric around $\tm=T/2$. This is a consequence of the nonequilibrium nature of the process.

\subsection{Criterion to detect nonequilibrium dynamics}

\label{sec:criterion}

\begin{figure}[t]
\includegraphics[scale=0.6]{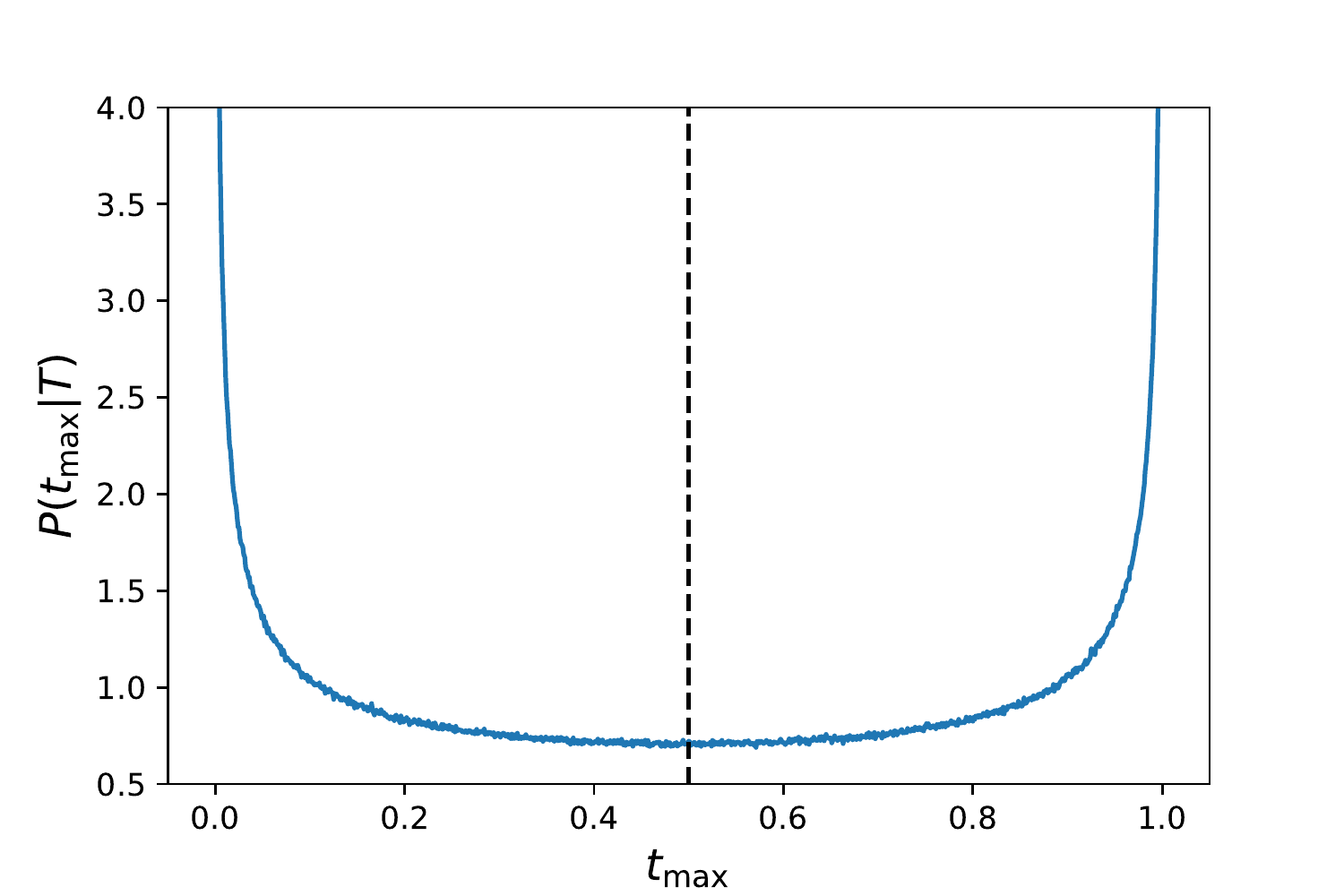}  
\caption{\label{fig:asym_V_} Probability density function $\PT$ as a function of $\tm$, obtained from numerical simulations of Brownian motion in an asymmetric potential $V(x)$, with $V(x)=x^2$ for $x>0$ and $V(x)=-x$ for $x<0$, $T=1$, and $D=1$. The distribution appears to be symmetric around the midpoint $\tm=T/2$ (vertical dashed line). }
\end{figure}

From the exact results of the previous sections, we have observed that for equilibrium systems corresponding to an overdamped Brownian particle in a confining potential $V(x)$ the probability distribution of the time $\tm$ of the maximum is symmetric around the midpoint $\tm=T/2$. Let us stress that this property is not related to the symmetry $V(x)=V(-x)$ of the potentials that we have investigated in Section \ref{sec:eq} (see Fig.~\ref{fig:asym_V_}, where we show that this property is valid even when $V(x)\neq V(-x)$). On the other hand, for the nonequilibrium processes we have considered, this symmetry is not present and $\PT\neq P(T-\tm |T)$. In this Section, we show that this symmetry is quite general and can be used to develop a technique to detect nonequilibrium fluctuations in steady states. In particular, we show that one has the property $\PT=P(T-\tm|T)$ for any equilibrium process. Note that the inverse implication is not true as there are nonequilibrium processes with this symmetry.

We consider a time series of duration $T$. For simplicity, we focus on a discrete-time process $x_i$ with $1\leq i\leq T$ -- where $i$ and $T$ are positive integers. This derivation immediately generalizes to the continuous-time case. We assume that this time series is generated from an equilibrium Markov process. We denote by $P(\{x_i\})$ the probability of observing a given trajectory $\{x_i\}=\{x_1\,,x_2\,,\ldots\,,x_T\}$ and by $\{\bar{x}_i\}=\{x_T,\ldots ,x_i\}$ the time-reversed trajectory associated with $\{x_i\}$. Note that if the system is at equilibrium it is easy to show that $P(\{x_i\})=P(\{\bar{x}_i\})$ (time-reversal symmetry).

The distribution of the time $\tm$ of the maximum can be written as
\begin{equation}
P(\tm|T)=\int_{-\infty}^{\infty}dx_1\ldots\int_{-\infty}^{\infty}dx_T~  \Theta_{\tm}(\{x_i\}) P\left(\{x_i\}\right)\,,
\label{discrete}
\end{equation}
where 
\begin{equation}
\Theta_{k}\left(\{x_i\}\right)=\prod_{i\neq k}\theta\left(x_k-x_i\right)\,.
\end{equation}
Here $\theta(z)$ is the Heaviside step function, i.e., $\theta(z)=1$ for $z>0$ and $\theta(z)=0$ otherwise. The function $\Theta_{k}\left(\{x_i\}\right)$ is one if the maximum of the trajectory $\{x_i\}$ is attained at step $k$ and is zero otherwise. Performing the change of variables $x_i\to\bar{x}_i=x_{T-i}$ in Eq.~\eqref{discrete}, we get 
\begin{equation}
P(\tm|T)=\int_{-\infty}^{\infty}d\bar{x}_1\ldots\int_{-\infty}^{\infty}d\bar{x}_T  ~\Theta_{\tm}(\{\bar{x}_{T-k}\}) P\left(\{\bar{x}_k\}\right)\,,
\label{discrete2}
\end{equation}
where we have used the relation $P(\{x_i\})=P(\{\bar{x}_i\})$. It is easy to show that $\Theta_{\tm}(\{\bar{x}_{T-i}\})=\Theta_{T-\tm}(\{\bar{x}_{i}\})$, meaning that if the maximum of the forward trajectory is reached at time $\tm$ then the maximum of the backward trajectory is reached at time $T-\tm$. Using this relation, we find
\begin{equation}
P(\tm|T)=\int_{-\infty}^{\infty}d\bar{x}_1\ldots\int_{-\infty}^{\infty}d\bar{x}_T  ~\Theta_{T-\tm}(\{\bar{x}_{i}\}) P\left(\{\bar{x}_i\}\right)\,.
\label{discrete3}
\end{equation}
We now recognize the expression of the right-hand side to be $P(T-\tm|T)$ (see Eq.~\eqref{discrete}). Therefore, we obtain $P(\tm|T)=P(T-\tm|T)$. To summarize we have shown that if a process is at equilibrium then $P(\tm|T)=P(T-\tm|T)$. As anticipated, this symmetry can be used to determine whether or not a stationary system is nonequilibrium.

Imagine that one has access to a long time series $x(\tau)$, for instance, obtained from some experiments, and one does not know the specific details of the underlying system. For example, this time series could represent the position of a molecular motor along a microtubule or a Brownian particle in an optical trap. This setup has become increasingly relevant due to recent developments in single-particle tracking \cite{shen2017single}. Then, one of the most fundamental questions that one can ask about the system is whether or not it is at equilibrium. In particular, in the context of biological systems, these questions are relevant since nonequilibrium fluctuations typically signal the active consumption of energy.

Throughout the last decades, several methods to determine the nonequilibrium nature of a system have been developed -- for a recent review, see \cite{GMG18}. Many of these techniques also quantify how much the system is out of equilibrium, usually as a bound on the entropy production  \cite{LH19,MGK20,manikandan2021quantitative,roldan2021quantifying,otsubo2022estimating}. A popular technique is based on the verification of the fluctuation-dissipation relation, which relates correlation and response for equilibrium systems \cite{cugliandolo1997fluctuation,martin2001comparison,mizuno2007nonequilibrium,TFA16}. If a violation of this relation is observed, one can immediately conclude that the system is nonequilibrium. Note that the main drawback of this method is that it requires perturbing the system to measure the response function. Many other techniques have been proposed, including the detection of violations of the detailed balance condition\cite{ZS07,BBF16,mura2018nonequilibrium} or the analysis of waiting-time distributions \cite{tu2008nonequilibrium,skinner2021estimating}.

Using the fact that $\PT$ is symmetric for equilibrium systems, we introduce a new method based on two steps. First, divide the time series $x(\tau)$ into $N$ blocks of duration $T$ (assuming that the time series is long enough such that $N\gg1$). Compute the time $\tm^i$ at which the maximum is reached within each block (where the index $i$ refers to the $i$-th block). From these $N$ values, build the empirical distribution $\PT$. If this distribution is not symmetric around $\tm=T/2$, the process is necessarily nonequilibrium. On the other hand, if $\PT=P(T-\tm|T)$ our test is inconclusive. Note that for a multidimensional system, one can apply the criterion to any of its components.

As anticipated, there exist nonequilibrium processes for which the distribution of $\tm$ is symmetric. 
As an example, we can consider a single one-dimensional active Ornstein-Uhlenbeck process (AOUP) in a harmonic potential $V(x)=\alpha x^2$ \cite{fodor2016far}. The system is described by the position $x(\tau)$ and the speed $v(\tau)$ of the particle. The system evolves according to
\begin{equation}
\frac{dx(t)}{dt}=-\alpha x(t)+v(t)+\sqrt{2D}\xi(t)\,,
\end{equation}
where $\xi(t)$ is a Gaussian white noise with zero mean and correlator $\langle \xi(t)\xi(t')\rangle=\delta(t-t')$ and $v(t)$ evolves as
\begin{equation}
\frac{dv(t)}{dt}=-\frac{v}{\tau_a}+\frac{\sqrt{2D_a}}{\tau_a}\zeta(t)\,,
\end{equation}
where $D_a>0$, $\tau_a>0$, and $\zeta(t)$ is a Gaussian white noise. We also assume that $\xi(t)$ and $\zeta(t)$ are uncorrelated. Note that since the equations of motion are linear, the process is Gaussian.

Even though $x(t)$ depends on the evolution of $v(t)$, there is no feedback from $v(t)$ to $x(t)$. This creates probability currents in the phase space $(x,v)$ and hence the system is out of equilibrium. Nevertheless, it can be shown analytically that the one-dimensional process describing the position of the particle $x(t)$ satisfies time reversal symmetry \cite{BO}. Thus, even though the process is nonequilibrium, the distribution $\PT$ of the time $\tmax$ at which the position $x(\tau)$ is maximal is symmetric around $\tm=T/2$. Interestingly, this is a consequence of the fact that this is a Gaussian stationary process. Indeed, as shown below, for any one-dimensional Gaussian stationary process the distribution of $\tm$ is always symmetric around $\tm=T/2$.

Let us consider a one-dimensional discrete-time Gaussian stationary process $x_k$ with $1\leq k \leq T$ (it is easy to generalize the following argument to continuous-time processes). Without loss of generality, we assume that the mean value of $x_k$ vanishes. The probability of a trajectory $\{x_k\}=\{x_1\,,\ldots\,,x_T\}$ is
\begin{equation}
P(\{x_k\})=\mathcal{N} ~\exp\left[-\frac12 \sum_{i,j} ~x_i \Sigma^{-1}_{i,j}x_j\right]\,,
\label{Pxk_1}
\end{equation}
where $\Sigma_{i,j}=\langle x_i x_j\rangle$ is the covariance matrix and $\mathcal{N}$ is a normalization constant. Since the process is stationary, the covariance $\Sigma_{i,j}$ only depends on $|i-j|$, yielding
\begin{equation}
P(\{x_k\})=\mathcal{N} ~\exp\left[-\frac12 \sum_{i,j} ~x_i \Sigma^{-1}(|i-j|)x_j\right]\,.
\label{Pxk_2}
\end{equation}
The probability of the time-reversed trajectory $\{\bar{x}_k\}=\{x_{T-k}\}$ is given by
\begin{equation}
P(\{\bar{x}_k\})=\mathcal{N} ~\exp\left[-\frac12 \sum_{i,j} ~x_{T-i} \Sigma^{-1}(|i-j|)x_{T-j}\right]\,.
\label{Pxk_3}
\end{equation}
Performing the change of variable $(i,j)\to(i'=T-i,j'=T-j)$, we obtain
\begin{equation}
P(\{\bar{x}_k\})=\mathcal{N} ~\exp\left[-\frac12 \sum_{i',j'} ~x_{i'} \Sigma^{-1}(|i'-j'|)x_{j'}\right]\,.
\label{Pxk_4}
\end{equation}
Using Eq.~\eqref{Pxk_2}, we get
\begin{equation}
P(\{\bar{x}_k\})=P(\{x_k\})\,,
\end{equation}
meaning that the process is symmetric under time reversal. As shown above, this implies that the distribution of $t_{\rm m}$ is symmetric around $t_{\rm m}=T/2$.

\section{Conclusions}

\label{sec:conclusion}

In summary, we have investigated the distribution of the time $\tm$ at which a stationary stochastic process reaches its global maximum within a time window $[0,T]$. Using a path decomposition technique, we have computed analytically the distribution $\PT$ of the time $\tm$ of the maximum for several processes, both at equilibrium and out of equilibrium.

The class of equilibrium processes that we have considered corresponds to an overdamped Brownian particle moving in a one-dimensional potential $V(x)$ such that $V(x)\approx\alpha |x|^p$ for large $|x|$, with $\alpha>0$ and $p>0$. We have computed the distribution $\PT$ exactly in the cases $V(x)=\alpha |x|$ (corresponding to $p=1$) and $V(x)=\alpha x^2$ (corresponding to the Ornstein-Uhlenbeck process with $p=2$). From these exact computations, we have observed that the distribution of $\tm$ is symmetric around $\tm=T/2$, i.e., $\PT=P(T-\tm|T)$, for any equilibrium process. This property is a consequence of the time-reversal symmetry of equilibrium systems. Moreover, we have shown that the distribution of $\PT$, once appropriately scaled, becomes completely universal for any $\alpha>0$ and $p>0$ in the late-time limit $T\gg 1$. 

We have also considered two models of nonequilibrium stationary processes for which we could compute exactly the distribution of $\tm$: a Brownian particle with stochastic resetting and a single RTP in a confining potential $V(x)=\mu |x|$. In both cases, we have shown that the distribution of $\tm$ is not symmetric around $\tm=T/2$. From this observation, we have presented a sufficiency test, based on the measurement of $\tm$, which allows detecting nonequilibrium fluctuations in stationary systems.

For future studies, it would be interesting to investigate the joint distribution of the time $\tm$ of the maximum and the time $t_{\min}$ of the minimum for a stationary process of duration $T$. For short times ($T\ll1$), we expect these two times to be strongly correlated, while we expect $\tm$ and $t_{\min}$ to become independent at late times. From the joint distribution of $\tm$ and $t_{\min}$ one can also obtain several relevant quantities, including the distribution of the time $\tau=t_{\min}-\tm$ between the global maximum and the global minimum \cite{MMS19,MMS20}. 

Another relevant direction would be to investigate the distribution of the time $\tm$ of the maximum for an overdamped Brownian particle in a potential that grows as $V(x)\approx\alpha \ln(|x|)$ for large $|x|$, with $\alpha>D$ where $D$ is the diffusion constant (for $\alpha<D$ the process does not reach a steady state). The distribution of the global maximum for this model was investigated in Refs.~\cite{OPR20,PRO20}, where it was shown that that the average maximum grows for late times as $\langle M\rangle\approx T^{1/(1+\alpha/D)}$.
Although it appears quite challenging to exactly compute the distribution of $\tm$ for this model, it would be relevant to investigate whether the universality of the distribution $\PT$, presented in Subsection \ref{sec:univ} remains valid in this case.

%tmax-tmin

%0<p<1

\appendix
\section{Computation of $\tilde{G}^M(x,s|x_0)$ and $\tilde{Q}^M(x,s)$ for $p=1$}

\label{app:G_p1}

In this appendix, we compute the constrained propagator $\tilde{G}^M(x,s|x_0)$ and the survival probability $\tilde{Q}^M(x,s)$ for an overdamped Brownian particle in a potential $V(x)=\alpha|x|$. To determine the constrained propagator, we need to solve the Fokker-Planck equation \eqref{forward_FP_p1_LT}. It is helpful to consider three cases depending on the signs of $x_0$ and $M$.

\noindent{\it The case where $M<0$ and $x_0<0$.} When $M$ and $x_0$ are both negative, we can solve Eq.~\eqref{forward_FP_p1_LT} in the two regions $-\infty\leq x\leq x_0$ and $x_0\leq x \leq M$ separately. In each of these regions the delta function in Eq.~\eqref{forward_FP_p1_LT} disappears and we obtain
\begin{equation}\label{eq:FP_V3}
D \partial^2_x \tilde{G}^M(x,s|x_0)-\alpha \partial_x \tilde{G}^M(x,s|x_0)-s \tilde{G}^M(x,s|x_0)=0\,.
\end{equation}
Solving this differential equation yields
\begin{equation}
 \tilde{G}^M(x,s|x_0)=\begin{cases}
\displaystyle A_{+}e^{(\alpha+k)(x-x_0)/(2 D)}+A_{-}e^{(\alpha-k)(x-x_0)/(2 D)}&\;\;\;{\rm if}\;x<x_0\;,\\
\displaystyle B_{+}e^{(\alpha+k)(x-x_0)/(2 D)}+B_{-}e^{(\alpha-k)(x-x_0)/(2 D)}&\;\;\;{\rm if}\;x_0<x<M\;,
\end{cases}\label{eq:G_1}
\end{equation}
where $A_+,\,A_-,\,B_+,$ and $B_-$ are arbitrary constants and $k=\sqrt{\alpha^2+4sD}$.
To fix these constants, we use the boundary conditions in Eqs.~\eqref{absorbing_condition_fw_LT} and \eqref{boundary_condition_fw_LT} and the following matching conditions
\begin{equation}\label{eq:continuity_x0}
\tilde{G}^M(x_0^+,s|x_0)=\tilde{G}^M(x_0^-,s|x_0)\,,
\end{equation}
and
\begin{equation}\label{eq:deriv_x0}
-1=D\partial_x\tilde{G}^M(x_0^+,s|x_0)-D\partial_x\tilde{G}^M(x_0^-,s|x_0)\,.
\end{equation}
The condition in Eq.~\eqref{eq:continuity_x0} imposes the continuity of the propagator $\tilde{G}^M(x,s|x_0)$ at $x=x_0$, while the condition in Eq.~\eqref{eq:deriv_x0} can be obtained by integrating both sides of Eq.~\eqref{forward_FP_p1_LT} over a small interval $[x_0-\delta,x_0+\delta]$ and then taking the limit $\delta\to 0$.

From these conditions, we find that for $x_0\leq x \leq M$
\begin{equation}
\tilde{G}^M(x,s|x_0)=\frac1k e^{\alpha (x-x_0)/(2D)}\left[e^{-k (x-x_0)/(2D)}-e^{-k(2M-x-x_0)/(2D)}\right]\,.
\end{equation}
Finally, for $x=M-\epsilon$ we obtain, expanding to leading order in $\epsilon$,
\begin{equation}\label{eq:G_1_expanded}
\tilde{G}^M(M-\epsilon,s|x_0)\approx    \frac{\epsilon}{D}e^{(\alpha-k)(M-x_0)/(2D)}\,,
\end{equation}
where we recall that $k=\sqrt{\alpha^2+4sD}$.

\noindent{\it The case where $M>0$ and $x_0>0$.} In this case, solving Eq. (\ref{forward_FP_p1_LT}) in the three regions $-\infty<x\leq0$, $0\leq x\leq x_0$, and $x_0\leq x \leq M$, we obtain
\begin{equation}
 \tilde{G}^M(x,s|x_0)=\begin{cases}
\displaystyle A_{+}e^{(\alpha+k)x/(2 D)}+A_{-}e^{(\alpha-k)x/(2 D)}&\;\;\;{\rm if}\;x<0\;,\\
\displaystyle B_{+}e^{(-\alpha+k)(x-x_0)/(2 D)}+B_{-}e^{(-\alpha-k)(x-x_0)/(2 D)}&\;\;\;{\rm if}\;0<x<x_0\;,\\
\displaystyle C_{+}e^{(-\alpha+k)(x-x_0)/(2 D)}+C_{-}e^{(-\alpha-k)(x-x_0)/(2 D)}&\;\;\;{\rm if}\;x_0<x<M\;,
\end{cases}
\end{equation} 
where $A_+,\,A_-,\,B_+,\,B_-,\,C_+,$ and $C_-$ are arbitrary constants. Note that, even though we use the same notation for $A_+,\,A_-,\,B_+,\,B_-$, these constants are different from the ones in Eq.~\eqref{eq:G_1}. To fix these constants we impose the boundary conditions in Eqs.~(\ref{absorbing_condition_fw_LT}) and (\ref{boundary_condition_fw_LT}) and the matching conditions at $x=x_0$ given in Eqs. (\ref{eq:continuity_x0}) and (\ref{eq:deriv_x0}). Moreover, in this case one also needs to impose the matching conditions at $x=0$, which are given by
\begin{equation}\label{eq:continuity_0}
\tilde{G}^M(0^+,s|x_0)=\tilde{G}^M(0^-,s|x_0)\,,
\end{equation}
\begin{equation}\label{eq:deriv_0}
D\partial_x\tilde{G}^M(0^+,s|x_0)-D\partial_x\tilde{G}^M(0^-,s|x_0)+2\alpha\tilde{G}^M(0,s|x_0)=0\,.
\end{equation}
The condition in Eq.~(\ref{eq:continuity_0}) imposes the continuity of $\tilde{G}^M(x,s|x_0)$ at $x=0$, while the condition in Eq.~(\ref{eq:deriv_0}) can be obtained by integrating Eq.~(\ref{forward_FP_p1_LT}) over a small interval $[-\delta,\delta]$ and then taking the limit $\delta\to 0$. Solving these conditions we obtain, to leading order in $\epsilon$,
\begin{equation}\label{eq:G_2_expanded}
\tilde{G}^M(M-\epsilon,s|x_0)\approx    \frac{\epsilon}{D}\frac{(k-\alpha)e^{k x_0 /D}+\alpha}{(k-\alpha)e^{k M /D}+\alpha}e^{(-\alpha+k)(M-x_0)/(2D)}\,,
\end{equation}
where $k=\sqrt{\alpha^2+4sD}$

\noindent{\it The case where $M>0$ and $x_0<0$.} This case is analogous to the previous one. We find that, to leading order in $\epsilon$, the constrained propagator is given by
\begin{equation}\label{eq:G_3_expanded}
\tilde{G}^M(M-\epsilon,s|x_0)\approx    \frac{k \epsilon}{D}\frac{e^{(k-\alpha) x_0 /(2D)}e^{(-k-\alpha) M /(2D)}}{k-\alpha+\alpha e^{-k M /D}}\,.
\end{equation}

We next focus on the survival probability $\tilde{Q}^M(x,s)$, which can be computed by solving the backward Fokker-Planck \eqref{backward_FP_LT}. It is useful to distinguish two cases depending on the sign of $M$.

\noindent{\it The case where $M>0$.}  

In this case, we solve Eq.~\eqref{backward_FP_LT} separately for $x>0$ and $x<0$. When $x>0$, Eq. (\ref{backward_FP_LT}) becomes
\begin{equation}
s \tilde{Q}^M(x,s)-1=D\partial^2_x \tilde{Q}^M(x,s)-\alpha \partial_x \tilde{Q}^M(x,s)\,.
\end{equation}
The most general solution of this equation is
\begin{equation}
\tilde{Q}^M(x,s)=\frac{1}{s}+A_+ e^{(\alpha+k)/(2D)x}+A_- e^{(\alpha-k)/(2D)x}\,,
\end{equation}
where we recall that $k=\sqrt{\alpha^2+4sD}$.
Similarly, when $x<0$, we obtain
\begin{equation}
\tilde{Q}^M(x,s)=\frac{1}{s}+B_+ e^{(-\alpha+k)/(2D)x}+B_- e^{(-\alpha-k)/(2D)x}\,.
\end{equation}
To fix the four constants $A_-$, $A_+$, $B_-$ and $B_+$, we need to impose the boundary conditions in Eqs. (\ref{absorbing_condition_bw_LT}) and (\ref{boundary_condition_bw_LT}), and the continuity of $\tilde{Q}^M(x,s)$ and $\partial_x \tilde{Q}^M(x,s)$ at the origin. Imposing these conditions one obtains 
\begin{eqnarray}
&& A_+=-\frac{k-\alpha}{s}\frac{e^{-\frac{\alpha M}{2D}}}{(k-\alpha)e^{\frac{kM}{2D}}+\alpha e^{-\frac{-kM}{2D}}}\,,\\
&& A_-=\frac{\alpha}{k-\alpha}A_+\,.
\end{eqnarray}
Then, the probability weight of the second segment is thus given by, expanding for small values of $\epsilon$,
\begin{equation}\label{eq:tilde_Q1}
\tilde{Q}^M(M-\epsilon,s)\approx\frac{\epsilon}{s}\frac{k-\alpha}{2D}\frac{(k+\alpha)e^{kM/D}-\alpha}{(k-\alpha)e^{kM/D}+\alpha}\,.
\end{equation}

\noindent{\it The case where $M<0$.} 

In this case one has to consider just the region $x<M$. Solving Eq. (\ref{backward_FP_LT_p1}) in this region, we find
\begin{equation}
\tilde{Q}^M(x,s)=\frac{1}{s}+A_+ e^{(-\alpha+k)/(2D)x}+A_- e^{(-\alpha-k)/(2D)x}\,.
\end{equation}
Using the boundary conditions in Eqs. (\ref{absorbing_condition_bw_LT}) and (\ref{boundary_condition_bw_LT}) one can fix the values of $A_+$ and $A_-$. Expanding to leading order in $\epsilon$, we obtain
\begin{equation}\label{eq:tilde_Q2}
\tilde{Q}^M(M-\epsilon,s)\approx\frac{\epsilon}{s}\frac{k-\alpha}{2D}\,.
\end{equation}

\section{Laplace inversion of Eq.~\eqref{G_LT}}
\label{app:Laplace inversion}

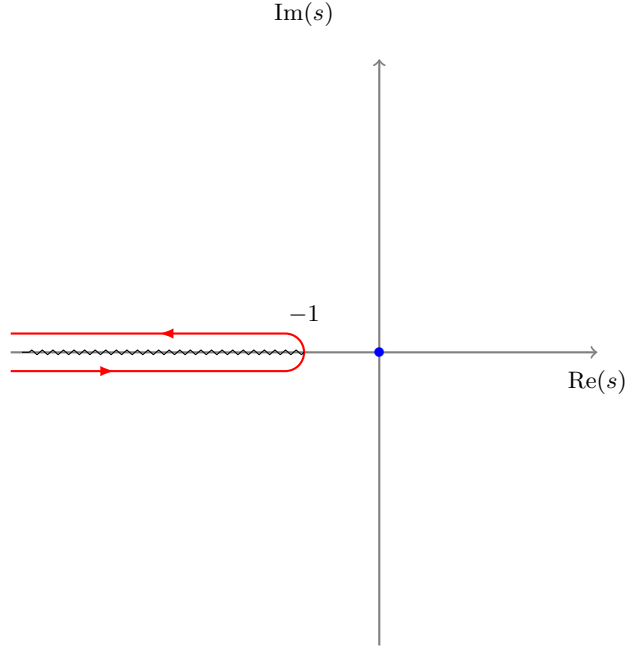
\begin{figure}
%%%
%%%
\centering 
\begin{tikzpicture}[xscale=-1]
% Configurable parameters 
\def\gap{0.3}
\def\bigradius{3}
\def\littleradius{0.25}

% Axes 
\draw [-> , help lines,thick] (1.3*\bigradius,0) -- (-1.3*\bigradius, 0);
\draw [help lines,->,thick] (-1, -1.3*\bigradius) -- (-1, 1.3*\bigradius);

\node [blue] at (-1,0) {\textbullet};

\draw[red, thick,   decoration={ markings,
    mark=at position 0.17 with {\arrow{latex}}, 
    mark=at position 0.755 with {\arrow{latex}}  
  }, 
  postaction={decorate}]  
let 
\n2 = {90},
\n3 = {\littleradius,\littleradius},
\n4 = {sin(-\n2)*\littleradius},
\n5 = {cos(-\n2)*\littleradius},
\n6 = {sin(-360+\n2)*\littleradius},
\n7 = {cos(-\n2)*\littleradius+\littleradius/2.0}
in 
(1.3*\bigradius,-\littleradius) -- (\littleradius,-\littleradius) 
(\littleradius,\littleradius) arc (-90+180:90+180:\littleradius) 
(\littleradius,\littleradius)  -- (1.3*\bigradius,\littleradius);

% The labels 
%%\node at (-1.3*\bigradius,0.3) {$r_2$};
%%\node at (-1.3*\bigradius,-0.3) {$r_1$};

\node at (0,0.5) {$-1$};

\node at (-1.3*\bigradius, -0.4) {Re($s$)};
\node at (0, 1.5*\bigradius) {Im($s$)};

\draw [-,
line join=round,
decorate, decoration={
    zigzag,
    segment length=4,
    amplitude=.9,post=lineto,
    post length=1pt 
}]  (0,0) -- (1.25*\bigradius,0);

\end{tikzpicture}
\caption{Integration contour used in Eq.~\eqref{integration}. The black wiggled line corresponds to the branch cut of the integrand, while the blue dot corresponds to the pole $s=0$.}
\label{fig:Bromwich}
\end{figure}

In this appendix, we derive the Laplace inversion
\begin{equation}
\mathcal{L}^{-1}_{s\to t}\left[\frac{1+\sqrt{1+s}}{s}\right]=1+\operatorname{erf}(\sqrt{t})+\frac{1}{\sqrt{\pi t}}e^{-t}\,.
\label{laplace_app_}
\end{equation}
Inverting the Laplace transform formally, we find
\begin{equation}
\mathcal{L}^{-1}_{s\to t}\left[\frac{1+\sqrt{1+s}}{s}\right]=\frac{1}{2\pi i}\int_{\Gamma}ds~\frac{1+\sqrt{1+s}}{s}\,,
\end{equation}
where $\Gamma$ is the imaginary-axis Bromwich contour in the complex $s$ plane. Using the relation $(1+\sqrt{1+s})(1-\sqrt{1-s})=s$, we can rewrite the integrand as
\begin{equation}
\mathcal{L}^{-1}_{s\to t}\left[\frac{1+\sqrt{1+s}}{s}\right]=\frac{1}{2\pi i}\int_{\Gamma}ds~\frac{1}{1-\sqrt{1+s}}\,.
\end{equation}
The integrand has a single pole at $s=0$ and a branch cut in the real-$s$ axis for $s<-1$. Evaluating the residue at $s=0$ and using the parametrization $s=-1+r e^{\pm i\pi}$ to evaluate the integral around the branch cut (see Fig.~\ref{fig:Bromwich}), we find
\begin{equation}
\mathcal{L}^{-1}_{s\to t}\left[\frac{1+\sqrt{1+s}}{s}\right]=2+\frac{1}{2\pi i}e^{-t}\int_{0}^{\infty}dr~e^{-rt}\frac{1}{1-i\sqrt{r}}-\frac{1}{2\pi i}e^{-t}\int_{0}^{\infty}dr~e^{-rt}\frac{1}{1+i\sqrt{r}}\,.
\label{integration}
\end{equation}
Regrouping the terms on the right-hand side, we obtain
\begin{equation}
\mathcal{L}^{-1}_{s\to t}\left[\frac{1+\sqrt{1+s}}{s}\right]=2+\frac{1}{\pi }e^{-t}\int_{0}^{\infty}dr~e^{-rt}\frac{\sqrt{r}}{1+r}\,.
\end{equation}
Computing the integral over $r$ with Mathematica, we obtain the result in Eq.~\eqref{laplace_app_}.

\section{Computation of $\tilde{G}^M(x,s|x_0)$ and $\tilde{Q}^M(x,s)$ for $p=2$}

\label{app:G_p2}

In this Appendix, we compute the constrained propagator $\tilde{G}^M(x,s|x_0)$ and the survival probability $\tilde{Q}^M(x,s)$ in the case of a Brownian particle in a harmonic potential $V(x)=\alpha x^2$.

To compute the constrained propagator, we need to solve the Fokker-Planck equation \eqref{eq:FP_LT_ou}. The most general solution of the differential equation \eqref{eq:FP_LT_ou} reads, for $x\neq x_0$
\begin{equation}
 \tilde{G}^M(x,s|x_0)=\begin{cases}
\displaystyle A_{+}e^{-2\alpha x^2 /(4D)} D_{-s/(2\alpha)}\left(\sqrt{\frac{2\alpha}{D}}x\right)+A_{-}e^{-2\alpha x^2 /(4D)}D_{-s/(2\alpha)}\left(\sqrt{\frac{2\alpha}{D}}x\right)&\;\;\;{\rm if}\;x<x_0\;,\\ \\
\displaystyle B_{+}e^{-2\alpha x^2 /(4D)} D_{-s/(2\alpha)}\left(\sqrt{\frac{2\alpha}{D}}x\right)+B_{-}e^{-2\alpha x^2 /(4D)}D_{-s/(2\alpha)}\left(\sqrt{\frac{2\alpha}{D}}x\right)&\;\;\;{\rm if}\;x_0<x<M\;,
\end{cases}\label{eq:G_ou}
\end{equation}
where $D_p(z)$ is the parabolic cylinder function. In order to fix the arbitrary constants $A_+,\,A_-,\,B_+,$ and $B_-$, we use the boundary conditions in Eqs. (\ref{absorbing_condition_fw_LT}) and (\ref{boundary_condition_fw_LT}) as well as the matching conditions
\begin{equation}\label{eq:match_1_ou}
\tilde{G}^M(x_0^+,s|x_0)=\tilde{G}^M(x_0^-,s|x_0)\,,
\end{equation}
\begin{equation}\label{eq:match_2_ou}
-1=D\partial_x \tilde{G}^M(x_0^+,s|x_0)-D\partial_x\tilde{G}^M(x_0^-,s|x_0)\,.
\end{equation}
These condition in Eq. (\ref{eq:match_1_ou}) imposes the continuity at $x=x_0$, while Eq. (\ref{eq:match_2_ou}) can be obtained from Eq. (\ref{eq:FP_LT_ou}) by integrating over a small interval $[x_0-\delta,x_0+\delta]$ and then taking the limit $\delta\to 0$. From this conditions we find
\begin{equation}\label{eq:A_+_ou}
A_+=0\,,
\end{equation}
\begin{equation}\label{eq:A_-_ou}
A_-=\left[1-\frac{D_{-s/(2\alpha)}\left(-\sqrt{2\alpha/D}M\right)}{D_{-s/(2\alpha)}\left(\sqrt{2\alpha/D}M\right)}\frac{D_{-s/(2\alpha)}\left(\sqrt{2\alpha/D}x_0\right)}{D_{-s/(2\alpha)}\left(-\sqrt{2\alpha/D}x_0\right)}\right]B_-\,,
\end{equation}
\begin{eqnarray}
\label{eq:B_-_ou}
B_{-}&&=-\frac{D_{-s/(2\alpha)}\left(\sqrt{2\alpha/D}M\right)}{D_{-s/(2\alpha)}\left(-\sqrt{2\alpha/D}M\right)}\\
&\times &\frac{\frac{1}{\sqrt{2\alpha D}}e^{2\alpha x_0^2 /(4D)}D_{-s/(2\alpha)}\left(-\sqrt{2\alpha/D}x_0\right)}{D_{1-s/(2\alpha)}\left(\sqrt{2\alpha/D}x_0\right)D_{-s/(2\alpha)}\left(-\sqrt{2\alpha/D}x_0\right)+D_{1-s/(2\alpha)}\left(-\sqrt{2\alpha/D}x_0\right)D_{-s/(2\alpha)}\left(\sqrt{2\alpha/D}x_0\right)}\,,
\end{eqnarray}
\begin{equation}\label{eq:B_+_ou}
B_{+}=\frac{\frac{1}{\sqrt{2\alpha D}}e^{2\alpha x_0^2 /(4D)}D_{-s/(2\alpha)}\left(-\sqrt{2\alpha/D}x_0\right)}{D_{1-s/(2\alpha)}\left(\sqrt{2\alpha/D}x_0\right)D_{-s/(2\alpha)}\left(-\sqrt{2\alpha/D}x_0\right)+D_{1-s/(2\alpha)}\left(-\sqrt{2\alpha/D}x_0\right)D_{-s/(2\alpha)}\left(\sqrt{2\alpha/D}x_0\right)}\,.
\end{equation}
Plugging the expressions for $B_-$ and $B_+$, given in Eqs. (\ref{eq:B_-_ou}) and (\ref{eq:B_+_ou}) into Eq. (\ref{eq:G_ou}) we obtain, to leading order in $\epsilon$
\begin{eqnarray}
\label{expression_GM_2_}
&&\tilde{G}^M(M-\epsilon,s|x_0)\approx    \frac{\epsilon}{D}e^{-(M^2-x_0^2)2\alpha/(4D)}\frac{D_{-s/(2\alpha)}\left(-\sqrt{2\alpha/D}x_0\right)}{D_{-s/(2\alpha)}\left(-\sqrt{2\alpha/D}M\right)}\\ &\times &\frac{D_{1-s/(2\alpha)}\left(\sqrt{2\alpha/D}M\right)D_{-s/(2\alpha)}\left(-\sqrt{2\alpha/D}M\right)+D_{1-s/(2\alpha)}\left(-\sqrt{2\alpha/D}M\right)D_{-s/(2\alpha)}\left(\sqrt{2\alpha/D}M\right)}{D_{1-s/(2\alpha)}\left(\sqrt{2\alpha/D}x_0\right)D_{-s/(2\alpha)}\left(-\sqrt{2\alpha/D}x_0\right)+D_{1-s/(2\alpha)}\left(-\sqrt{2\alpha/D}x_0\right)D_{-s/(2\alpha)}\left(\sqrt{2\alpha/D}x_0\right)}\nonumber\,.
\end{eqnarray}
Using the relation (see Appendix \ref{app:G_p2})
\begin{equation}\label{eq:wron}
D_{1-p}(y)D_{-p}(-y)+D_{1-p}(-y)D_{-p}(y)=\frac{\sqrt{2\pi}}{\Gamma(p)}\,,
\end{equation}
the expression in Eq.~\eqref{expression_GM_2_} simplifies to
\begin{eqnarray}
\label{bw_LT_p2}
\tilde{G}^M(M-\epsilon,s|x_0)\approx    \frac{\epsilon}{D}e^{-(M^2-x_0^2)2\alpha/(4D)}\frac{D_{-s/(2\alpha)}\left(-\sqrt{2\alpha/D}x_0\right)}{D_{-s/(2\alpha)}\left(-\sqrt{2\alpha/D}M\right)}\,.
\end{eqnarray}

We now focus on the survival probability, which can be obtained by solving the backward Fokker-Planck equation \eqref{backward_FP_LT_p2}. The most general solution of Eq.~(\ref{backward_FP_LT_p2}) is given by
\begin{equation}\label{eq:Q_ou_LT}
\tilde{Q}^M(x,s)=\frac{1}{s}+e^{2\alpha x^2 /(4D)}\left[A_+ D_{-s/(2\alpha)}\left(\sqrt{\frac{2\alpha}{D}}x\right)+A_- D_{-s/(2\alpha)}\left(-\sqrt{\frac{2\alpha}{D}}x\right)\right]\,.
\end{equation}
Imposing the boundary conditions in Eqs. (\ref{absorbing_condition_bw_LT}) and (\ref{boundary_condition_bw_LT}), we obtain 
\begin{equation}\label{eq:A_-_ou_Q}
A_-=0\,,
\end{equation}
\begin{equation}\label{eq:A_+_ou_Q}
A_+=-\frac1s \frac{e^{-2\alpha M^2}}{D_{-s/(2\alpha)}\left(-\sqrt{\frac{2\alpha}{D}}M\right)}\,.
\end{equation}
Finally, using Eqs. (\ref{eq:Q_ou_LT})-(\ref{eq:A_+_ou_Q}) and expanding for small $\epsilon$, we obtain
\begin{equation}
\tilde{Q}^M(M-\epsilon,s)\approx    \frac{\epsilon}{s}\left[\frac{M2\alpha}{D}+\sqrt{\frac{2\alpha}{D}}\frac{D_{1-s/(2\alpha)}\left(-\sqrt{\frac{2\alpha}{D}}M\right)}{D_{-s/(2\alpha)}\left(-\sqrt{\frac{2\alpha}{D}}M\right)}\right]\,.
\end{equation}

\section{Derivation of Eqs. (\ref{eq:wron}), (\ref{eq:relation2}), and \eqref{relation_large_p}}

\label{app:wronskian}

\noindent{\it Derivation of Eq. (\ref{eq:wron})}

Let us define the function
\begin{equation}
W(y)=D_{1+p}(y)D_{p}(-y)+D_{1+p}(-y)D_{p}(y)\,,
\end{equation}
We will first show that $W(y)$ is constant with respect to $y$ and then we will evaluate it in the special case $y=0$. First of all, we notice that
using the following recursive equation \cite{Gradshteyn}
\begin{equation}
\frac{dD_p(z)}{dz}=\frac{z}{2}D_p(z)-D_{p+1}(z)\,,
\end{equation}
the function $W(y)$ can be rewritten as the Wronskian
\begin{equation}\label{eq:Wy}
W(y)=f(y)g'(y)-g(y)f'(y)\,,
\end{equation}
where
\begin{equation}
f(y)=D_p(y)
\end{equation}
and
\begin{equation}
g(y)=D_p(-y)
\end{equation}
are two independent solutions of the differential equation \cite{Gradshteyn}
\begin{equation}\label{eq:diff_u}
u''(y)+(p+\frac12-\frac{y^2}{4})u(y)=0\,.
\end{equation}
Thus, differentiating both sides of Eq. (\ref{eq:Wy}) with respect to $y$ and using Eq. (\ref{eq:diff_u}), we get
\begin{equation}
W'(y)=0\,.
\end{equation}

In the special case $y=0$, we obtain
\begin{equation}
W(0)=2D_{1+p}(0)D_{p}(0)\,.
\end{equation}
Using the relation \cite{Gradshteyn}
\begin{equation}
D_{p}(0)=\frac{2^{p/2}\sqrt{\pi}}{\Gamma\left(\frac{1-p}{2}\right)}\,,
\end{equation}
we obtain
\begin{equation}
W(0)=2\pi\frac{ 2^{p+1/2}}{\Gamma\left(-\frac{p}{2}\right)\Gamma\left(\frac{1-p}{2}\right)}=\frac{\sqrt{2\pi}}{\Gamma(-p)}\,.
\end{equation}
Thus, we have shown that for any $y$
\begin{equation}
D_{1+p}(y)D_{p}(-y)+D_{1+p}(-y)D_{p}(y)=\frac{\sqrt{2\pi}}{\Gamma(-p)}\,.
\end{equation}
Finally, with the change of variable $p\to-p$ we recover Eq. (\ref{eq:wron}).

\noindent{\it Derivation of Eq. (\ref{eq:relation2})}

We want to derive the following relation
\begin{equation}
\int_{-\infty}^{z}dw\,e^{-w^2 /4}D_{-p}(-w)=-\frac{1}{p}e^{-z^2/4}\left[z D_{-p}(-z)+D_{1-p}(-z)\right]\,.
\end{equation}
First, we define
\begin{equation}
u(w)=e^{-w^2/4}D_{-p}(-w)\,.
\end{equation}
It can be shown that the function $u(w)$ satisfies the following differential equation \cite{Gradshteyn}
\begin{equation}
u''(w)+wu'(w)+(1-p)u(w)=0\,,
\end{equation}
from which we obtain
\begin{equation}\label{eq:diff_u_2}
u(w)=-\frac{1}{1-p}\left(u''(w)+wu'(w)\right)\,.
\end{equation}
Integrating Eq. (\ref{eq:diff_u_2}) over $w\in (-\infty,z)$, we obtain
\begin{equation}
\int_{-\infty}^{z}dw\,u(w)=-\frac{1}{1-p}\left(u'(z)+\int_{-\infty}^{z}dw\,wu'(w)\right)\,,
\end{equation}
Integrating by parts, we find
\begin{equation}
\int_{-\infty}^{z}dw\,u(w)=-\frac{1}{1-p}\left(u'(z)+zu(z)-\int_{-\infty}^{z}dw\,u(w)\right)\,,
\end{equation}
from which we obtain 
\begin{equation}
\int_{-\infty}^{z}dw\,u(w)=\frac{1}{p}\left(u'(z)+z u(z)\right)\,.
\end{equation}
Finally, using the definition of $u(z)$, we obtain after few steps of algebra, Eq. (\ref{eq:relation2}).

\noindent{\it Derivation of Eq. (\ref{relation_large_p})}

In order to derive the asymptotic behavior in Eq.~\eqref{relation_large_p}, we use the integral representation
of $D_{-p}(-z)$ \cite{Gradshteyn}
\begin{equation}
D_{-p}(-z)= \frac{e^{-z^2/4}}{\Gamma(p)} \int_0^{\infty}dx\, e^{-\frac{x^2}{2}+ z\, x}\, x^{p-1}\, .
\label{int_trans.1}
\end{equation}
Hence the ratio is given by
\begin{equation}
\frac{D_{-(p+1)}(-z)}{D_{-p}(-z)} =  \frac{1}{p} \frac{\int_0^{\infty} dx\, e^{-\frac{x^2}{2}+ z\, x}\, 
x^p}{\int_0^{\infty} dx\, e^{-\frac{x^2}{2}+ z\, x}\, x^{p-1}} 
 =  \frac{1}{p} \frac{I_p(z)}{I_{p-1}(z)}\, ,
\label{ratio.1}
\end{equation}
where we used $\Gamma(p+1)= p\, \Gamma(p)$ and defined
\begin{equation}
I_p(z)= \int_0^{\infty}dx\,  e^{-\frac{x^2}{2}+ z\, x+ p\, \ln x}\, .
\label{Ipz.1}
\end{equation} 
Setting $z= \sqrt{p}\, u$ in \eqref{Ipz.1} gives
\begin{equation}
I_p(\sqrt{p}\, u)= \int_0^{\infty} dx\,  e^{-\frac{x^2}{2}+ u\,\sqrt{p}\, x+ p\, \ln x} \, .
\label{Ipz.2}
\end{equation}
We first rescale $x= \sqrt{p}\, y$ and re-write \eqref{Ipz.2} as
\begin{equation}
I_p(\sqrt{p}\, u)= \sqrt{p}\, e^{\frac{1}{2}\, p\,\ln p}\, \int_0^{\infty} dy\, e^{p\, 
\left[ u\, y- \frac{y^2}{2}+\ln y\right]}\, .
\label{Ipz.3}
\end{equation}
Repeating the same exercise for $I_{p-1}(\sqrt{p}\, u)$ gives
\begin{equation}
I_{p-1}(\sqrt{p}\, u) = \sqrt{p}\, e^{\frac{1}{2}\, (p-1)\, \ln p}\, \int_0^{\infty} \frac{dy}{y}\,
e^{p\, \left[ u\, y- \frac{y^2}{2}+\ln y\right]} \, .
\label{Ipz.4}
\end{equation}
Taking the ratio and substituting in \eqref{ratio.1} we get an exact relation (note that we still haven't made
the large $p$ aproximation)  
\begin{equation}
\frac{D_{-(p+1)}(-\sqrt{p}\, u)}{D_{-p}(-\sqrt{p}\, u)} = \frac{1}{\sqrt{p}}\, \frac{ \int_0^{\infty} dy\, e^{p\,
\left[ u\, y- \frac{y^2}{2}+\ln y\right]}}{\int_0^{\infty} \frac{dy}{y}\,
e^{p\,\left[ u\, y- \frac{y^2}{2}+\ln y\right]}}\, .
\label{ratio.2}
\end{equation}

We now take the large $p$ limit keeping $u$ fixed. In this limit, we can evaluate the integrals
in the numerator and the denominator of \eqref{ratio.2} by the saddle point method. For example,
in the numerator, the saddle point $y^*$ occurs at
\begin{equation}
u-y^*+\frac{1}{y^*}=0\, , \quad {\rm implying}\quad y^*= \frac{u\pm \sqrt{u^2+4}}{2}\, .
\label{saddle.1}
\end{equation}
We need to choose the $+$ sign (since the integral is over $y\ge 0$) and get
\begin{equation}
y^*=  \frac{u+ \sqrt{u^2+4}}{2}\, .
\label{saddle.2}
\end{equation}
Actually, from \eqref{ratio.2} we see that for the ratio, we do not even need to compute the 
saddle point actions explicitly (as they cancel out between the numerator and the denominator), leading to
the leading term for large $p$
\begin{equation}
\frac{D_{-(p+1)}(-\sqrt{p}\, u)}{D_{-p}(-\sqrt{p}\, u)} \approx \frac{y^*}{\sqrt{p}}= \frac{1}{\sqrt{p}}\,
\frac{u+ \sqrt{u^2+4}}{2}\, .
\label{ratio_largep.3}
\end{equation}

\section{Survival probability of Brownian motion with drift}
\label{app:surv_drift}

In this appendix, we compute the survival probability $Q^M(x,t)$ of a Brownian particle in the presence of a constant drift, defined as the probability that the process remains below position $M$ up to time $t$, having started from position $x$ at the initial time. A derivation of this quantity $Q^M(x,t)$ can be found in Ref.~\cite{Redner_book}. For completeness, we present an alternative computation based on a path-integral technique.

The Langevin equation of the process reads
\begin{equation}
\frac{dx(\tau)}{d\tau}=\mu+\eta(\tau)\,,
\end{equation}
where $\mu$ is the constant (positive or negative) drift. It is useful to consider the process $y(\tau)=M-x(\tau)$, which satisfies the Langevin equation
\begin{equation}
\frac{dy(\tau)}{d\tau}=-\mu+\eta(\tau)\,.
\end{equation}
Then the survival probability $Q^M(x,t)$ is just the probability that $y(\tau)$ is always positive for $\tau\in [0,t]$, with $y(0)= M-x_0$. Then, the survival probability can be written as
\begin{equation}
Q^M(x_0,t)=\int_{0}^{\infty}dy~\int_{y(0)=M-x_0}^{y(t)=y}\mathcal{D}y(\tau)~\exp\left[-\frac{1}{4D}\int_{0}^{t}d\tau~\left(\dot{y}+\mu\right)^2\right]\prod_{\tau=0}^{t}\theta\left[y(\tau)\right]\,,
\end{equation}
where $\int_{y(0)=M-x_0}^{y(t)=y}\mathcal{D}y(\tau)$ indicates the integral over all trajectories from $y(0)=M-x_0$ to $y(t)=y$ and $\theta(z)$ is the Heaviside theta function. The product of theta function selects only those trajectories that remain above the origin up to time $t$. This expression can be rewritten as
\begin{equation}
Q^M(x_0,t)=e^{-\mu^2 t/(4D)}\int_{0}^{\infty}dy~e^{-\mu(y-M+x_0)/(2D)}\left\{\int_{y(0)=M-x_0}^{y(t)=y}\mathcal{D}y(\tau)~\exp\left[-\frac{1}{4D}\int_{0}^{t}d\tau~\left(\dot{y}\right)^2\right]\prod_{\tau=0}^{t}\theta\left[y(\tau)\right]\right\}\,.
\end{equation}
The expression inside curly brackets is simply the constrained propagator of Brownian motion without drift, i.e., the probability that the process goes from $M-x_0$ to $y$ in time $t$, while always remaining above the origin. This quantity is easy to compute and reads \cite{Redner_book}
\begin{equation}
Q^M(x_0,t)=e^{-\mu^2 t/(4D)}\int_{0}^{\infty}dy~e^{-\mu(y-M+x_0)/(2D)}\left\{\frac{1}{\sqrt{4\pi Dt}}\left[e^{-(y-M+x_0)^2/(4Dt)}-e^{-(y+M-x_0)^2/(4Dt)}\right]\right\}\,.
\end{equation}
Computing the integral over $y$, we finally obtain
\begin{equation}
Q^M(x_0,t)=\frac{1}{2}\left[\operatorname{erfc}\left(-\frac{M-x_0-\mu t}{\sqrt{4Dt}}\right)-e^{\mu (M-x_0)/D}\operatorname{erfc}\left(\frac{M-x_0+\mu t}{\sqrt{4Dt}}\right)\right]\,.
\end{equation}

\section{Laplace inversion of Eq.~\eqref{FR_LT_der}}
\label{app:LI_2}

Using the relation
\begin{equation}
(a+b)^{\alpha}=\sum_{k=0}^{\infty} \binom{\alpha}{k}a^k b^{\alpha-k}\,,
\end{equation}
where
\begin{equation}
\binom{\alpha}{k}=\frac{(\alpha)(\alpha-1)\ldots(\alpha-k+1)}{k!}\,,
\end{equation}
we rewrite Eq.~\eqref{FR_LT_der}
\begin{eqnarray}
&&\int_{0}^{\infty}d\tilde{T}~e^{-s \tilde{T}}\langle T_1(\tilde{T})\rangle =  
 \frac{1}{4(1+s)(1+\sqrt{1+s})^2}+\frac{1}{4\sqrt{1+s}(\sqrt{1+s}-1)^2}\\ &\times &\sum_{k=0}^{\infty}\binom{-3}{k} \int_{0}^{1}du~u^{1/\sqrt{1+s}}\left[s\left(3+s/u-2\sqrt{1+s}+(\sqrt{1+s}-1)\ln(u)\right)-2(\sqrt{1+s}-1)\right]u^k s^{-3-k}\,.\nonumber
\end{eqnarray}
Computing the integral over $u$ yields
\begin{eqnarray}
\int_{0}^{\infty}d\tilde{T}~e^{-s \tilde{T}}\langle T_1(\tilde{T})\rangle =  
 \frac{1}{4(1+s)(1+\sqrt{1+s})^2}+ \frac{(1+s)(s-2+2\sqrt{1+s})}{4s^2}+\frac{1+s}{4} \sum_{k=1}^{\infty}\binom{-3}{k}\frac{1}{ s^{3+k}}G_k(1+s)\,,
 \label{LT_T1_G}
\end{eqnarray}
where 
\begin{equation}
G_k(p)=\frac{3+(4+3k)\sqrt{p}+(1+k)^2p}{4(1+k\sqrt{p})(1+(1+k)\sqrt{p})^2}\,.
\label{G_k}
\end{equation}
Moreover, for $\alpha=-3$ we find
\begin{equation}
\binom{-3}{k}=\frac{(-3)(-4)\ldots(-2-k)}{k!}=(-1)^k\frac12 (k+1)(k+2)\,.
\end{equation}

We next invert the Laplace transform in Eq.~\eqref{LT_T1_G} term by term. The first term on the right-hand side of Eq.~\eqref{LT_T1_G} can be written as
\begin{equation}
\mathcal{L}^{-1}_{s\to t}\left[\frac{1}{4(1+s)(1+\sqrt{1+s})^2}\right]=\mathcal{L}^{-1}_{s\to t}\left[\frac{2-2\sqrt{1+s}+s}{4(1+s)s^2}\right]=\frac14 \left[2t (1-\erf(\sqrt{t}))+e^{-t}-\frac{2}{\sqrt{\pi}}\Gamma\left(\frac32,t\right)\right]\,,
\label{first_term}
\end{equation}
where we performed the last Laplace inversion using Mathematica and $\Gamma(a,t) = \int_t^\infty x^{a-1}\, e^{-x}\, dx$ is the upper incomplete Gamma function. Similarly, the second term in Eq.~\eqref{LT_T1_G} can be inverted as
\begin{eqnarray}
&&\mathcal{L}^{-1}_{s\to t}\left[\frac{(1+s)(s-2+2\sqrt{1+s})}{4s^2}\right]\\
&=&\frac{1}{96}\left[-4t(2t^2+3t-18)+\frac{2}{\sqrt{\pi}}\sqrt{t}(3+16t+4t^2)e^{-t}+(-3-30t+36t^2+8t^3)\operatorname{erf}(\sqrt{t})\right]\,.\nonumber
\label{second_term}
\end{eqnarray}

It is more challenging to invert the terms with $k\geq1$ in Eq.~\eqref{LT_T1_G}. We start by computing the inverse Laplace transform with respect to $p$ of $G_k(p)$. Multiplying and dividing the right-hand side of Eq.~\eqref{G_k} by the factor $(-1+k\sqrt{p})(-1+(1+k)\sqrt{p})^2$ we rewrite $G_k(p)$ as
\begin{equation}
G_k(p)=\frac{-3 + 2 (1 + 3 k) \sqrt{
  p} + (4 + 4 k - k^2) p + (-2 - 9 k - 10 k^2 - 3 k^3) p^{
  3/2} + (-1 - 2 k - k^2) p^2 + k (1 + k)^4 p^{5/2}}{4k^2 (k+1)^4(p-1/k^2)(p-1/(k+1)^2)^2}\,.
\end{equation}
This quantity can now be inverted with Mathematica yielding
\begin{eqnarray}
h_k(t)&\equiv &\mathcal{L}^{-1}_{s\to t}\left[G_k(s+1)\right]=e^{-t}\mathcal{L}^{-1}_{p\to t}\left[G_k(p)\right]=\frac{1}{k^2}\left\{-e^{-t+t/k^2}k(1-k)^2+e^{-t}\frac{k\left[k(1+k)^3-2k^3 t\right]}{\sqrt{\pi t}(1+k)^3}\right\} \\
&+ &\frac{1}{k^2}\left[\operatorname{erf}\left(\frac{\sqrt{t}}{k}\right)e^{-t+t/k^2}(1-k)^2\right]\frac{1}{(1+k)^4}e^{-t+t/(1+k)^2}\left[(1+k)^2 (k^2-2)+2kt\right]\left[1-\operatorname{erf}\left(\frac{\sqrt{t}}{(k+1)}\right)\right]\,.\nonumber
\label{G_k_inv}
\end{eqnarray}
Thus, the terms with $k\geq 1$ in Eq.~\eqref{LT_T1_G} can be inverted by combining the Laplace inversion formulae in Eqs.~\eqref{inv_lapl_sqrt} and \eqref{G_k_inv} with convolution theorem, yielding
\begin{eqnarray}
g_k(t)& \equiv &(-1)^k\frac12 (k+1)(k+2)\mathcal{L}^{-1}_{s\to t}\left[\frac{1+s}{s^{3+k}} G_k(1+s)\right] \\
&= &(-1)^k\frac12 (k+1)(k+2)\int_{0}^{t}d\tau\,h_k(t-\tau)\tau^{k+1}\left(\frac{1}{(k+1)!}+\frac{\tau}{(k+2)!}\right)\,.
\label{g_k_LT}
\end{eqnarray}

Finally, plugging the results in Eqs.~\eqref{first_term}, \eqref{second_term}, and \eqref{g_k_LT} into Eq.~\eqref{LT_T1_G}, we obtain
\begin{equation}
\langle T_1(\tilde{T})\rangle=\tilde{T} f(\tilde{T})\,,
\end{equation}
where the scaling function $f(t)$ is given by
\begin{eqnarray} \label{foft_app}
f(t)&=&\frac{1}{96}\left[-4(2t^2+3t-18)+\frac{2}{\sqrt{\pi}}\frac{1}{\sqrt{t}}(3+16t+4t^2)e^{-t}+(-3-30t+36t^2+8t^3)\frac1t \operatorname{erf}(\sqrt{t})\right]\nonumber\\ &+&\frac{1}{2t}\left[e^{-t}-\frac{2}{\sqrt{\pi}}\Gamma\left(\frac32,t\right)\right]+\sum_{k=1}^{\infty} \frac1t g_k(t)\,,
\end{eqnarray}
where $g_k(t)$ is given in Eq.~\eqref{g_k_LT}.

\section{Computation of the constrained propagator for a single RTP in a potential $V(x)=\mu |x|$}
\label{app:prop}

In this appendix we compute the constrained propagator $G_M^{\pm}(x,t|x_0,+)$, defined as the probability that a single RTP in a potential $V(x)=\mu |x|$ starting from position $x_0$ and positive direction arrives at position $x$ with velocity $\pm v_0$ at time $t$. The Laplace transform $\tilde{G}_M^{\pm}(x,s|x_0,+)$ with respect to $t$ of this quantity satisfies the following system of ordinary differential equations (see Eq.~\eqref{LT_FP_RTP_G})
\begin{equation}
\begin{cases}
s\tilde{G}_M^{+}(x,s|x_0,+)-\delta(x-x_0)=- \partial_x\left[\left(-\mu \operatorname{sign}(x)+v_0\right)\tilde{G}_M^{+}(x,s|x_0,+)\right]-\gamma ~\tilde{G}_M^{+}(x,s|x_0,+)+\gamma ~\tilde{G}_M^{-}(x,s|x_0,+)\\
\\
s\tilde{G}_M^{-}(x,s|x_0,+)=- \partial_x\left[\left(-\mu \operatorname{sign}(x)-v_0\right)\tilde{G}_M^{-}(x,s|x_0,+)\right]-\gamma~ \tilde{G}_M^{-}(x,s|x_0,+)+\gamma~ \tilde{G}_M^{+}(x,s|x_0,+)
\end{cases}\,,
\label{LT_FP_RTP_G_APP}
\end{equation}
with boundary conditions
\begin{equation}
\begin{cases}
\tilde{G}_M^{\pm}(-\infty,s|x_0,+)=0\\
\\
\tilde{G}_M^{-}(M,s|x_0,+)=0\,.
\label{boundary_condition_RTP_APP}
\end{cases}
\end{equation}
It is useful to distinguish three cases, depending on the sign of the starting point $x_0$ and of the maximum $M$:
\begin{itemize}
\item the case where $x_0>0$ and $M>0$,
\item the case where $x_0<0$ and $M>0$,
\item the case where $x_0<0$ and $M<0$.
\end{itemize}
Note that the case where $x_0>0$ and $M<0$ is not considered because by definition $M>x_0$.
Below, we will present the details of the solution of Eq.~\eqref{LT_FP_RTP_G_APP} only in the first case, i.e., where $x_0>0$ and $M>0$. It is easy to extend our derivation to the other cases.

In the case where $M>x_0>0$, we distinguish three different regions: $-\infty<x<0$ (I), $0<x<x_0$ (II) and $x_0<x<M$ (III). Our goal is to exactly solve Eq.~\eqref{LT_FP_RTP_G_APP} separately in the three regions and then to properly match the obtained solutions. In the first region ($x<0$), Eq.~\eqref{LT_FP_RTP_G_APP} reads
\begin{eqnarray}
\begin{cases}
\left[\gamma+s+(v_0+\mu)\partial_x\right]\tilde{G}_M^{+}(x,s|x_0,+)=\gamma ~\tilde{G}_M^{-}(x,s|x_0,+)\,,\\
\\
\left[\gamma+s-(v_0-\mu)\partial_x\right]\tilde{G}_M^{-}(x,s|x_0,+)=\gamma ~\tilde{G}_M^{+}(x,s|x_0,+)\,.
\end{cases}\
\label{LT_FP_RTP_G_APP2}
\end{eqnarray}
Applying the operator $\left[\gamma+s+(v_0+\mu)\partial_x\right]$ to the second equation in \eqref{LT_FP_RTP_G_APP2} and then using the first equation, we find
\begin{equation}
\left[\gamma+s+(v_0+\mu)\partial_x\right]\left[\gamma+s-(v_0-\mu)\partial_x\right]\tilde{G}_M^{-}(x,s|x_0,+)=\gamma^2 ~\tilde{G}_M^{-}(x,s|x_0,+)\,,
\label{LT_FP_G-}
\end{equation}
which depends only on the function $\tilde{G}_M^{-}(x,s|x_0,+)$. We use the exponential ansatz $\tilde{G}_M^{-}(x,s|x_0,+)\sim e^{\lambda x}$, yielding the condition
\begin{equation}
\left[\gamma+s+(v_0+\mu)\lambda\right]\left[\gamma+s-(v_0-\mu)\lambda\right]=\gamma^2\,.
\label{lambda_eq}
\end{equation}
Hence, we obtain
\begin{equation}
\lambda_{\pm}=\frac{\mu(\gamma+s)\pm k}{v_0^2-\mu^2}\,,
\end{equation}
where 
\begin{equation}
k=\sqrt{s^2v_0^2+2sv_0^2\gamma+\gamma^2\mu^2}\,.
\label{eq:k_app}
\end{equation}
Thus, for $x<0$, we find
\begin{equation}
\tilde{G}_M^{-}(x,s|x_0,+)=A~ e^{(\mu(\gamma+s)+ k)x/(v_0^2-\mu^2)}+B~ e^{(\mu(\gamma+s)- k)x/(v_0^2-\mu^2)}\,,
\end{equation}
where $A$ and $B$ are two constants to be determined. Using the second equation in \eqref{LT_FP_RTP_G_APP2}, we obtain
\begin{equation}
\tilde{G}_M^{+}(x,s|x_0,+)=A~\frac{v_0(\gamma+s)-k}{\gamma(v_0+\mu)} e^{(\mu(\gamma+s)+ k)x/(v_0^2-\mu^2)}+B~\frac{v_0(\gamma+s)+k}{\gamma(v_0+\mu)} e^{(\mu(\gamma+s)- k)x/(v_0^2-\mu^2)}\,.
\end{equation}

Applying the same technique, it is easy to solve Eq.~\eqref{LT_FP_RTP_G_APP} in the regions II (where $0<x<x_0$) and III (where $x_0<x<M$), yielding
\begin{equation}
\tilde{G}_M^{-}(x,s|x_0,+)=
\begin{cases}
A~ e^{(\mu(\gamma+s)+ k)x/(v_0^2-\mu^2)}+B~ e^{(\mu(\gamma+s)- k)x/(v_0^2-\mu^2)}~~&\text{ for }~~ x<0\,,\\
\\
C~ e^{(-\mu(\gamma+s)+ k)x/(v_0^2-\mu^2)}+D~ e^{(-\mu(\gamma+s)- k)x/(v_0^2-\mu^2)}~~&\text{ for }~~ 0<x<x_0\,,\\
\\
E~ e^{(-\mu(\gamma+s)+ k)x/(v_0^2-\mu^2)}+F~ e^{(-\mu(\gamma+s)- k)x/(v_0^2-\mu^2)}~~&\text{ for }~~ x_0<x<M\,,\\
\\
\end{cases}
\end{equation}
and 
\begin{equation}
\tilde{G}_M^{+}(x,s|x_0,+)=
\begin{cases}
A~\frac{v_0(\gamma+s)-k}{\gamma(v_0+\mu)} e^{(\mu(\gamma+s)+ k)x/(v_0^2-\mu^2)}+B~\frac{v_0(\gamma+s)-k}{\gamma(v_0+\mu)} e^{(\mu(\gamma+s)- k)x/(v_0^2-\mu^2)}~~&\text{ for }~~ x<0\,,\\
\\
C~ \frac{v_0(\gamma+s)-k}{\gamma(v_0-\mu)}e^{(-\mu(\gamma+s)+ k)x/(v_0^2-\mu^2)}+D~\frac{v_0(\gamma+s)-k}{\gamma(v_0-\mu)} e^{(-\mu(\gamma+s)- k)x/(v_0^2-\mu^2)}~~&\text{ for }~~ 0<x<x_0\,,\\
\\
E~\frac{v_0(\gamma+s)-k}{\gamma(v_0-\mu)} e^{(-\mu(\gamma+s)+ k)x/(v_0^2-\mu^2)}+F~\frac{v_0(\gamma+s)-k}{\gamma(v_0-\mu)} e^{(-\mu(\gamma+s)- k)x/(v_0^2-\mu^2)}~~&\text{ for }~~ x_0<x<M\,,\\
\\
\end{cases}
\end{equation}
where $A,~B,~C,~D,~E~,$ and $F$ are constants to be determined.

To fix these six constants, we need to impose as many conditions. First of all, from Eq.~\eqref{LT_FP_RTP_G_APP2} we know that $\tilde{G}_M^{-}(x,t|x_0,+)$ is continuous at $x=x_0$ and thus
\begin{equation}
\tilde{G}_M^{-}(x_0^{+},t|x_0+)=\tilde{G}_M^{-}(x_0^{-},t|x_0,+)\,.
\label{cond_1}
\end{equation}
Moreover, integrating the first equation in \eqref{LT_FP_RTP_G_APP} over the interval $x\in (x_0-\epsilon,x_0+\epsilon)$ and then taking the limit $\epsilon\to 0$, we obtain the following condition for $\tilde{G}_M^{+}(x,t|x_0+)$
\begin{equation}
(\mu-v_0)\left[\tilde{G}_M^{+}(x_0^{+},t|x_0+)-\tilde{G}_M^{+}(x_0^{-},t|x_0,+)\right]=-1\,.
\label{cond_2}
\end{equation}
Similarly, integrating both equations in \eqref{LT_FP_RTP_G_APP} in a small interval around $x=0$, we obtain
\begin{equation}
(-\mu+v_0)\tilde{G}_M^{+}(x_0^{+},t|x_0+)-(\mu+v_0)\tilde{G}_M^{+}(x_0^{-},t|x_0,+)=0\,.
\label{cond_3}
\end{equation}
and
\begin{equation}
(-\mu-v_0)\tilde{G}_M^{-}(x_0^{+},t|x_0+)-(\mu-v_0)\tilde{G}_M^{-}(x_0^{-},t|x_0,+)=0\,.
\label{cond_4}
\end{equation}
Finally, also imposing the two boundary conditions in Eq.~\eqref{boundary_condition_RTP_APP}, the six constants can be exactly determined, leading to the result in Eq.~\eqref{G_RTP_A}.

\section{Computation of the survival probability for a single RTP in a potential $V(x)=\mu |x|$}
\label{app:surv}

In this appendix, we compute the survival probability $Q_M^{\pm}(x,t)$ for a single RTP in a potential $V(x)=\mu |x|$. This quantity $Q_M^{\pm}(x,t)$ is defined as the probability that the particle remains below position $M$ up to time $t$, having started from position $x$ and with direction $\pm$. The Laplace transform of this probability satisfies the following coupled differential equations (see Eq.~\eqref{FP_RTP_Q_LT})
\begin{equation}
\begin{cases}
s\tilde{Q}_M^{+}(x,s)-1= \left(-\mu \operatorname{sign}(x)+v_0\right)\partial_x\tilde{Q}_M^{+}(x,s)+\gamma ~\tilde{Q}_M^{-}(x,s)-\gamma ~\tilde{Q}_M^{+}(x,s)\,,\\
\\
s~\tilde{Q}_M^{-}(x,s)-1= \left(-\mu \operatorname{sign}(x)-v_0\right)\partial_x\tilde{Q}_M^{-}(x,s)+\gamma ~\tilde{Q}_M^{+}(x,s)-\gamma ~\tilde{Q}_M^{-}(x,s)\,,\end{cases}
\label{FP_RTP_Q_LT_APP}
\end{equation}
with boundary conditions
\begin{equation}
\begin{cases}
\tilde{Q}_M^{\pm}(-\infty,s)=1/s\\
\tilde{Q}_M^{+}(M,s)=0\,.
\end{cases}
\label{boundary_condition_RTP_LT_APP}
\end{equation}

To proceed, we distinguish two cases, depending on the sign of $M$. If $M<0$, we just need to consider the region $x<M$, where Eq.~\eqref{FP_RTP_Q_LT_APP} reads
\begin{equation}
\begin{cases}
\left[ s+\gamma-(v_0+\mu)\partial_x\right] \tilde{Q}_M^{+}(x,s)= \gamma ~\tilde{Q}_M^{-}(x,s)+1\,,\\
\\
\left[ s+\gamma+(v_0-\mu)\partial_x\right] \tilde{Q}_M^{-}(x,s)= \gamma ~\tilde{Q}_M^{+}(x,s)+1\,.\end{cases}
\label{FP_RTP_Q_LT_APP2}
\end{equation}
The solution of the differential equation above is given by the sum of two terms. The first term is a the constant solution $\tilde{Q}_M^{\pm}(x,s)=1/s$ and the second term solves the associated homogeneous equation
\begin{equation}
\begin{cases}
\left[ s+\gamma-(v_0+\mu)\partial_x\right] \tilde{Q}_M^{+}(x,s)= \gamma ~\tilde{Q}_M^{-}(x,s)\,,\\
\\
\left[ s+\gamma+(v_0-\mu)\partial_x\right] \tilde{Q}_M^{-}(x,s)= \gamma ~\tilde{Q}_M^{+}(x,s)\,.\end{cases}
\label{FP_RTP_Q_LT_APP3}
\end{equation}
We now apply the operator $\left[ s+\gamma-(v_0+\mu)\partial_x\right]$ to both sides of the second equation in \eqref{FP_RTP_Q_LT_APP3} and, using the first equation, we find
\begin{equation}
\left[ s+\gamma-(v_0+\mu)\partial_x\right] \left[ s+\gamma+(v_0-\mu)\partial_x\right] \tilde{Q}_M^{-}(x,s)= \gamma^2 ~\tilde{Q}_M^{-}(x,s)\,,
\end{equation}
which only depends on $\tilde{Q}_M^{-}(x,s)$. Using the exponential ansatz $\tilde{Q}_M^{-}(x,s)\sim e^{\lambda x}$, we obtain the following condition for $\lambda$
\begin{equation}
\left[ s+\gamma-(v_0+\mu)\gamma\right] \left[ s+\gamma+(v_0-\mu)\gamma\right] = \gamma^2\,,
\end{equation}
which has the two solutions
\begin{equation}
\lambda_{\pm}=\frac{-\mu(s+\gamma)\pm k}{v_0^2-\mu^2}\,,
\end{equation}
where
\begin{equation}
k=\sqrt{s^2v_0^2+2sv_0^2\gamma+\gamma^2\mu^2}\,.
\end{equation}
Thus, we find that the solution of Eq.~\eqref{FP_RTP_Q_LT_APP2} is
\begin{equation}
\tilde{Q}_M^{-}(x,s)=\frac{1}{s}+A ~e^{(-\mu(s+\gamma)+k)x/(v_0^2-\mu^2)}+B~ e^{(-\mu(s+\gamma)-k)x/(v_0^2-\mu^2)}\,,
\label{tilde_Q_1}
\end{equation}
where $A$ and $B$ are arbitrary constants. Using the second line of Eq.~\eqref{FP_RTP_Q_LT_APP2}, we also find that
\begin{equation}
\tilde{Q}_M^{+}(x,s)=\frac{1}{s}+A \frac{v_0(\gamma+s)+k}{\gamma(v_0+\mu)}~e^{(-\mu(s+\gamma)+k)x/(v_0^2-\mu^2)}+B\frac{v_0(\gamma+s)-k}{\gamma(v_0+\mu)}~ e^{(-\mu(s+\gamma)-k)x/(v_0^2-\mu^2)}\,.
\label{tilde_Q_2}
\end{equation}
Finally, the constants $A$ and $B$ can be fixed by using the boundary conditions in Eq.~\eqref{boundary_condition_RTP_LT_APP} and one obtains the final result in the first line of Eq.~\eqref{Q_RTP}.

In the case $M>0$ one has to solve the Eq.~\eqref{FP_RTP_Q_LT_APP} in the regions $x<0$ and $x>0$ separately and then match the two solutions. In the region $x<0$, the solution of Eq.~\eqref{FP_RTP_Q_LT_APP} is given in Eqs.~\eqref{tilde_Q_1} and \eqref{tilde_Q_2}. On the other hand, when $x>0$, it is easy to show that the most general solution of Eq.~\eqref{FP_RTP_Q_LT_APP} can be written as
\begin{equation}
\tilde{Q}_M^{-}(x,s)=\frac{1}{s}+C ~e^{(\mu(s+\gamma)+k)x/(v_0^2-\mu^2)}+D~ e^{(\mu(s+\gamma)-k)x/(v_0^2-\mu^2)}\,,
\label{tilde_Q_3}
\end{equation}
\begin{equation}
\tilde{Q}_M^{+}(x,s)=\frac{1}{s}+C \frac{v_0(\gamma+s)+k}{\gamma(v_0-\mu)}~e^{(\mu(s+\gamma)+k)x/(v_0^2-\mu^2)}+D\frac{v_0(\gamma+s)-k}{\gamma(v_0-\mu)}~ e^{(\mu(s+\gamma)-k)x/(v_0^2-\mu^2)}\,,
\label{tilde_Q_4}
\end{equation}
where $C$ and $D$ are arbitrary constants. Finally, the four constants $A$, $B$, $C$, and $D$ can be determined by imposing the boundary conditions \eqref{boundary_condition_RTP_LT_APP} and the continuity of $\tilde{Q}_M^{\pm}(x,s)$ at $x=0$. Applying these conditions, we find the result in Eq.~\eqref{Q_RTP}.


\begin{thebibliography}{10}
\bibitem{D81} B. Derrida, Phys. Rev. B {\bf 24}, 2613 (1981).

\bibitem{BBP07} G. Biroli, J-P. Bouchaud, and M. Potters, J. Stat. Mech. P07019 (2007).

\bibitem{KM00} P. L. Krapivsky and S. N. Majumdar, Phys. Rev. Lett. {\bf 85}, 5492 (2000).

\bibitem{MK02} S. N. Majumdar, and P. L. Krapivsky, Phys. Rev. E {\bf 65}, 036127 (2002).

\bibitem{MK03}  S. N. Majumdar and P. L. Krapivsky, Physica A {\bf 318}, 161 (2003).

\bibitem{SBA98} P. Sibani, M. Brandt, and P. Alstrom, 	Int. J. Mod. Phys. B {\bf 12}, 361 (1998).

\bibitem{KJ05} J. Krug and K. Jain, Phys. A: Stat. Mech. Appl. {\bf 358}, 1 (2005).

\bibitem{MP20} S. N. Majumdar, A. Pal, G. Schehr, Phys. Rep. {\bf 840}, 1 (2020).

\bibitem{Gumbel_book} E. J. Gumbel, \emph{Statistics of extremes} (Columbia university press, 1958).

\bibitem{DW80} C. Dale and R. Workman, Financ. Anal. J. {\bf 36}, 71 (1980).

\bibitem{BC04} J. Baz, and G. Chacko, \emph{Financial derivatives: pricing, applications, and mathematics.} (Cambridge University Press, 2004).

\bibitem{RM07} J. Randon-Furling and S.N. Majumdar, J. Stat. Mech. P10008 (2007).

\bibitem{MB08} S.~N. Majumdar and J.-P. Bouchaud, Quant. Fin. \textbf{8}, 753 (2008).

\bibitem{CK15} A. Clauset, M. Kogan, and S. Redner, Phys. Rev. E {\bf 91}, 062815 (2015).

\bibitem{Levy}  P. L\'evy, \emph{Sur certains processus stochastiques homog\'enes},Compos. Math. 7, 283 (1940).

\bibitem{Feller} W. Feller, \emph{Introduction to Probability Theory and Its Applications} (John Wiley \& Sons, New York, 1950).

\bibitem{SA53} E. Sparre Andersen, {\it On the fluctuations of sums of random variables}, Math. Scand. {\bf 1}, 263 (1954).

\bibitem{She79}L. A. Shepp, J.  Appl. Proba. {\bf 16}, 423 (1979).%The joint density of the maximum and its location for a Wiener process with drift.

\bibitem{B03} E. Buffet, J. Appl. Math. Stoch. Anal. \textbf{16}, 201 (2003).

\bibitem{RFM07} J. Randon-Furling and S. N. Majumdar, J. Stat. Mech. 10008 (2007).

\bibitem{MRK08} S. N. Majumdar, J. Randon-Furling, M.~J. Kearney, and M. Yor, J. Phys. A: Math. Theor. \textbf{41}, 365005 (2008).

\bibitem{SLD10} G. Schehr and P. Le Doussal, J. Stat. Mech. 01009 (2010).

\bibitem{MY10}P. M\"orters and P. Yuval, \textit{Brownian motion}, Vol. \textbf{30}, Cambridge University Press, (2010).

\bibitem{MMS19} F. Mori, S. N. Majumdar, and G. Schehr, Phys. Rev. Lett. {\bf 123}, 200201 (2019).

\bibitem{MLD20} F. Mori, P. Le Doussal, S. N. Majumdar, and G. Schehr, Phys. Rev. E {\bf 102}, 042133 (2020).

\bibitem{SP21} P. Singh and A. Pal, Phys. Rev. E {\bf 103}, 052119 (2021).

\bibitem{MMSS21}S. N. Majumdar, F. Mori, H. Schawe, and G. Schehr, Phys. Rev. E {\bf 103}, 022135 (2021).

\bibitem{DW16} M. Delorme and K. J. Wiese, Phys. Rev. E {\bf 94}, 052105 (2016).

\bibitem{SDW18} T. Sadhu, M. Delorme, and K. J. Wiese, Phys. Rev. Lett. {\bf 120}, 040603 (2018).

\bibitem{M10}  S. N. Majumdar, Physica A {\bf 389}, 4299 (2010).

\bibitem{MRZ10}  S. N. Majumdar, A. Rosso, and A. Zoia, J. Phys. A {\bf 43}, 115001 (2010).

\bibitem{S22} P. Singh, Phys. Rev. E {\bf 105}, 024113 (2022).

\bibitem{SK19}P. Singh and A. Kundu, J. Stat. Mech. 083205 (2019).%preprint arXiv:1906.09442 (2019).%Generalised Arcsine'laws for run-and-tumble particle in one dimension.

\bibitem{MLD20a} F. Mori, P. Le Doussal, S. N. Majumdar, and G. Schehr, Phys. Rev. Lett. {\bf 124}, 090603 (2020).

\bibitem{RS11} J. Rambeau and G. Schehr, Phys. Rev. E {\bf 83}, 061146 (2011).

\bibitem{RMC09} J. Randon-Furling, S. N. Majumdar, and A. Comtet, Phys. Rev. Lett. {\bf 103}, 140602 (2009). 



\bibitem{MCR10} S. N. Majumdar, A. Comtet and J. Randon-Furling, J. Stat. Phys. {\bf 138}, 955 (2010).

\bibitem{DMR13} E. Dumonteil, S. N. Majumdar, A. Rosso, and A. Zoia, Proc. Natl. Acad. Sci. U.S.A. {\bf 110}, 4239 (2013).

\bibitem{HMSS20} A. K. Hartmann, S. N. Majumdar, H. Schawe, and G. Schehr, J. Stat. Mech. 053401 (2020).

\bibitem{SKMS22} P. Singh, A. Kundu, S. N. Majumdar, H. Schawe, J. Phys. A.: Math. Theor. {\bf 55}, 225001 (2022).

\bibitem{MMS21} F.  Mori, S. N. Majumdar, and G. Schehr, Europhys. Lett. {\bf 135}, 30003 (2021).

\bibitem{D97} A. R. Dean, Science {\bf 276}, 917 (1997).%. "Thermodynamics and kinetics of a Brownian motor."

\bibitem{WFM20} J. B. Weiss, B. Fox-Kemper, D. Mandal, A. D. Nelson, and R. K. P. Zia, J. Stat. Phys. {\bf 179}, 1010 (2020).%Nonequilibrium oscillations, probability angular momentum, and the climate system.

\bibitem{jarzynski} C. Jarzynski, Phys. Rev. Lett. {\bf 78}, 2690 (1997).%Nonequilibrium equality for free energy differences. 

\bibitem{K98} J. Kurchan, J. Phys. A: Math. Gen. {\bf 31}, 3719 (1998).%Fluctuation theorem for stochastic dynamics.

\bibitem{C1999} G. E. Crooks, Phys. Rev. E {\bf 60}, 2721 (1999).%Entropy production fluctuation theorem and the nonequilibrium work relationfor free energy differences

\bibitem{seifert05} U. Seifert, Phys. Rev. Lett. {\bf 95}, 040602 (2005).%Entropy production along a stochastic trajectory and an integral fluctuation theorem.

\bibitem{seifert12} U. Seifert, Rep. Prog. Phys. {\bf 75}, 126001 (2012).%Stochastic thermodynamics, fluctuation theorems and molecular machines. 

\bibitem{HG20} J. M. Horowitz and T. R. Gingrich, Nat. Phys. {\bf 16}, 15 (2020).%Thermodynamic uncertainty relations constrain non-equilibrium fluctuations.

\bibitem{GMG18} F. S. Gnesotto, F. Mura, J. Gladrow, and C. P. Broedersz, Rep. Prog. Phys. {\bf 81}, 066601 (2018).%Broken detailed balance and non-equilibrium dynamics in living systems: a review.

\bibitem{EM_2011} M.R. Evans and S.N. Majumdar,  Phys. Rev. Lett. , {\bf 106}, 160601 (2011).

\bibitem{MMSS21b} S. N. Majumdar, P. Mounaix, S. Sabhapandit, and G. Schehr, J. Phys. A {\bf 55}, 034002 (2021).

\bibitem{DLMF} F. W. J. Olver, A. B. Olde Daalhuis, D. W. Lozier, B. I. Schneider, R. F. Boisvert, C. W. Clark, B. R. Miller, B. V. Saunders, H. S. Cohl, and M. A. McClain, NIST Digital Library of Mathematical Functions. http://dlmf.nist.gov/, Release 1.1.6 of 2022-06-30 (2022). 

\bibitem{EMS20} M .R. Evans, S. N. Majumdar, and G. Schehr, J. Phys. A: Math. Theor. {\bf 53}, 193001 (2020).

\bibitem{SKM_19} A. Dhar, A. Kundu, S. N. Majumdar, S. Sabhapandit and G. Schehr, Phys. Rev. E {\bf 99}, 032132 (2019).

\bibitem{MMS20} F. Mori, S. N. Majumdar, and G. Schehr, Phys. Rev. E {\bf 101}, 052111 (2020).

\bibitem{Gradshteyn} I. S. Gradshteyn, and I. M. Ryzhik, \emph{Table of integrals, series, and products.}, Academic press (1965).

\bibitem{SM20} S. Sabhapandit, S. N. Majumdar, Phys. Rev. Lett. {\bf 125}, 200601 (2020).

\bibitem{montanari2002optimizing} A. Montanari and R. Zecchina, Phys. Rev. Lett. {\bf 88}, 178701 (2002).

\bibitem{reuveni2014role} S. Reuveni, M. Urbakh, and J. Klafter, Proc. Natl. Acad. Sci. U.S.A. {\bf 111}, 4391 (2014).

\bibitem{MSS15} S.~N. Majumdar, S. Sabhapandit, and G. Schehr, Phys. Rev. E {\bf 91}, 052131 (2015).

\bibitem{BBPM20} B. Besga, A. Bovon, A Petrosyan, S. N. Majumdar, S. Ciliberto, Phys. Rev. Res. {\bf 2}, 032029 (R) (2020).

\bibitem{FBPC21} F. Faisant, B. Besga, A. Petrosyan, S. Ciliberto, and S. N. Majumdar, J. Stat. Mech. 113203 (2021).

\bibitem{MVB20} G. Marcado-V\'asquez, D. Boyer, S. N. Majumdar and G. Schehr, J. Stat. Mech. 113203 (2020).

\bibitem{MVB22} G. Mercado-V\'asquez, D. Boyer and S. N. Majumdar,J. Stat. Mech. 063203 (2022).

\bibitem{kusmierz2014first} L. Kusmierz, S. N. Majumdar, S. Sabhapandit, and G. Schehr, Phys. Rev. Lett. {\bf 113}, 220602 (2014).

\bibitem{kusmierz2015optimal} L. Kusmierz and E. Gudowska-Nowak, Phys. Rev. E, {\bf 92}, 052127 (2015).

\bibitem{campos2015phase} D. Campos, and V. M\'endez, Phys. Rev. E {\bf 92}, 062115 (2015).

\bibitem{evans2018run} M. R. Evans and S. N. Majumdar, J. Phys. A {\bf 51}, 475003 (2018).

\bibitem{masoliver2019telegraphic} J. Masoliver, Phys. Rev. E {\bf 99}, 012121 (2019).

\bibitem{GMS14} S. Gupta, S.~N. Majumdar, G. Schehr, Phys. Rev. Lett. {\bf 112}, 220601 (2014).

\bibitem{magoni2020ising} M. Magoni, S. N. Majumdar, and G. Schehr, Phys. Rev. Res. {\bf 2}, 033182 (2020).

\bibitem{TFPS20} O. Tal-Friedman, A. Pal, A. Sekhon, S. Reuveni, and Y. Roichman, J. Phys. Chem. Lett. {\bf 11}, 7350 (2020).

\bibitem{Redner_book}  S. Redner, A Guide to First-Passage Processes, (Cambridge University Press, 2001)

\bibitem{M05} S. N. Majumdar, Curr. Sci. {\bf 89}, 2076 (2005).

\bibitem{BMS13} A. J. Bray, S. N. Majumdar, and G. Schehr, Adv. in Phys. {\bf 62}, 225 (2013). 

\bibitem{kac1974stochastic} M. Kac, Rocky Mt. J. Math. {\bf 4}, 497 (1974).

\bibitem{stadje1987exact} W. Stadje, J. Stat. Phys. {\bf 46}, 207 (1987).

\bibitem{orsingher1990probability} E. Orsingher, Adv. Appl. Probab. {\bf 22}, 915 (1990).

\bibitem{weiss2002some} G. H. Weiss, Phys. A: Stat. Mech. Appl. {\bf 311}, 381 (2002).

\bibitem{berg2004coli} H. C. Berg, \emph{E. coli in Motion} (Springer New York, 2004).

\bibitem{bechinger2016active} C. Bechinger, R. Di Leonardo, H. Löwen, C. Reichhardt, G. Volpe, G., and G. Volpe, Rev. Mod. Phys. {\bf 88}, 045006 (2016).

\bibitem{sevilla2019stationary} F. J. Sevilla, A. V. Arzola, and E. P. Cital, Phys. Rev. E {\bf 99}, 012145 (2019).

\bibitem{dhar2019run} A. Dhar, A., Kundu, S. N. Majumdar, S. Sabhapandit, and G. Schehr, Phys. Rev. E {\bf 99}, 032132 (2019).

\bibitem{slowman2016jamming} A. B. Slowman, M. R. Evans, and R. A. Blythe, Phys. Rev. Lett. {\bf 116}, 218101 (2016).

\bibitem{metson2020jamming} M. J. Metson, M. R. Evans, and R. A. Blythe, J. Stat. Mech.: Theory Exp. 103207 (2020).

\bibitem{shen2017single} H. Shen,  L. J. Tauzin, R. Baiyasi, W. Wang, N. Moringo, B. Shuang, and C. F. Landes, Chem. Rev. {\bf 117}, 7331 (2017).

\bibitem{LH19} J. Li, J. M. Horowitz, T. R. Gingrich, and N. Fakhri, Nat. Comm. {\bf 10}, 1 (2019).%Quantifying dissipation using fluctuating currents.

\bibitem{MGK20} S. K. Manikandan, D. Gupta, and S. Krishnamurthy, Phys. Rev. Lett. {\bf 124}, 120603 (2020).%Inferring entropy production from short experiments.

\bibitem{manikandan2021quantitative} S. K. Manikandan, S. Ghosh, A. Kundu, B. Das, V. Agrawal, D. Mitra, A. Banerjee, and S. Krishnamurthy, Commun. Phys. {\bf 4}, 1 (2021).

\bibitem{roldan2021quantifying} \'E. Rold\`an, J. Barral, P. Martin, J. M. Parrondo, and F. J\:ulicher, New J. Phys. {\bf 23}, 083013 (2021).

\bibitem{otsubo2022estimating} S. Otsubo, S. K. Manikandan, T. Sagawa, and S. Krishnamurthy, Commun. Phys.  {\bf 5}, 1 (2021).

\bibitem{cugliandolo1997fluctuation} L.F. Cugliandolo,  D. S. Dean, and J. Kurchan, Physi. Rev. Lett. {\bf 79}, 2168 (1997).

\bibitem{martin2001comparison} P. Martin, P., A. J. Hudspeth, and F. Jülicher, Proc. Natl. Acad. Sci. U.S.A. {\bf 98}, 14380 (2001).

\bibitem{mizuno2007nonequilibrium} D. Mizuno, C. Tardin, C. F. Schmidt, and F. C. MacKintosh, Science {\bf 315}, 370 (2007). 

\bibitem{TFA16} H. Turlier, D. A. Fedosov, B. Audoly, T. Auth, N. S. Gov, C. Sykes, J-F. Joanny, G. Gompper, and T. Betz, Nat. Phys. {\bf 12}, 513 (2016).%Equilibrium physics breakdown reveals the active nature of red blood cell flickering.

\bibitem{ZS07} R. K. P. Zia and B. Schmittmann, J. Stat. Mech. 07012 (2007).%Probability currents as principal characteristics in the statistical mechanics of non-equilibrium steady states.

\bibitem{BBF16} C. Battle, C. P. Broedersz, N. Fakhri, V. F. Geyer, J. Howard, C. F. Schmidt, and F. C. MacKintosh, Science {\bf 352}, 604 (2016).%Broken detailed balance at mesoscopic scales in active biological systems

\bibitem{mura2018nonequilibrium} F. Mura, G. Grzegorz, and C. P. Broedersz, Phys. Rev. Lett. {\bf 121}. 038002 (2018).

\bibitem{tu2008nonequilibrium} Y. Tu, Proc. Natl. Acad. Sci. U.S.A. {\bf 105}, 11737 (2008).

\bibitem{skinner2021estimating} D. J. Skinner and J. Dunkel, Phys. Rev. Lett. {\bf 127}, 198101 (2021).

\bibitem{fodor2016far} \'E. Fodor, C. Nardini, M. E. Cates, J. Tailleur, P. Visco, and F. Van Wijland, Phys. Rev. Lett. {\bf 117}, 038103 (2016).

\bibitem{BO} L. Dabelow, S. Bo and R. Eichhorn, J. Stat. Mech. 033216 (2021).

\bibitem{OPR20} M. Onofri, G. Pozzoli, M. Radice, and R. Artuso, J. Stat. Mech. Theory Exp. 113201 2020, .

\bibitem{PRO20} G. Pozzoli, M. Radice, M. Onofri, and R. Artuso, Entropy {\bf 22}, 1431 (2020).

\end{thebibliography}
\end{document}